\documentclass[11pt]{article}
\pdfoutput=1
\usepackage[margin=1.0in]{geometry}
\usepackage{amsmath,amssymb,graphicx,bbm}
\usepackage{hyperref}
\usepackage{slashed}

\usepackage[T1]{fontenc}
\usepackage[p,osf]{baskervillef}
\usepackage[varqu,varl,var0]{inconsolata}
\usepackage[scale=.95,type1]{cabin}
\usepackage[baskerville,vvarbb]{newtxmath}
\usepackage[cal=boondoxo]{mathalfa}

\usepackage[font=small,labelfont=bf]{caption}
\usepackage{cite}
\usepackage{enumerate}

\newcommand{\be}{\begin{equation}}
\newcommand{\ee}{\end{equation}}

\usepackage{xcolor}
\usepackage{bm}
\newcommand{\nospace}{\hspace{1sp}}

\newcommand{\df}{\mathrel{:=}}
\newcommand{\dfr}{\mathrel{=:}}

\DeclareMathOperator{\Tr}{Tr}

\DeclareMathOperator{\Aut}{Aut}

\newcommand{\Sp}{\mathrm{U\!Sp}}
\newcommand{\SO}{\mathrm{SO}}
\newcommand{\Spin}{\mathrm{Spin}}
\newcommand{\SU}{\mathrm{SU}}
\newcommand{\U}{\mathrm{U}}

\newcommand{\mathdash}{\ensuremath{-\!\!-}}

\newcommand{\iu}{{\mathrm i}}
\newcommand{\E}{{\mathrm e}}

\numberwithin{equation}{section}

\title{\bf  The Weak Gravity Conjecture \\ and Emergence from an Ultraviolet Cutoff}
\author{Ben Heidenreich$^a$, Matthew Reece$^b$, Tom Rudelius$^{c}$\\
{\small \color{gray} \texttt{bheidenreich~(@perimeterinstitute.ca),~mreece~(@physics.harvard.edu),~rudelius~(@ias.edu)}}\\
{\small $^a$ Perimeter Institute for Theoretical Physics, Waterloo, Ontario, Canada N2L 2Y5}\\
{\small $^b$ Department of Physics, Harvard University, Cambridge, MA, 02138}\\
{\small $^c$ School of Natural Sciences, Institute for Advanced Study, Princeton, NJ 08540, USA}}

\begin{document}
\maketitle

\begin{abstract}
We study ultraviolet cutoffs associated with the Weak Gravity Conjecture (WGC) and Sublattice Weak Gravity Conjecture (sLWGC). There is a magnetic WGC cutoff at the energy scale $e G_N^{-1/2}$ with an associated sLWGC tower of charged particles. A more fundamental cutoff is the scale at which gravity becomes strong and field theory breaks down entirely. By clarifying the nature of the sLWGC for nonabelian gauge groups we derive a parametric upper bound on this strong gravity scale for arbitrary gauge theories. Intriguingly, we show that in theories approximately saturating the sLWGC, the scales at which loop corrections from the tower of charged particles to the gauge boson and graviton propagators become important are parametrically identical. This suggests a picture in which gauge fields emerge from the quantum gravity scale by integrating out a tower of charged matter fields. We derive a converse statement: if a gauge theory becomes strongly coupled at or below the quantum gravity scale, the WGC follows. We sketch some phenomenological consequences of the UV cutoffs we derive.
\end{abstract}

\section{Introduction}

\subsection{The Weak Gravity Conjecture} \label{subsec:introWGC}

The Weak Gravity Conjecture~\cite{Arkanihamed:2006dz} is an interesting proposal for a universal feature of all quantum gravities, and is one of the most concrete and falsifiable observations of the swampland program~\cite{Vafa:2005ui, Ooguri:2006in}. In its most minimal form, the conjecture states that in any theory of quantum gravity with a massless gauge boson there is a charged particle with charge-to-mass ratio greater than or equal to that of a large semiclassical extremal black hole.

If quantum gravities exist which violate the Weak Gravity Conjecture (WGC), they will have unusual properties. In particular, large near-extremal black holes in these theories cannot completely evaporate, but instead evolve slowly towards extremality, resulting in a tower of stable extremal black holes. However, unlike stable black hole remnants in theories with global symmetries, the mass of these stable extremal black holes increases in proportion to their charge, hence the sharpest contradictions (e.g., an infinite density of states in violation of the covariant entropy bound \cite{banks:2010zn}) do not occur, and these observations fall short of a compelling argument for the conjecture.

On the other hand, strong circumstantial evidence for the WGC comes from the absence of counterexamples in string theory, which provides many examples of consistent quantum gravities. While in some of the more involved string constructions it may be difficult to check the WGC explicitly, there are many examples, such as compactifications of the perturbative heterotic string, where the conjecture is both non-trivial and verified by explicit calculation.

As yet there is no convincing proof of the Weak Gravity Conjecture (WGC) in a general setting, although recently there have been two interesting claims to derive a version of the WGC from entropy bounds~\cite{Hod:2017uqc, Fisher:2017dbc} in specific simple gravitational effective field theories, the latter closely related to (but disagreeing with) earlier work that highlighted unusual features of loop corrections to black hole entropy in the presence of WGC-violating particles~\cite{Cottrell:2016bty,Shiu:2017toy}. (An even earlier argument based on entropy bounds was sketched in~\cite{banks:2006mm}.) Other recent work, based on detailed evidence from numerical GR, has suggested that the WGC could be a consequence of the Weak Cosmic Censorship Conjecture~\cite{Horowitz:2016ezu, Crisford:2017zpi, Crisford:2017gsb}. An incomplete sampling of other recent work related to the WGC includes \cite{rudelius:2015xta,Montero:2015ofa,Brown:2015iha,Bachlechner:2015qja,Hebecker:2015rya,Brown:2015lia,junghans:2015hba,Heidenreich:2015wga,Ibanez:2015fcv,Hebecker:2015zss,Baume:2016psm,Klaewer:2016kiy,Ooguri:2016pdq,Freivogel:2016qwc,Dolan:2017vmn,Hebecker:2017wsu,Hebecker:2017uix,Montero:2017yja,Palti:2017elp,Hebecker:2017lxm,Brennan:2017rbf,Furuuchi:2017upe}. In this paper, we will not attempt to sort out the precise status of these various arguments, but merely note that circumstantial evidence in favor of the WGC has been steadily increasing, and it is plausible that some proof of the conjecture will eventually become generally accepted. 

The WGC generalizes in a straightforward manner to $p$-form gauge symmetries and their corresponding charged $p$-branes, as well as to theories with massless scalars (see, e.g.,~\cite{Heidenreich:2015nta}) and/or with multiple massless gauge bosons. In the latter case, the conjecture states that for any rational direction in charge space there is a superextremal state in the theory (possibly a multiparticle state). Here ``rational direction'' means any ray which intersects a site in the charge lattice and ``superextremal'' means that the charge-to-mass ratio of the state is greater than or equal to that of a large semiclassical extremal black hole with a parallel charge vector $\vec{Q} \propto \vec{Q}_{\rm BH}$.\footnote{In the case of a multiparticle state, the ``mass'' is simply the sum of the particle masses (equal to the ADM energy in the limit of large spatial separation between the particles).} This is equivalent to the graphical ``convex hull condition'' (CHC) of~\cite{cheung:2014vva}.  It is important to note that the black hole extremality bound (and hence the weak gravity bound) can be modified by dilatonic couplings \cite{gibbons:1982ih,myers:1986un,gibbons:1987ps,garfinkle:1990qj}, but for the purposes of this paper we will ignore this possibility.

Since its inception, the WGC has been considered alongside a number of stronger variants of the conjecture. This is in part because it is difficult to test the minimal WGC stated above if we only have access to the low-energy effective field theory (EFT); the superextremal charged particles which satisfy the conjecture could be very heavy, and not part of the EFT. However, we must be cautious in considering stronger conjectures. While the WGC has no known counterexamples in string theory, many stronger variants proposed in the literature do~\cite{Heidenreich:2016aqi}! The earliest of these variants is the ``strong WGC'' of~\cite{Arkanihamed:2006dz}, proposed alongside the original WGC itself. For theories with a single photon, the strong WGC conjectures that the lightest charged particle is superextremal, which implies the WGC. For multiple photons, the same statement \emph{does not} imply the WGC, hence there have been various attempts to formulate a generalization which does---see, e.g.,~\cite{Brown:2015iha,Heidenreich:2015nta}---all of which imply that the lightest charged particle is superextremal. However, there are well-understood supersymmetric examples in string theory for which this is not the case~\cite{Heidenreich:2016aqi}, hence the strong WGC and all of its variants are false.

In this paper we will be primarily concerned with a strong variant of the WGC for which there is substantial evidence, the ``Sublattice WGC" (sLWGC) \cite{Heidenreich:2016aqi}.  In more than four dimensions, the sLWGC holds that in any theory of quantum gravity with massless gauge fields, there must exist a sublattice of the charge lattice (of finite index) with a superextremal particle at every site. The sLWGC has been shown to hold in toroidal orbifolds of type II and heterotic string theory, and (up to some subtleties) it follows from modular invariance in tree level string theory \cite{Heidenreich:2016aqi} (also see \cite{Montero:2016tif} for a closely analogous AdS$_3$/CFT$_2$ argument).

There are some issues which arise in interpreting this conjecture. For one, what do we mean by ``particle''? We have previously argued that multiparticle states are insufficient, because if the sLWGC is satisfied by multiparticle states it may fail to be satisfied after dimensional reduction \cite{Heidenreich:2015nta}. (Related work on the WGC in different dimensions appeared in \cite{Brown:2015iha}.) Furthermore, evidence from perturbative string theory supports the sLWGC for single particles \cite{Heidenreich:2016aqi}. On the other hand, it is clear in examples that at many lattice sites the superextremal particles required by the sLWGC are unstable, so the statement as formulated is only clearly defined at parametrically weak coupling. In this case ``particle'' can mean an unstable (though narrow) resonance.
 
Secondly, we have purposefully excluded the four dimensional case above because the conjecture, strictly as stated, cannot be true in four dimensions: there are examples of four-dimensional quantum gravities with photons coupled to massless charged particles.\footnote{For instance, this occurs in type II string theory compactified on a Calabi-Yau manifold with a conifold singularity~\cite{Strominger:1995cz}.} In such a theory, the gauge coupling of the photon runs to zero in the deep infrared, implying that parametrically large black holes can have parametrically large charge-to-mass ratios, and only massless particles are superextremal; however, there cannot be massless charged particles everywhere on a sublattice because this would imply an infinite number of massless particles. The issue is quantum in nature: for instance, in heterotic orbifolds of this type the sLWGC is naively satisfied at tree level but fails due to the one-loop running of the gauge coupling; a tower of charged resonances is still present, but the resonances are now subextremal due to the running. The same issue does not arise in higher dimensional theories as massless charged particles do not renormalize the gauge coupling to zero.\footnote{Quantum corrections may still be important in higher dimensions---the evidence from~\cite{Heidenreich:2016aqi} is entirely at tree-level, which is sufficient to establish a superextremal sublattice in cases where the particles are BPS but otherwise not---but no counterexamples to the strict statement of the conjecture are known in $D>4$.}

Nonetheless, if the sLWGC is true in higher dimensions then likely some analogous statement should hold in four dimensions, perhaps with the notion of superextremal replaced by a renormalized version. If there are no very light charged particles then we expect that the sLWGC should be satisfied up to order-one factors in the charge-to-mass ratios. Throughout the paper we will discuss four-dimensional examples on the same footing as higher dimensional ones with this assumption in mind. Cases with very light charged particles are interesting in their own right, but in this paper we consider them only very briefly in~\S\ref{subsec:conifolds}.\footnote{Note that for our purposes, the Standard Model electron is not ``very light'', because the renormalized photon coupling near the WGC scale $e M_{\rm Pl}$ differs only by an order-one amount from the infrared coupling.}

Occasionally in this paper we will refer to the ``LWGC''~\cite{Heidenreich:2015nta}, which is a criterion similar to the sLWGC but with superextremal states across the entire charge lattice. The LWGC holds in some but not all quantum gravities~\cite{Heidenreich:2016aqi}, but---at least in simple examples---theories which violate the LWGC have superextremal states at an order-one fraction of sites in the charge lattice, hence many consequences of the LWGC are robust against its violations.

\subsection{Ultraviolet cutoffs}

It has long been appreciated that the WGC has implications for the energy scales of new physics.  In particular, the ``magnetic version'' of the WGC holds that an abelian gauge theory of coupling constant $e$ should have superextremal magnetic monopoles.  Assuming that the mass of the magnetic monopole is not much less than the energy $\Lambda/e^2$ stored in its magnetic field yields \cite{Arkanihamed:2006dz}
\be
\Lambda \lesssim e M_{\rm Pl}.
\label{eq:magneticWGC}
\ee
Here $1/\Lambda$ is the radius at which the semiclassical computation of the field energy breaks down. The magnetic WGC requires new physics at or below this scale, but the nature of this new physics varies in different examples, and does not necessarily signal a breakdown in effective field theory in general.
The scale of quantum gravity---at which local quantum field theory breaks down entirely---may be much higher, and is not directly constrained by the magnetic WGC.

 For instance, in the case of a 't Hooft monopole, $\Lambda$ is roughly the scale at which the abelian gauge theory completes to a nonabelian gauge group.  Above this scale, gravity remains weakly coupled, and the nonabelian gauge theory description is valid. The sLWGC postulates a tower of particles arising at a scale of order $e M_{\rm Pl}$, strengthening the magnetic WGC argument that this is a new physics scale. Nonetheless, the particles in the tower may remain weakly coupled and be treated in an effective theory. For example, a tower of Kaluza-Klein particles can signal the breakdown of 4d effective theory but be treated within a 5d effective theory.
 
In this paper, we will see that once we impose the sLWGC, we can also make statements about a more fundamental cutoff: the quantum gravity scale where gravity is strongly coupled and QFT breaks down entirely.  We will argue that theories satisfying the sLWGC obey a nontrivial property: if we consider energy scales far up the tower of charged states, i.e., large compared to $e G_N^{-1/2}$, loop corrections imply that both gravity and the gauge theory become increasingly strongly coupled. A theory that saturates an sLWGC-like bound has the property that gravity and gauge theory become strongly coupled at {\em the same parametric energy scale}. This is a highly suggestive property, and offers the possibility of answering some of the interpretational questions about the meaning of the sLWGC. As we approach strong coupling and the charged particles become increasingly broad, it suggests that it is the density of states of different charges that must behave nicely in order that the evolving strengths of gravity and electromagnetism become strong at the same scale. It also suggests that we can think of the sLWGC as giving a sufficient condition for us to be able to think of a gauge theory as {\em emergent}: the smallness of the coupling at low energies is a consequence of the dynamics of heavy particles in the ultraviolet. This fits very comfortably with Harlow's proposal that the WGC is a property of emergent gauge fields needed to enforce factorizability of the Hilbert space in quantum gravity with multiple asymptotic boundaries \cite{Harlow:2015lma}.

The sLWGC may be thought of as saying that, in effect, all gauge theories in the context of quantum gravity share properties of Kaluza-Klein theories, with associated towers of charged particles. If we compactify a $D+1$ dimensional gravity theory on a circle of radius $R$, both the gauge theory coupling $e_{\rm KK}$ and the gravitational coupling are obtained by tree-level matching in terms of the higher-dimensional Planck scale:
\begin{align}
\frac{1}{e_{\rm KK}^2} &= \pi R^3 M_{\rm{Pl}; D+1}^{D-1}, \\
M_{\rm{Pl};D}^{D-2} &= 2\pi R M_{\rm{Pl};D+1}^{D-1},
\end{align}
with $M_{\rm{Pl}; D}$ the $D$-dimensional Planck scale. The higher-dimensional Planck scale $M_{\rm{Pl}; D+1}$ may be interpreted as the scale at which quantum gravity necessarily becomes strong, $\Lambda_{\rm QG} \lesssim M_{\rm{Pl}; D+1}$, and the matching ensures that this is well below the $D$-dimensional Planck scale. Counting Kaluza-Klein modes shows that this parametrically agrees with the ``species bound''~\cite{ArkaniHamed:2005yv,Distler:2005hi,Dimopoulos:2005ac,Dvali:2007hz,Dvali:2007wp} 
\be
\Lambda_{\rm QG}^{D-2}  \lesssim N_{\rm d.o.f.} M_{\rm{Pl};D}^{D-2},
\ee
where $N_{\rm d.o.f.}$ is the number of degrees of freedom with mass below $\Lambda_{\rm QG}$. The species bound and its gauge theory analog will play a significant role in this paper. 

For a general quantum gravity theory, such a simple tree-level matching argument may not apply. However, in theories that satisfy the sLWGC, there will always be a tower of charged particles, and these particles affect gravitational and gauge interactions through loops.
 We will see that these loop effects generically lower the scale $\Lambda_{\rm QG}$ as well as the dynamical scale of the gauge theory, and under certain assumptions they naturally match these two scales. Furthermore, there is a sort of converse statement: if gauge theory and gravity become strongly coupled at (parametrically) the same energy scale, there must be a particle satisfying the WGC (up to order one factors).

\bigskip

Our paper is organized as follows. In \S\ref{sec:warmup}, we will examine the familiar case of a U($1$) theory in 4d, showing that an LWGC-saturating tower of particles leads to a coincidence of the U($1$) Landau pole scale $\Lambda_{{\rm U}(1)}$ and the species bound scale $\Lambda_{\rm QG}$. In \S\ref{sec:loops}, we discuss the form of loop corrections to the photon and graviton propagators in a general $D$-dimensional theory and generalize the argument to $D$ dimensions; some details are left to appendix~\ref{app:moreloops}.
In \S\ref{sec:nonabelian}, we revisit our arguments for the sLWGC from \cite{Heidenreich:2016aqi} to show that for nonabelian gauge theories the WGC-obeying particles at each sublattice site should be taken to be the highest-weight state in their representation, so that different sublattice sites correspond to different representations of the nonabelian group. In \S\ref{sec:generalgroup}, we apply this newfound understanding of the nonabelian sLWGC to give a general argument for the UV cutoff scale $\Lambda_{\rm QG}$ demanded by the nonabelian sLWGC. We also show that the coincidence of strong coupling scales for gauge theory and gravity persists for arbitrary gauge groups.

In \S\ref{sec:unification}, we consider converse statements. In particular, assuming that a gauge theory becomes strong below the quantum gravity scale is sufficient to derive the original WGC. Assuming that the parametric fractional size of loop corrections to the gauge boson and graviton propagators are the same over a range of energies allows a stronger sLWGC-like statement to be derived. In this section we also consider the case of Higgsed gauge theories, clarifying some arguments from earlier literature. In  \S\ref{sec:caveats} we consider some examples of quantum gravity theories that do {\em not} fit in the paradigm we have discussed elsewhere in this paper. For instance, string theories with $g_s \ll 1$ have a Hagedorn density of states that invalidates some of our arguments. In these cases our use of simple EFT loop calculations is no longer valid, so modified arguments may carry over. In \S\ref{sec:pheno} we briefly discuss some possible phenomenological applications of our UV cutoffs to nonabelian theories with very small gauge couplings. Finally, in \S\ref{sec:conclusions} we summarize our conclusions and discuss some open questions.

\section{Warmup: Landau pole and species bound for a 4d LWGC spectrum}
\label{sec:warmup}

In theories with a tower of charged particles, both gauge interactions and gravity become strongly coupled in the ultraviolet. Let us begin with the familiar case of four dimensions, where it is well-known that charged particles lead to a Landau pole for abelian gauge theories due to the running of the coupling. At one loop, the gauge coupling $e_{\rm UV}$ at a scale $\Lambda_{\rm UV}$ is related to the low-energy gauge coupling $e$ according to:
\be
\frac{1}{e_{\rm UV}^2} = \frac{1}{e^2} - \sum_i \frac{b_i}{8\pi^2} q_i^2 \log \frac{\Lambda_{\rm UV}}{m_i}, \label{eq:Landaupole}
\ee
with $m_i$ and $q_i$ the mass and charge of the particles in the tower and $b_i$ a beta function coefficient. For gravity, the UV cutoff can be understood in terms of the ``species bound,'' which can be thought of as a result of divergent quantum corrections to the Einstein-Hilbert term cut off at $\Lambda_{\rm QG}$~\cite{ArkaniHamed:2005yv,Distler:2005hi,Dimopoulos:2005ac,Dvali:2007hz,Dvali:2007wp}
\be
M_{\rm Pl}^2 \gtrsim N_{\rm d.o.f.} \Lambda_{\rm QG}^2.
   \label{eq:4dspeciesbound}
\ee
Beyond this perturbative argument, there are various other motivations of the species bound, for instance based on demanding that semiclassical black holes of radius $\Lambda_{\rm QG}^{-1}$ not evaporate too quickly \cite[\S3.1]{Dvali:2007wp}. 

Suppose now that we have a tower of particles with masses approximately saturating the LWGC bound; that is, there is a particle of every charge $q$ with $m \sim e q M_{\rm Pl}$. Then the number of particles below a mass scale $\Lambda$ is $N(\Lambda) \geq \Lambda/(e M_{\rm Pl})$, which implies the species bound
\be
M_{\rm Pl}^2 \gtrsim N(\Lambda_{\rm QG}) \Lambda_{\rm QG}^2 \geq \frac{\Lambda_{\rm QG}}{e M_{\rm Pl}} \Lambda_{\rm QG}^2 \,,
\ee
and hence
\be
\Lambda_{\rm QG} \lesssim e^{1/3} M_{\rm Pl}.
\ee
Compare this to the gauge theory Landau pole $\Lambda_{\rm U(1)}$: if we treat the logarithms and numerical prefactors as parametrically order one,\footnote{See~\S\ref{subsec:conifolds} for a discussion of large logarithms.} and ask for the scale at which $e_{\rm UV} \to 0$ according to (\ref{eq:Landaupole}), we find
\be
\frac{1}{e^2} \sim \sum_{q = 1}^{Q} q^2 \sim Q^3,
\ee 
where $Q$ is the largest charge in the tower, $Q \sim \Lambda_{\rm U(1)}/(e M_{\rm Pl})$. Again, this leads to the conclusion
\be
\Lambda_{\rm U(1)} \sim e^{1/3} M_{\rm Pl}.  \label{eq:4dLandauBound}
\ee
Thus we see that a tower of charged particles implies UV cutoffs on both gauge theory and gravity. If the spectrum consists solely of a tower of near-extremal particles, then parametrically {\em both} the gauge theory and gravity cutoffs are at the scale $e^{1/3} M_{\rm Pl}$. We can think of this, loosely speaking, as a form of ``gauge-gravity unification.'' We do not mean that gravity and gauge theory are unified in the same way that different gauge groups are unified in GUTs, but simply that we can think of the weakness of the two forces as having emerged in the infrared from integrating out a tower of states starting at a common scale $\Lambda_{\rm QG}$ in which all kinetic terms have their naive, order-one size (in appropriate units).

It is straightforward to generalize the above argument to the case where the LWGC is violated but the sLWGC is satisfied on a sublattice of index $k>1$. In this case, we obtain
\be
\Lambda_{\rm QG}, \Lambda_{\rm U(1)} \lesssim (k e)^{1/3} M_{\rm Pl},
\ee
with the two scales again parametrically the same when the spectrum is dominated by near-extremal particles. In string theory examples $k$ cannot be parametrically large, thus---at least in these cases---the consequences of the sLWGC are similar to those of the LWGC.

If the spectrum differs greatly from our assumptions---for instance, if there are many more neutral particles that enter in the species bound but do not affect the running of the gauge theory---then the sLWGC does not necessarily imply gauge-gravity unification. However, the sLWGC {\em always} implies a cutoff on the quantum gravity scale that goes to zero as $e \to 0$.

\section{Loops and UV cutoffs for gauge theory and gravity in $D$ dimensions}
\label{sec:loops}

The discussion in the previous section focused on the familiar case of four dimensions, where the Landau pole and species bound arguments for UV cutoffs are familiar. Similar results hold in a general $D$-dimensional theory, where both gauge theory and gravity are generically non-renormalizable. If many particles run in loops, the loop expansion can break down at prematurely low scales. To explain this point it is useful to adopt a somewhat different language that is suitable for both gauge theory and gravity.

\subsection{Growth of amplitudes with energy} \label{sec:ampgrowth}

From the kinetic terms for gauge theory and gravity,
\be
S = \int d^D x \sqrt{-g} \left( \frac{1}{16\pi G_N} {\cal R} - \frac{1}{4e^2} F_{\mu \nu}F^{\mu \nu}\right),
\ee
we read off that $G_N$ has mass dimension $2-D$ and $e^2$ has mass dimension $4-D$. We customarily introduce a reduced Planck mass $M_{\rm Pl}^{D-2} = 1/(8 \pi G_N)$, and we could likewise introduce a mass scale associated with the gauge theory, $M_{\rm U(1)}^{D-4} = 1/e^2$. We might guess that in $D > 4$, where both gravity and gauge theory are nonrenormalizable, effective field theory breaks down at the scale $M_{\rm Pl}$ or $M_{\rm U(1)}$. However, in a theory with a large number of degrees of freedom $N$ we know that this naive dimensional analysis can be modified by powers of $N$.

In four dimensions we discussed the gauge theory cutoff in terms of a logarithmically running coupling constant. In higher dimensions, we should be more cautious. For $p^2 \ll m_i^2$ we can integrate out a heavy particle $i$, expanding in powers of $p^2$ to obtain a threshold correction to $1/e^2$ together with an infinite sum of higher-derivative operators. For $p^2 \gg m_i^2$, however, the result is that the size of the loop correction grows with momentum---and faster than logarithmically, when $D > 4$. Although this may sometimes be referred to as a ``power-law running'' of the coupling, there is no straightforward sense in which the momentum dependence of loops can be absorbed in a running coupling in a process-independent manner in a general non-renormalizable field theory \cite{Anber:2011ut}.

Nonetheless, the lack of a well-defined renormalized coupling does not prevent us from estimating the energy scale at which loop amplitudes become large and perturbation theory breaks down. Consider, for example, the two-point function of the photon.
The sum of iterated 1PI loop corrections to the photon propagator, with the leading one-loop 1PI graph, has the form
\be
\left<{\tilde A}_\mu(p){\tilde A}_\nu(-p)\right> = \frac{\eta^{\mu \nu} - p^\mu p^\nu/p^2}{p^2+\iu \epsilon} \frac{1}{1 + \Pi(p^2)}.
\ee
The function $\Pi(p^2)$ can be read off from the standard one-loop QED vacuum polarization calculation.
For example, for a set of charged scalars of charge $q_i$ and mass $m_i$, we compute
\be
\Pi_{\rm unreg}(p^2) = \frac{2e^2}{(4\pi)^{D/2}} \Gamma(2-D/2) \int_0^1 dx\, x(2x-1)\sum_i q_i^2 \left[m_i^2 - p^2 x (1-x)\right]^{D/2-2}. 
\label{eq:vacuumpol}
\ee
For $D$ odd this expression is finite as written, while for $D$ even we can use dimensional regularization $D \to D - \epsilon$ to see that it contains additional logarithmic dependence on $p^2$. The divergent piece in even dimensions can be absorbed by counterterms (including certain higher-derivative operators in $D>4$). By rescaling the photon field we impose the renormalization condition $\Pi(0) = 0$.\footnote{With this renormalization condition, the gauge coupling which appears in (\ref{eq:vacuumpol}) is the infrared gauge coupling. This condition cannot be imposed in $D=4$ with massless charged particles due to infrared divergences, which corresponds to the fact that the gauge coupling flows to zero in the deep infrared; this special case is discussed further in~\S\ref{subsec:conifolds}.} If all the charged scalars are light, $m_i^2 \lesssim p^2$, then we can estimate
\be
|\Pi(p^2)| \sim e^2 p^{D-4} \sum_i q_i^2 \,,
\ee
up to order-one factors and logarithms.

The lesson from this is that loop amplitudes grow with momentum $p$ at a rate that depends on both the spacetime dimension and the spectrum of charged particles with $m \lesssim p$. In the above example, strong coupling arises when $|\Pi(p^2)| \sim 1$. For a tower of approximately LWGC-saturating particles, that is a tower for which $m^2 \sim e^2 q^2 M_{\rm Pl}^{D-2}$, if we sum up to energy $p$ we reach a maximum charge $Q \sim p/(e M_{\rm Pl}^{(D-2)/2})$ and find
\be
|\Pi(p^2)| \sim e^2 p^{D-4} \sum_{q = 1}^{Q} q^2 \sim e^2 \left(\frac{p}{e M_{\rm Pl}^{(D-2)/2}}\right)^3 p^{D-4}.
\ee
We denote the scale $p$ at which this becomes order one by $\Lambda_{\rm U(1)}$, given by:
\be
\Lambda_{\rm U(1)} \sim \left(e M_{\rm Pl}^{3(D-2)/2}\right)^{1/(D-1)}. \label{eq:DdLandauBound}
\ee
This is the $D$-dimensional analogue of the 4d Landau pole bound (\ref{eq:4dLandauBound}). Notice that we are {\em not} saying that the gauge coupling becomes strong at this scale, since the meaning of a running gauge coupling is unclear away from four dimensions. Rather, we are saying that the loop expansion breaks down at this scale because the large number of degrees of freedom causes amplitudes to become large at energies well below $M_{\rm Pl}$. 

The above calculation is rather naive, and there are several possible objections one might raise. Firstly, the contribution to $\Pi(p^2)$ that we have computed is not the only one, or even necessarily the largest one. In the spirit of Wilsonian effective field theory, we should include higher-dimensional operators (including higher-derivative operators) suppressed by the cutoff $\Lambda_{\rm U(1)}$.\footnote{Some of these operators appear as counterterms in even dimension $D>4$, in which case the remaining finite contribution is what we consider here.}
This will generate power-law corrections of the form $\Pi(p^2) \to \Pi(p^2)+ p^2 / \Lambda_{\rm U(1)}^2 + \ldots$, but the contribution of the light particles from above is of the form $p^{D-1} / \Lambda_{\rm U(1)}^{D-1}$ which is subleading to $p^2 / \Lambda_{\rm U(1)}^2$ for $p \ll \Lambda_{\rm U(1)}$ and $D\ge 4$.

In fact, this is far less problematic than it sounds. We are mainly interested in estimating an \emph{upper bound} on the cutoff $\Lambda_{\rm U(1)}$. It is always possible that higher dimensional operators appear at a lower scale and ruin the effective field theory, but even if they do not, the light charged states will eventually cause the loop expansion to break down. This ``highest possible cutoff'' due to the light spectrum is what we are attempting to estimate.

Similarly, we have neglected charged particles with masses between $p$ and the cutoff. In appendix~\ref{app:heavyloops} we estimate their contribution, which turns out to be roughly of the form $\Pi(p^2) \to \Pi(p^2) + p^2 / \Lambda_{\rm U(1)}^2$ in the above example, once we have summed of the entire LWGC-saturating heavy spectrum. While this is again larger than the contribution of light charged particles (and similar in form to corrections from higher derivative operators), it doesn't parametrically change the scale at which the loop expansion breaks down: as we approach the scale at which $|\Pi(p^2)|$ becomes order-one the heavy-particle contribution starts to go away for the simple reason that we are approaching the cutoff, hence there are not many particles with $p \lesssim m \lesssim \Lambda$. For this reason, neglecting heavy particles will never change our estimate of where the loop expansion breaks down.

A second objection is that we have only considered the photon two-point function at one-loop, which is moreover an off-shell quantity (meaning that it may not be well-defined outside of the effective field theory description). In appendix~\ref{app:higherloops} we briefly discuss higher loop diagrams and on-shell S-matrix elements, arguing that the loop expansion breaks down at parametrically the same scale as above.

To capture the heuristics discussed above, we find it convenient to define
\be
\lambda_{\rm gauge}(E) \df e^2 E^{D-4} \sum_{i:\, m_i < E} I(i) \,,  \label{eq:lambdagauge}
\ee
where $I(i)$ is the Dynkin index of the representation of particle $i$ (simply $q_i^2$ in the U($1$) case). Here we have purposefully thrown out all numerical factors (which are process dependent), neglected logarithmic factors, and assumed no significant degree of cancellation between the terms. $\lambda_{\rm gauge}(E)$ estimates the contribution of particles with mass $m < E$ to the size of loop corrections. Effective field theory breaks down due to loops of light particles when $\lambda_{\rm gauge} \sim 1$, unless it breaks down at a lower energy scale for other reasons.

A similar analysis applies to gravity. Although the species bound is often phrased in terms of loop corrections to the Planck scale, the relevant aspect is not so much threshold corrections per se as the growth of typical scattering amplitudes with energy. A similar calculation of the off-shell graviton propagator can be carried out to see this growth explicitly in a class of diagrams, but again we will capture the parametric dependence with a simple function
\be
\lambda_{\rm grav}(E) \df G_N E^{D-2} \sum_{i:\, m_i < E} \dim(R_i),  \label{eq:lambdagrav}
\ee
where $\dim(R_i)$ is the dimension of the gauge representation of particle $i$ (that is, the total number of degrees of freedom in the multiplet). We see immediately that the condition $\lambda_{\rm grav}(E) \lesssim 1$ reproduces the familiar species bound:
\be
M_{\rm Pl}^{D-2} \geq N_{\rm d.o.f.} \Lambda_{\rm QG}^{D-2}.
   \label{eq:speciesbound}
\ee

\subsection{U($1$) gauge theory in $D$ dimensions}
\label{sec:U1example}

We now revisit the example of \S\ref{sec:warmup} in a general $D$ dimensional theory satisfying the sLWGC. We will choose a sublattice with spacing $k$ so that for each natural number $n$ there should exist a particle of charge $kn$ and mass
\be
G_N m^2 \leq c_{\rm WGC} e^2 k^2 n^2,
\label{eq:WGC}
\ee
with $c_{\rm WGC}$ a fixed order-one number determined by the extremal black hole solutions in the low energy effective theory. Below we will systematically neglect this and other order-one factors. 

To begin, we consider the effect of these particles on the strong coupling scale of gravity. There is a tower of particles up to charge $k n_{\rm max}$ with masses below the cutoff $\Lambda_{\rm QG}$, with
\be
n_{\rm max} \sim \frac{\Lambda_{\rm QG}}{e k M_{\rm Pl}^{(D-2)/2}}. \label{eq:towerN}
\ee
Below the scale $\Lambda_{\rm QG}$, there are therefore $N_{\rm d.o.f.} \sim n_{\rm max}$ such particles with mass below $ \Lambda_{\rm QG}$.  Plugging into (\ref{eq:speciesbound}) and solving for $\Lambda_{\rm QG}$, we find
\be
{\rm U}(1): \quad \Lambda_{\rm QG} \lesssim (e k)^{\frac{1}{D-1}} M_{\rm Pl}^{\frac{3(D-2)}{2(D-1)}},   \label{eq:LambdaQGU1}
\ee
as we previously derived in \cite{Heidenreich:2016aqi}. For the case $D = 4$, this reads $\Lambda_{\rm QG} \lesssim (ek)^{1/3} M_{\rm Pl}$.

Next, consider the Landau pole of the U($1$) gauge theory. In this case, we have 
\be
\lambda_{\rm gauge}(E) \sim e^2 E^{D-4} \sum_{n=1}^{n_{\rm max}} (k n)^2 \sim e^2 k^2 E^{D-4} \left(\frac{E}{e k M_{\rm Pl}^{(D-2)/2}}\right)^3. \label{eq:lambdagaugeU1simple}
\ee
In the second step we used equation (\ref{eq:towerN}). We see then that the condition $\lambda_{\rm gauge}(\Lambda) \sim 1$ leads to $\Lambda$ parametrically matching the scale $\Lambda_{\rm QG}$ in (\ref{eq:LambdaQGU1}). Notice that in this analysis we have ignored various constant factors as well as logarithms. We have also assumed a tower of approximately equally-spaced states. We have also assumed that the density of states of a given mass is dominated by these charged states; if this assumption is violated, e.g., if there is a large number of light uncharged states, then the quantum gravity scale $\Lambda_{\rm QG}$ could be significantly below the scale of the Landau pole.  Under these assumptions, and as a {\em parametric} statement (rather than a precise quantitative one), this suggests that theories that approximately saturate the sLWGC bounds will tend to exhibit the phenomenon of gauge-gravity unification as defined in \S\ref{sec:warmup}.

\section{The Nonabelian sLWGC}
\label{sec:nonabelian}

The sLWGC can be applied to commuting generators within a nonabelian gauge group. In~\cite{Heidenreich:2015nta,Heidenreich:2016jrl,Heidenreich:2016aqi} we have essentially ignored the nonabelian nature of the group and discussed the (s)LWGC as a statement about the Cartan generators. This can be motivated, for example, by compactifying the theory on a circle with a Wilson line that breaks the gauge group to a product of U($1$)s. In this way, much of the evidence we have found for the sLWGC in string theory applies to nonabelian groups.

\begin{figure}[!h]\begin{center}
\includegraphics[width=0.35\textwidth]{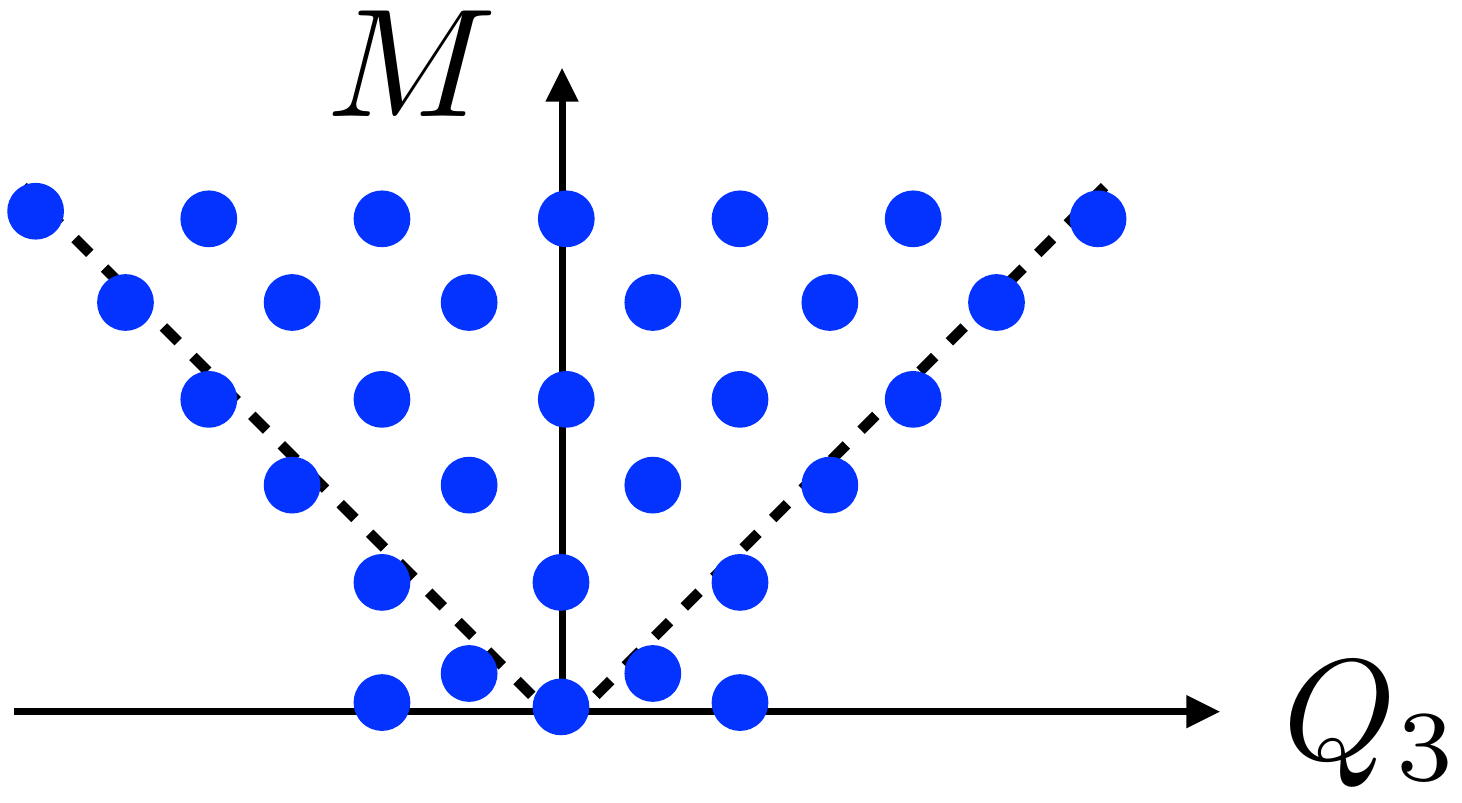}
\hspace*{2cm}
\includegraphics[width=0.35\textwidth]{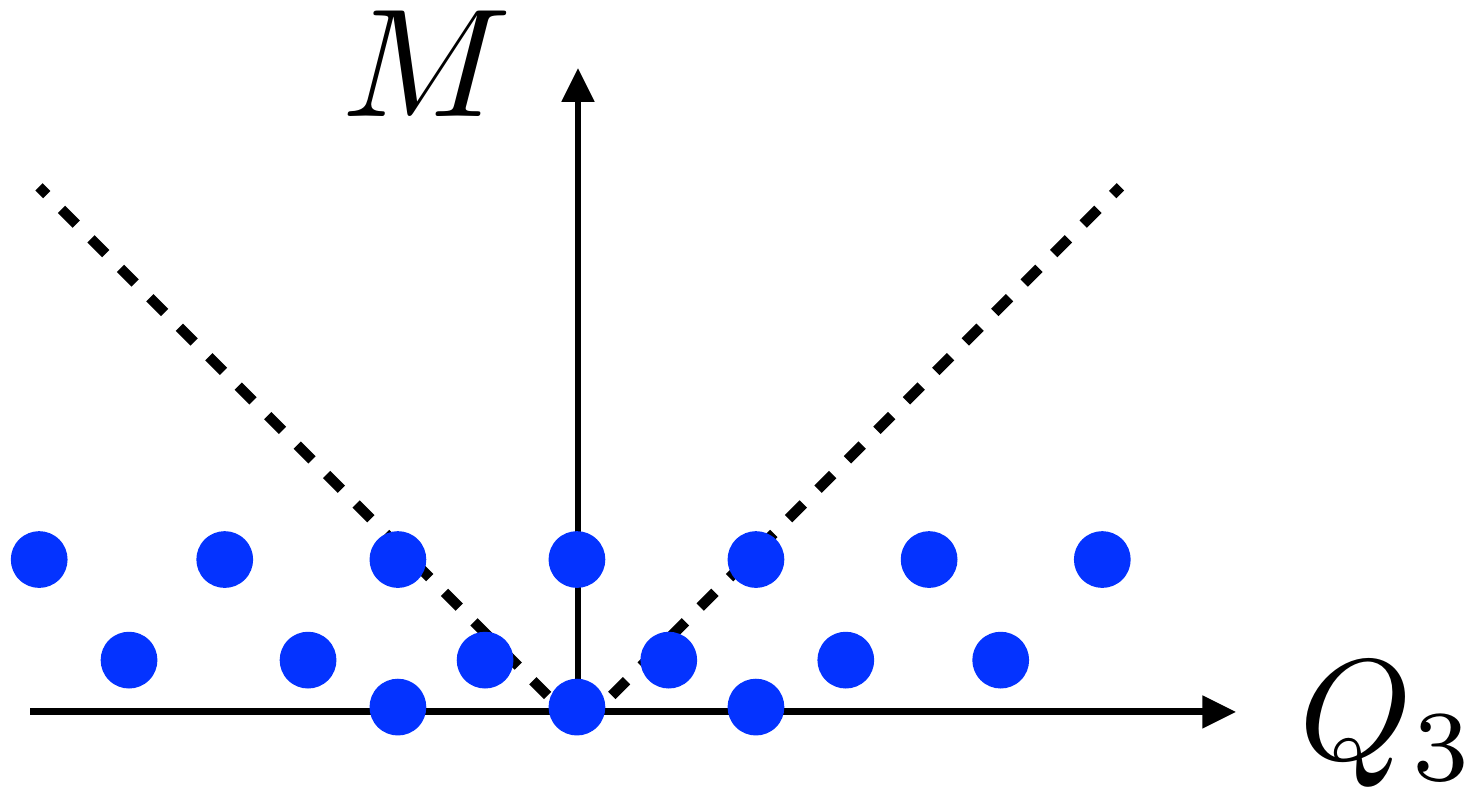}
\end{center}
\caption{The nonabelian sLWGC (left) and abelian sLWGC (right) for an $\SU(2)$ gauge group.  For a sublattice of fixed index, the nonabelian sLWGC requires many more particles charged under the $\U(1)$ Cartan below a given mass scale than the abelian sLWGC does, as the latter can be satisfied by particles charged under a sparse set of representations, provided they are sufficiently light.}
\label{fig:nonabeliansLWGC}
\end{figure}

However, we would like to postulate a slightly stronger statement: there should exist superextremal particles in each {\em representation} of the group rather than simply with each Cartan charge. For instance, in an SU($2$) gauge theory there is a lattice site with Cartan charge $1$, but states of this charge exist in all representations with positive integer spin. We would postulate that the correct nonabelian sLWGC cannot be satisfied at this lattice site with representations of higher spin, but requires a spin $1$ representation with superextremal particles of charge $1$, as shown in figure \ref{fig:nonabeliansLWGC}. This stronger statement is satisfied, for instance, in the $SO(32)$ and $E_8 \times E_8$ heterotic string theories.

To make a precise conjecture, we first review some basic facts about a compact nonabelian Lie group $G$. Let
$\Phi$ denote the set of roots of $G$, each designated by a weight vector
$\vec{Q} \in \Phi$, i.e., a set of Cartan charges. We choose a set of
positive roots $\Phi^+$ such that for any root $\vec{Q}$, $\vec{Q} \in \Phi^+$ iff $- \vec{Q} \notin
\Phi^+$ and for $\vec{Q}_{1, 2} \in \Phi^+$, $\vec{Q}_1 + \vec{Q}_2$ is in
$\Phi^+$ if it is a root. Simple roots are positive roots which are not the
sum of two other positive roots. The simple roots are linearly independent and
span the space of roots, hence the number of simple roots equals the rank of
$G$ minus the rank of its center $Z (G)$.

Given a set of positive roots, there is a partial ordering on weights with
$\vec{Q}_1 \geqslant \vec{Q}_2$ if $\vec{Q}_1 - \vec{Q}_2$ is a non-negative
linear combination of positive (equivalently, simple) roots. The highest
weight $\vec{Q}_R$ of representation $R$ (if it it exists) is the unique
weight which satisfies $\vec{Q}_R \geqslant \vec{Q}$ for all weights $\vec{Q}$
in $R$. A weight $\vec{Q}$ is dominant if $\vec{Q} \cdot \vec{Q}_{\alpha}
\geqslant 0$ for all simple roots $\vec{Q}_{\alpha}$.\footnote{Here the Cartan generators are normalized so that within each simple subalgebra $\Tr H_i H_j \propto \delta_{i j}$ in any representation. This doesn't complete fix the inner product, which encodes additional information in the worldsheet argument to follow.} An irreducible
representation (irrep) $R$ has a highest weight $\vec{Q}_R$ which is
dominant. Moreover, for any dominant $\vec{Q}$ in the weight
lattice\footnote{The weight lattice is the set of all possible weights in
finite dimensional unitary representations of $G$, which form a lattice since
$G$ is compact. While all weights $\vec{Q}$ must be algebraically integral,
i.e.,
$
  2 \vec{Q}_{\alpha} \cdot \vec{Q} / \vec{Q}_{\alpha}^2 \in \mathbbm{Z} \,,
$
only for $G$ simply connected does this completely fix the weight lattice.
Otherwise there are further conditions; for instance, there are no spin-$1 /
2$ representations of $\SO(3)$, only of its simply connected double
cover $\SU(2)$. Moreover, when $Z (G)$ has non-zero rank the abelian
charges are quantized as well. The weight lattice need not factor between the
abelian and semi-simple components of the Lie algebra, as demonstrated by,
e.g., $G = \U(N) = (\SU(N) \times \U(1)) /\mathbbm{Z}_N$, for which the
$\U(1)$ charge mod $N$ is fixed to be equal to the charge under the
$\mathbbm{Z}_N$ center of $\SU(N)$.} $\Gamma_G$ of $G$ there is a
unique irrep $R$ of $G$ and all finite dimensional representations of $G$ are
direct sums of these.

The hyperplanes orthogonal to the roots divide the space of weights into Weyl
chambers, which are permuted by the Weyl group (acting freely and
transitively on them). The dominant weights lie in a Weyl chamber $\vec{Q}
\cdot \vec{Q}_{\alpha} \geqslant 0$, known as the fundamental Weyl chamber. A
choice of positive roots is equivalent to a choice of fundamental Weyl
chamber.

We can now state the nonabelian version of the sublattice Weak Gravity
Conjecture:\\

\noindent
{\bf The nonabelian sLWGC:} For any quantum gravity in $D \geqslant 5$ with zero cosmological constant
  and unbroken gauge group $G$, there is a finite-index Weyl-invariant
  sublattice $\Gamma_0$ of the weight lattice $\Gamma_G$ such that for every
  dominant weight $\vec{Q}_R \in \Gamma_0$ there is a superextremal resonance
  transforming in the $G$ irrep $R$ with highest weight $\vec{Q}_R$.\\

\noindent
Here ``Weyl-invariant'' means that $\Gamma_0$ is invariant under the action of
the Weyl group $W (G)$, which ensures that the conjecture is independent of
the choice of fundamental Weyl chamber. ``Superextremal'' means the same as in
the abelian case: the charge-to-mass ratio of the resonance is greater than or
equal to that of a large extremal black hole with a parallel weight vector.
Note that in the special case where $G$ is abelian this reduces to the
abelian sLWGC of \cite{Heidenreich:2016aqi}; otherwise the nonabelian sLWGC is strictly stronger.

We will make two arguments in favor of this conjecture. First, we show that it
holds in the NSNS sector of tree-level string theory (with caveats similar to
those for the abelian sLWGC). Second, we show that it is preserved upon
compactification of a higher dimensional theory which
satisfies the conjecture on a Ricci flat manifold. Based on these arguments, we conclude that the evidence for the nonabelian sLWGC is similar to that for the abelian sLWGC of~\cite{Heidenreich:2016aqi}.

NSNS sector gauge bosons correspond to worldsheet conserved currents, with
OPEs
\begin{align}
  J^a (z) J^b (0) &\sim \frac{k^{a b}}{z^2} + \frac{i c^{a b}_{\quad c}}{z} J^c
  (0) + \ldots \,, &  J^{\tilde{a}} (\bar{z}) J^{\tilde{b}} (0) &\sim
  \frac{\tilde{k}^{\tilde{a}  \tilde{b}}}{\bar{z}^2} + \frac{i
  \tilde{c}^{\tilde{a}  \tilde{b}}_{\quad \tilde{c}}}{\bar{z}} J^{\tilde{c}}
  (0) + \ldots \,,
\end{align}
corresponding to the Kac-Moody algebra. We fix $k^{a b} = \delta^{a b}$ and
$\tilde{k}^{\tilde{a}  \tilde{b}} = \delta^{\tilde{a}  \tilde{b}}$ by
normalizing the currents. The $c^{a b}_{\quad c}$ and $\tilde{c}^{\tilde{a} 
\tilde{b}}_{\quad \tilde{c}}$ are then structure constants, with
normalizations depending on the level for the nonabelian current algebra of
each simple factor of $G$. Note in particular that each simple factor of $G$
is either purely left-moving or purely right-moving (though the weight lattice
need not factor between left and right movers), hence simple roots have either
$\vec{Q}_L = 0$ or $\vec{Q}_R = 0$.

We introduce a chemical potential for the Cartan, as in the abelian case:
\begin{equation}
  Z = \Tr (q^{\Delta_L}  \bar{q}^{\Delta_R} y^{Q_L}  \bar{y}^{Q_R}) \,.
\end{equation}
By the same arguments as before, the spectrum is invariant under $Q
\rightarrow Q + \rho$ for $\rho \in \Gamma^{\ast}_Q$ with $T_{L, R}$, defined
by
\begin{equation}
  \Delta_{L, R} = T_{L, R} + \frac{1}{2} Q_{L, R}^2 \,,
\end{equation}
held fixed. The norm $Q^2 = Q_L^2 - Q_R^2$ is invariant under the Weyl group,
as is the weight lattice $\Gamma_Q$, hence so is $\Gamma_Q^{\ast}$. Starting
with the graviton state $\Delta_L = \Delta_R = 0$ and $Q_{L, R} = 0$, we
produce a state with $\Delta_{L, R} = \frac{1}{2} Q_{L, R}^2$ for every $Q \in
\Gamma_Q^{\ast}$.

To show that this state is the highest weight in its $G$ irrep, we proceed by
contradiction. If not, there is at least one simple root $\vec{Q}_+$ (a
left-mover for definiteness) such that there is a state with charge $(\vec{Q}_L
+ \vec{Q}_+, \vec{Q}_R)$ and $\Delta_L = \frac{1}{2} Q_L^2 = \frac{1}{2} 
(\vec{Q}_L + \vec{Q}_+)^2 - \vec{Q}_L \cdot \vec{Q}_+ - \frac{1}{2} 
\vec{Q}_+^2$ and $\Delta_R = \frac{1}{2} Q_R^2$. Thus, there is a
corresponding state with charge $(\vec{Q}_+, 0)$ and $\Delta_L = - \vec{Q}_L
\cdot \vec{Q}_+$ and $\Delta_R = 0$, where $\Delta_L$ is a non-positive
integer because $Q$ is dominant and lies in $\Gamma_Q^{\ast}$. Since the
graviton has spin $2$, this state also has spin 2, but then either (1)
$\Delta_L = 0$ and there are additional (charged) massless spin 2 particles,
or (2) $\Delta_L < 0$, and by turning on left-moving oscillators in the
non-compact directions we obtain massless particles of spin greater than 2. In
either case, the low energy limit is not Einstein gravity.\footnote{A similar
argument can be made for a unitary CFT$_2$, relevant for the AdS$_3$ WGC of \cite{Montero:2016tif}. In this case, there is a \emph{unique} operator (the identity)
with $\Delta_L = -c_L/24$ and $\Delta_R = -c_R/24$. All other operators have $\Delta_L > -c_L/24$
or $\Delta_R >  -c_R/24$. If the charge $(\vec{Q}_+, 0)$ state constructed above has
$\Delta_L = -c_L/24$ it contradicts the uniqueness of the identity operator (which
is uncharged), whereas if it has $\Delta_L < -c_L/24$ it violates unitarity.}

The rest of the argument for the nonabelian sLWGC from modular invariance goes
the same way as in the abelian case, with the same caveats as in~\cite{Heidenreich:2016aqi}.

We now consider compactification. As in the abelian case, the non-trivial
ingredients that can lift some of the KK modes are Wilson lines on torsion
cycles.\footnote{Wilson lines on non-torsion cycles are moduli whose values affect the masses of KK modes, but do not make them subextremal, since they affect the masses of extremal black holes in exactly the same way.} These can be viewed as a coming from a quotient
\begin{equation}
  \frac{G \times \hat{M}}{G_0} \,,
\end{equation}
where $M = \hat{M} / G_0$ is a Ricci-flat compact manifold, $G_0$ acts freely
and transitively on $\hat{M}$, and $G_0$ is a finite subgroup of $G$. To leave $G$
unbroken, we need $G_0 \subseteq Z (G)$ (otherwise, replace $G$ with its
unbroken subgroup in the following argument). If $\Gamma_0 \subseteq \Gamma_G$ is an extremal sublattice before
compactification then the intersection of $\Gamma_1 \df \Gamma_0 \times
\Gamma_{\rm KK}$ with the $G_0$-invariant sublattice $\Gamma_2 \subseteq
\Gamma_G \times \Gamma_{\rm KK}$ is an extremal sublattice after
compactification. Note that $\Gamma_{1, 2}$ are full dimensional sublattices,
hence so is $\Gamma_1 \cap \Gamma_2$ (each has finite coarseness). $\Gamma_1$
is Weyl-invariant by assumption, whereas since $G_0 \subset Z (G)$ is
Weyl-invariant, $\Gamma_2$ is also. Thus $\Gamma_0' \df \Gamma_1 \cap
\Gamma_2$ satisfies the nonabelian sLWGC in the lower-dimensional theory.

Note that the condition that $\Gamma_0$ is a Weyl-invariant sublattice of the weight lattice $\Gamma_G$ turns out to be rather restrictive. For instance, when $G$ is simple $\Gamma_0$ must be a multiple of one of a finite list of ``primitive'' Weyl-invariant sublattices, see appendix~\ref{app:WeylInv}.

\section{UV cutoffs for general gauge groups}
\label{sec:generalgroup}

In this section we show that the phenomenon that we have seen in \S\ref{sec:U1example} for a $\U(1)$ gauge theory coupled to gravity holds for a general gauge group at weak coupling: the species bound on $\Lambda_{\rm QG}$ and the generalized Landau pole bound from loop corrections to gauge couplings coincide. We will first work out the case of an SU($2$) gauge theory in detail, then generalize to arbitrary gauge groups including product groups. We then give a very general argument for why this coincidence of scales occurs.

\subsection{$\SU(2)$ gauge theory}

We now consider an SU($2$) gauge theory coupled to gravity, applying the nonabelian sLWGC from \S\ref{sec:nonabelian}. SU($2$) invariance implies that states of large charge come in large representations, which leads to an interesting new result compared with the U($1$) gauge theories we have considered so far (which have the same Cartan subalgebra).

The SU($2$) weight lattice is $\mathbb{Z}/2$ in conventions where the roots have charge $\pm 1$. An arbitrary finite-index sublattice takes the form $k \mathbb{Z}/2$ for $k$ a positive integer. The Weyl group is $\mathbb{Z}_2$, generated by charge conjugation, hence all of these are Weyl invariant. Thus, the minimal sLWGC-satisfying spectrum consists of a spin $k n/2$ multiplet for each $n\in \mathbb{Z}_{> 0}$. The dimension of the spin $j$ representation is $2j+1$, so the total number of states up to level $n_{\rm max}$ is
\be
N = \sum_{n=1}^{n_{\rm max}} (k n + 1) = k \frac{n_{\rm max} (n_{\rm max}+1)}{2} + n_{\rm max}\,.
\ee
Applying the species bound and the counting (\ref{eq:towerN}) we obtain
\be
{\rm SU}(2): \quad \Lambda_{\rm QG} \lesssim k^\frac{1}{D} g^\frac{2}{D} M_{\rm Pl}^{\frac{2(D-2)}{D}}\,,  \label{eq:LambdaQGSU2}
\ee
for SU($2$) gauge coupling $g \ll 1$.
In particular, in the four-dimensional case we have $\Lambda_{\rm QG} \lesssim k^{1/4} g^{1/2} M_{\rm Pl}$. 

The tower of SU($2$) charged states makes the gauge theory strongly coupled in the ultraviolet. As before, we compare the quantum gravity scale with the strong coupling scale of the gauge theory $\Lambda_{\rm SU(2)}$, which generalizes the Landau pole in four dimensions. As explained in~\S\ref{sec:ampgrowth}, this is the scale at which $\lambda_{\rm gauge} \sim 1$. We find
\begin{align}
\lambda_{\rm gauge}(E) =  g^2 E^{D-4} \sum_{n=1}^{n_{\rm max}} I(kn) \sim g^2 E^{D-4} k^3 n_{\rm max}^4 \sim \frac{E^D}{g^2 k M_{\rm Pl}^{2(D-2)}} \,,
\end{align}
where we use the Dynkin index $I(j)$ of the spin-$j$ representation,
\be
I(j) = \frac{2}{3} j (j + 1) (2 j + 1) \sim j^3 \,,
\ee
as well as (\ref{eq:towerN}). Solving $\lambda_{\rm gauge}(\Lambda_{\rm SU(2)}) \sim 1$, 
we find parametric agreement with equation (\ref{eq:LambdaQGSU2}). That is, the parametric scaling with $g$ and $k$ of the quantum gravity cutoff and the gauge theory cutoff are the same, and we again find the phenomenon we refer to as gauge-gravity unification.

In the above, we have assumed that resonances which approximately saturate the sLWGC bound dominate the spectrum. If there are many more subextremal resonances, or if the tower of resonances is parametrically superextremal, this would affect the coincidence of scales, but the upper bound~(\ref{eq:LambdaQGSU2}) still applies. Notice also that the SU($2$) bound is less sensitive to the sublattice spacing $k$ than the U($1$) bound: the former depends on the combination $g\sqrt{k}$ while the latter depends on the combination $ek$. 

\subsection{Larger groups}

We generalize to larger groups, beginning with $\SU(3)$. $\SU(3)$ has two Cartan generators, and thus irreps are labeled by two non-negative integers $p$, $q$. Each irrep has dimension $(p+1)(q+1)(p+q+2)/2$ and a highest-weight state with $Q^2 = (2/3)(p^2 + p q + q^2)$ in conventions where $Q^2 = 2$ for the roots. By the results of appendix~\ref{app:WeylInv}, the Weyl-invariant sublattices are $k$ times the weight lattice and $k$ times the root lattice. We focus on the former case for definiteness, the latter being similar. Thus, $p, q \in k \mathbb{Z}_{\ge 0}$ for irreps whose highest-weight states fall on this sublattice.

We now estimate the species bound on the quantum gravity scale. The total number of states in irreps whose highest weights lie on $k$ times the weight lattice with charge $Q^2 \le Q^2_{\rm max}$ is
\begin{equation}
N = \sum_{p, q \in \mathbb{Z}_{\ge 0}}^{(2/3) k^2 (p^2 + p q + q^2) \le Q_{\rm max}^2} \frac{(k p+1)(k q + 1) (k p + k q + 2)}{2} \,.
\end{equation}
Approximating the sum with an integral, we find $N \sim Q_{\rm max}^5 / (5 \sqrt{6} k^2)$ asymptotically at large $Q_{\rm max}$. Roughly speaking, the fifth power of $Q$ appears here because $\SU(3)$ has three negative roots (half the total number of roots, which equals the dimension of the group minus its rank). The negative roots act as lowering operators on the highest-weight state, and lead to representations of size $\sim Q^3$. Summing over the Cartan then gives $\sim Q^5$ total states. (We generalize this argument below).

Thus, for an superextremal sLWGC tower $m^2 \lesssim g^2 Q^2 M_{\rm Pl}^{D-2}$ we estimate
\begin{equation}
\lambda_{\rm grav}(E) \gtrsim \frac{E^{D-2}}{M_{\rm Pl}^{D-2}} \frac{\Bigl(E / g M_{\rm Pl}^{\frac{D-2}{2}}\Bigr)^5}{k^2} \,,
\end{equation}
which gives
\begin{equation} \label{eqn:SU3bound}
\SU(3): \quad \Lambda_{\rm QG} \lesssim k^{\frac{2}{D+3}} g^{\frac{5}{D+3}} M_{\rm Pl}^{\frac{7(D-2)}{2(D-3)}} \,.
\end{equation}
In four dimensions this is $\Lambda_{\rm QG} \lesssim k^{\frac{2}{7}} g^{\frac{5}{7}} M_{\rm Pl}$.

We now consider the ultraviolet behavior of the gauge theory. The Dynkin index for the $(p,q)$ irrep of $\SU(3)$ is
\begin{equation}
I(p,q) = \frac{(p+1)(q+1)(p+q+2)(p^2+p q+ q^2 +3 (p+q))}{24} \,.
\end{equation}
The total index for irreps with highest weights on $k$ times the weight lattice and $Q^2 \le Q^2_{\rm max}$ is then
\begin{equation}
I_{\rm tot} = \sum_{p, q \in \mathbb{Z}_{\ge 0}}^{(2/3) k^2 (p^2 + p q + q^2) \le Q_{\rm max}^2} \frac{(k p+1)(k q+1)(k p+k q+2)(k^2 (p^2+p q+ q^2)+3 k (p+q))}{24} \sim \frac{Q_{\rm max}^7}{56 \sqrt{6} k^2} \,,
\end{equation}
where we use an integral approximation at large $Q_{\rm max}$ as before. Thus, for a superextremal tower
\begin{equation}
\lambda_{\rm gauge}(E) \gtrsim g^2 E^{D-4} \frac{\Bigl(E / g M_{\rm Pl}^{\frac{D-2}{2}}\Bigr)^7}{k^2} \,,
\end{equation}
which gives parametrically the same bound as (\ref{eqn:SU3bound}).

As before, if the spectrum is dominated by a tower of near extremal resonances then gauge and gravitational loops become large at parametrically the same scale. In other cases this coincidence of scales may not occur, but (\ref{eqn:SU3bound}) still applies if the sLWGC holds.

It is straightforward to generalize these arguments to an arbitrary simple gauge group $G$, as follows. The dimension of an arbitrary irrep $R$ with highest weight $\vec{Q}_R$ is determined by the Weyl dimension formula
\begin{equation}
\dim(R) = \frac{\prod_{\vec{Q} \in \Phi_+} \vec{Q}\cdot (\vec{Q}_R + \vec{Q}_0)}{\prod_{\vec{Q} \in \Phi_+} \vec{Q}\cdot \vec{Q}_0} \,, \qquad \vec{Q}_0 \df \frac{1}{2} \sum_{\vec{Q} \in \Phi_+} \vec{Q} \,,
\end{equation}
where $\Phi_+$ denotes the set of positive roots, as in \S\ref{sec:nonabelian}. Asymptotically for large $\vec{Q}_R$ we find
\begin{equation} \label{eqn:dimapprox}
\dim(R) \sim f(\hat{Q}_R) |\vec{Q}_R|^{\ell_G} \,, \qquad f(\hat{Q}_R) \df \frac{\prod_{\vec{Q} \in \Phi_+} \vec{Q}\cdot \hat{Q}_R }{\prod_{\vec{Q} \in \Phi_+} \vec{Q}\cdot \vec{Q}_0} \,,
\end{equation}
where $\ell_G \df |\Phi_+| = |\Phi|/2$ is the number of positive roots and $f(\hat{Q}_R)$ is an order-one function which depends only on the direction of $\vec{Q}_R$ within the fundamental Weyl chamber. This makes precise the intuition from above that more raising/lowering operators (positive/negative roots) leads to larger representations.

We take $k$ times the weight lattice as a representative example of the Weyl-invariant sLWGC sublattice.\footnote{As shown in appendix~\ref{app:WeylInv}, since $G$ is simple any Weyl-invariant sublattice is a multiple of one of a finite number of ``primitive'' Weyl-invariant sublattices. The parametric dependence on $k$ will be the same regardless of which primitive Weyl-invariant sublattice we start with.} Thus, there are superextremal irreps for all
\begin{equation}
\vec{Q}_R(n) = k \sum_i n_i \vec{Q}_i \,, \qquad n_i \in \mathbb{Z}_{\ge 0} \,,
\end{equation}
with mass $m^2 \lesssim g^2 \vec{Q}_R^2 M_{\rm Pl}^{D-2}$, where $\vec{Q}_i$ are the fundamental weights. The number of states below some scale $E$ is then
\begin{equation}
N(E) \gtrsim \sum_{n_1, \ldots, n_{r_G} \ge 0}^{\vec{Q}_R(n)^2 \le Q_{\rm max}^2} f(\hat{Q}_R(n))\, |\vec{Q}_R(n)|^{\ell_G} \sim \frac{Q_{\rm max}^{r_G + \ell_G}}{k^{r_G}} \,,
\end{equation}
where $Q_{\rm max}^2 \simeq E^2 / (g^2 M_{\rm Pl}^{D-2})$ and $r_G$ is the rank of $G$. Here we again use an integral approximation but drop all numerical factors including the angular integral over $f(\hat{Q}_R)$ and the volume of the fundamental domain of the weight lattice. Putting this into the species bound, we obtain
\begin{equation} \label{eqn:simpleGbound}
\mbox{simple $G$}: \quad \Lambda_{\rm QG} \lesssim k^{\frac{r_G}{n_G + D-2}} g^{\frac{n_G}{n_G + D-2}} M_{\rm Pl}^{\frac{n_G+2}{n_G + D-2} \frac{D-2}{2}} \,,
\end{equation}
up to numerical factors, where $n_G \df r_G + \ell_G$ is the rank plus half the number of roots, equal to $(r_G + d_G)/2$ where $d_G$ is the dimension of $G$. In four dimensions, this takes the simpler form $\Lambda_{\rm QG} \lesssim k^{r/(n+2)} g^{n/(n+2)} M_{\rm Pl}$.

For instance, for $\SU(N)$, $d_G = N^2 - 1$ and $r_G = N-1$, implying that
\begin{equation}
n_{\SU(N)} = \frac{N^2 + N - 2}{2} \,.
\end{equation}
Notice that as $N \to \infty$, the bound~(\ref{eqn:simpleGbound}) asymptotically brings the quantum gravity cutoff close to the ``magnetic WGC cutoff'' of (\ref{eq:magneticWGC}), i.e., the scale at which the tower of charged states appears,
\be
\lim_{N \to \infty} \Lambda_{\rm QG} \to g M_{\rm Pl}^{(D-2)/2} \,,
\ee
where the dependence on $k$ goes away because $r_G \ll d_G$. Similar results hold for other large rank simple groups.
Thus for larger nonabelian groups, small gauge couplings become increasingly powerful constraints on the validity of effective field theory.

The matching of gauge theory and quantum gravity cutoffs continues to hold, as in the cases we have already considered. In particular, the quadratic Casimir $C_2(R)$ and the Dynkin index $I(R)$ of $R$ are given by
\begin{equation} \label{eqn:generalC2}
C_2(R) = \vec{Q}_R \cdot (\vec{Q}_R + 2 \vec{Q}_0) \,, \qquad I(R) = \frac{C_2(R)\, \dim(R)}{d_G} \,,
\end{equation}
hence $C_2(R) \simeq |\vec{Q}_R|^2$ and $I(R) \simeq \frac{f(\hat{Q}_R)}{d_G} |\vec{Q}_R|^{\ell_G+2}$ at large $\vec{Q}_R$. Thus,
\begin{equation}
\lambda_{\rm gauge}(E) \gtrsim g^2 E^{D-4} \sum_{n_1, \ldots, n_{r_G} \ge 0}^{\vec{Q}_R(n)^2 \le Q_{\rm max}^2} \frac{f(\hat{Q}_R(n))}{d_G} \, |\vec{Q}_R(n)|^{\ell_G+2} \sim g^2 E^{D-4} \frac{Q_{\rm max}^{r_G + \ell_G+2}}{k^{r_G}} \sim \frac{E^{n_G+D-2}}{k^{r_G} g^{n_G} M_{\rm Pl}^{(n_G+2)\frac{D-2}{2}}} \,,
\end{equation}
which gives parametrically the same bound as above. We give a simpler argument for this in~\S\ref{sec:generalargument}.

\subsection{Product groups}\label{sec:productgroup}

Next, we consider the case of product groups. For a product group we can no longer describe the Weyl-invariant sublattice $\Gamma_0 \subseteq \Gamma_G$ satisfying the sLWGC by a single integer (plus a finite number of choices) as above. While it is possible to proceed carefully and catalog the possibilities, we will assume the full LWGC in this section for simplicity. Unless $\Gamma_0$ is very sparse within $\Gamma_G$, the effect of the sublattice index is competitive with other numerical factors that we consistently ignore.

As a simple example, we consider an $\SU(2)\times\U(1)$ gauge group with small gauge couplings $g$ and $e$ respectively.
The irreps are labeled by $(j,q)$ for $j \in \mathbb{Z}_{\ge 0}/2$ and $q \in \mathbb{Z}$. The LWGC requires a particle in each irrep with mass at most
\be
m_{(j,q)}^2 \lesssim \left(g^2 j^2 + e^2 q^2\right) M_{\rm Pl}^{D-2}.
\ee
Since the dimension of each irrep is $2j+1$, the number of states below some mass scale $E$ is at least
\be
N(E) \gtrsim \sum_{j, q}^{g^2 j^2 + e^2 q^2 \le E^2 / M_{\rm Pl}^{D-2}} (2j+1) \sim \frac{1}{e g^2} \left(\frac{E}{M_{\rm Pl}^{(D-2)/2}}\right)^3 \,,
\ee
using an integral approximation for $E \gg g M_{\rm Pl}^{(D-2)/2}$ and $E \gg e M_{\rm Pl}^{(D-2)/2}$. Thus, applying the species bound, we obtain
\be
\SU(2) \times \U(1): \quad \Lambda_{\rm QG} \lesssim e^\frac{1}{D+1} g^\frac{2}{D+1} M_{\rm Pl}^\frac{5(D-2)}{2(D+1)}\,,   \label{eq:LambdaSU2U1}
\ee
up to order one factors and the dependence on the sublattice index.
Note that if $e \sim g$, we can understand this result using the logic of the previous subsection: we have one raising operator (from the $\SU(2)$ factor) and two Cartan generators (one from each factor), so the total number of states up to the $n^{\rm th}$ rung of the ladder scales as $n^3$, and the bound is given by setting $n_G=3$ in (\ref{eqn:simpleGbound}).

As before, the Landau pole bounds from the tower of charged states parametrically coincide with $\Lambda_{\rm QG}$. For instance, for the $\U(1)$
\be
\lambda_{\rm U(1)}(E) \sim e^2 E^{D-4} \sum_{j, q}^{g^2 j^2 + e^2 q^2 \le E^2 / M_{\rm Pl}^{D-2}} q^2 (2j+1) \sim e^2 E^{D-4} \frac{1}{e^3 g^2} \Biggl(\frac{E}{M_{\rm Pl}^{(D-2)/2}}\Biggr)^5 \sim \frac{E^{D+1}}{e g^2 M_{\rm Pl}^{5(D-2)/2}},
\ee
which reproduces equation (\ref{eq:LambdaSU2U1}). A similar result holds for the SU($2$) factor.

Note that the above discussion assumes $g M_{\rm Pl}^{(D-2)/2} \ll \Lambda_{\rm QG}$ and $e M_{\rm Pl}^{(D-2)/2} \ll \Lambda_{\rm QG}$. Even if both gauge couplings are small, this need not be true if one is much smaller than the other. Only gauge group factors with WGC scale $g M_{\rm Pl}^{(D-2)/2}$ below the quantum gravity scale contribute to our bounds on $\Lambda_{\rm QG}$. This is discussed further in~\S\ref{subsec:heavyspectra}.

More generally, we consider a gauge group $G=U(1)^{r_0} \times \prod_{i=1}^p G_i$ for simple $G_i$ with gauge couplings $g_i$ and an $r_0 \times r_0$ abelian gauge kinetic matrix $\tau_{i j}$ (generalizing $1/e^2$ for a single $\U(1)$). Irreps are labeled by $(\vec{q}_0,R_1, \ldots, R_p)$ for $\U(1)^{r_0}$ charges $\vec{q}_0$ and $G_i$ representation $R_i$, corresponding to the highest weight vector $\mathbf{Q}_R = (\vec{q}_0, \vec{q}_1, \ldots, \vec{q}_p)$.
Superextremal irreps must satisfy
\be
m_{\mathbf{Q}}^2 \lesssim \Biggl[  \sum_{\alpha, \beta} (q_0)_{\alpha} \tau^{\alpha \beta} (q_0)_{\beta} +\sum_i g_i^2 q_{i}^2 \Biggr] M_{\rm Pl}^{D-2} \dfr ||\mathbf{Q} ||^2 \,,
\ee
where $\tau^{\alpha \beta} \df (\tau^{-1})^{\alpha \beta}$. The LWGC requires that the total number of states below a mass scale $E$ is at least
\begin{equation}
N(E) \gtrsim \sum^{  || \mathbf{Q}_R ||  < E }_{R}  \dim(R) \sim \int_{  || \mathbf{Q} ||  < E } d^{r_0} \vec{q}_0 \prod_{i=1}^p f_i(\hat{q}_{i})\, |\vec{q}_{i}|^{\ell_i}\, d^{r_i} \vec{q}_{i} \,,
\end{equation}
where we use~(\ref{eqn:dimapprox}) to estimate $\dim(R)$, and $r_i$ and $\ell_i$ are the rank and half the number of roots of $G_i$, respectively, as above. Carrying out the integral,\footnote{To perform the integral, it is useful to reimagine the integral over the $r_i$ components of $q_{(i)}$ as an integral over $n_i = r_i + \ell_i$ components of some fictitious vector by passing to spherical coordinates, factoring out the angular integral (which we drop along with other numerical factors) and passing back to rectangular coordinates. In this way, the integral over $\vec{Q}$ reduces to a straightforward spherical integral.} we obtain
\begin{equation}
N(E) \gtrsim \frac{\sqrt{\det \tau} E^n}{\bigl(\prod_i g_i^{n_i}\bigr) M_{\rm Pl}^{n (D-2)/2}} \,,
\end{equation}
where $n = \sum_{i=0}^p n_i$, $n_i = r_i + \ell_i$ (so that $n_0 = r_0$), and we ignore the angular integrals along with all other numerical factors. Thus, the species bound gives
\begin{equation} \label{eqn:productbound}
\Lambda_{\rm QG} \lesssim \biggl( (\det \tau)^{-1/2} \prod_i g_i^{n_i} \biggr)^{\frac{1}{n+D-2}} M_{\rm Pl}^{\frac{(n+2)(D-2)}{2(n+D-2)}} \,,
\end{equation}
which generalizes (\ref{eqn:simpleGbound}).\footnote{It is relatively straightforward to see how the sublattice data should appear in this expression. For instance, if $\Gamma_0$ includes only $k_i$ times the weight lattice of $G_i$, then we should replace $g_i^{n_i}$ with $k_i^{r_i} g_i^{n_i}$. More generally, the index $|\Gamma_0|/|\Gamma_G|$ of the sublattice $\Gamma_0$ within $\Gamma_G$, i.e., the fraction of $\Gamma_G$ sites which lie on $\Gamma_0$, will appear as an extra factor inside the parentheses.}

Estimating the size of loop corrections in the gauge theory leads to parametrically the same bound, as in all our previous examples. For instance, focusing on one of the non-abelian factors $G_i$, we have
\be
I_i (R) = I(R_i) \prod_{j \ne i} \dim R_j \sim \frac{f_i(\hat{q}_i)}{d_i} |\vec{q}_i|^{\ell_i + 2} \prod_{j\ne i} f_j(\hat{q}_j)\, |\vec{q}_j|^{\ell_j} \,,
\ee
using the estimate below (\ref{eqn:generalC2}), so that below the mass scale $E$
\be
I_i^{\rm (tot)} \gtrsim \int_{  || \mathbf{Q} ||  < E } d^{r_0} \vec{q}_0\, \biggl( \frac{f_i(\hat{q}_{i})}{d_i} |\vec{q}_{i}|^{\ell_i+2}\, d^{r_i} \vec{q}_{i} \biggr) \prod_{j\ne i} f_j(\hat{q}_{j})\, |\vec{q}_{j}|^{\ell_j}\, d^{r_j} \vec{q}_{j} \sim \frac{\sqrt{\det \tau} E^{n+2}}{g_i^{n_i+2} \bigl(\prod_{j\ne i} g_j^{n_j}\bigr) M_{\rm Pl}^{(n+2)\frac{D-2}{2}}} \,.
\ee
This gives
\be
\lambda_i \gtrsim \frac{\sqrt{\det \tau} E^{n+D-2}}{\bigl(\prod_j g_j^{n_j}\bigr) M_{\rm Pl}^{(n+2)\frac{D-2}{2}}} \,,
\ee
leading to the same bound as in (\ref{eqn:productbound}).

\subsection{A general argument}
\label{sec:generalargument}

Now that we have checked a variety of examples, let us give a general argument for why we consistently find that $\lambda_{\rm gauge}(E)$ and $\lambda_{\rm grav}(E)$ become ${\cal O}(1)$ at parametrically the same energy. We focus on a particular $\U(1)$, which might be either an abelian factor in the gauge group or a Cartan generator of a nonabelian factor. Let $n_E(q)$ be the number of charge $q$ particles with mass less than $E$, which for energies $E \gg e M_{\rm Pl}^{(D-2)/2}$ we will approximate as a continuous function of $q$. Thus,
\begin{align}
\lambda_{\rm gauge}(E) &\sim e^2 E^{D-4} \int_0^{Q(E)} dq\, q^2 n_E(q)\,, &
\lambda_{\rm grav}(E) &\sim G_N E^{D-2} \int_0^{Q(E)} dq\, n_E(q)\,,
\end{align}
where $Q(E)$ is the largest charge in the spectrum for masses below $E$. The average charge of all the particles with mass less than $E$ is
\be
\langle q^2 \rangle_E = \frac{\int_0^{Q(E)} dq\, q^2 n_E(q)}{\int_0^{Q(E)} dq\, n_E(q)}.
\ee
We will see that for a large family of smooth functions $n_E(q)$ which cut off at $q=Q(E)$ the average charge $\langle q^2 \rangle_E$ is parametrically of the same order as $Q(E)^2$. This means that
\begin{equation}
\lambda_{\rm gauge}(E) \sim \frac{e^2}{G_N E^2} \langle q^2 \rangle_E \lambda_{\rm grav}(E) \sim \frac{e^2 Q(E)^2}{G_N E^2} \lambda_{\rm grav}(E) \,. 
 \label{eq:gaugegravEagree}
\end{equation}
However, the (s)LWGC requires $Q(E) \gtrsim E/\bigl(e M_{\rm Pl}^{(D-2)/2}\bigr)$, and in particular if the constraint is nearly saturated for $E \gg e M_{\rm Pl}^{(D-2)/2}$ then $Q(E) \sim E/\bigl(e M_{\rm Pl}^{(D-2)/2}\bigr)$. By (\ref{eq:gaugegravEagree}) this implies $\lambda_{\rm gauge}(E) \sim \lambda_{\rm grav}(E)$ for $E \gg e M_{\rm Pl}^{(D-2)/2}$, and in particular gauge theory and gravitational loop corrections become large at parametrically the same scale $\Lambda_{\rm QG}$.

Having understood the consequences, we now give arguments why typically $\langle q^2 \rangle_E \sim Q(E)^2$ up to order-one factors. We begin with a simple example: suppose that all particles of a given mass have the same $|q|$ and $|q| = E/E_0$ increases linearly with energy, as in an (s)LWGC saturating tower for a single $\U(1)$ gauge group. Let $\rho(E) \df \frac{dN}{dE}$ be the density of states. We then have
\be
n_E(q) = E_0 \rho(|q| E_0)\, \Theta(E-|q| E_0) \,,
\ee
where $\Theta(x)$ is the Heaviside step function, so that
\be
\langle q^2 \rangle_E = \frac{\int_0^{E/E_0} dq\, q^2 \rho(|q| E_0)}{\int_0^{E/E_0} dq \rho(|q| E_0)} = \frac{1}{E_0^2} \frac{\int_0^{E} dE'\, E'^2 \rho(E')}{\int_0^{E} dE' \rho(E')}.
\ee
For a minimal sLWGC saturating spectrum, and in real quantum gravities that behave like this, such as Kaluza Klein theory, $\rho(E)$ is asymptotically a constant and we obtain $\langle q^2 \rangle_E \simeq \frac{1}{3} (E/E_0)^2 = \frac{1}{3} \,Q(E)^2$. Even if $\rho(E)$ grows asymptotically the large $E$ part of the integral is enhanced and $\langle q^2 \rangle_E$ is yet closer to $Q(E)^2$. Only if $\rho(E)$ falls off as $1/E$ or faster does this conclusion change, but the sLWGC sets a lower bound $\rho(E) \gtrsim 1/(k E_0)$ for a sublattice spacing $k$, so such a falloff is incompatible with it.

For more complicated gauge groups, the spectrum is different
 because near-extremal particles which are charged under other gauge group factors can have $q \ll E / \bigl(e M_{\rm Pl}^{(D-2)/2}\bigr)$, as can the lower weights in nonabelian irreps. For instance, consider $n$ $\U(1)$s with gauge couplings $e_1, \ldots, e_n$ and an LWGC saturating spectrum. We find
\be
n_E(q_1) = \sum_{q_2, \ldots, q_n}^{\sum e_i^2 q_i^2 \le E^2/M_{\rm Pl}^{D-2}} 1 \sim \frac{1}{e_2 \ldots e_n} \biggl(\frac{E^2}{M_{\rm Pl}^{D-2}} - e_1^2 q_1^2 \biggr)^{\frac{n-1}{2}} \propto (Q_1(E)^2 - q_1^2)^{\frac{n-1}{2}} \,,
\ee
where $Q_1(E)^2 \simeq E^2 / \bigl(e_1^2 M_{\rm Pl}^{D-2}\bigr)$. This gives $\langle q_1^2 \rangle_E \simeq \frac{1}{n+2} Q_1(E)^2$. The same result still holds when there are nonabelian factors in the rest of the gauge group---with $n$ equal to the total rank plus half the total number of roots, as above---as well as with arbitrary sublattice spacings.\footnote{If $q_1$ is a Cartan charge of a nonabelian factor then $n_E(q_1)$ can take a more complicated functional form, but the qualitative behavior is similar.} As in our first example, higher multiplicities for the near-extremal states should not change the conclusion.

More generally, $\langle q^2 \rangle_E \sim Q(E)^2$ when $n_E(q)$ is not too sharply peaked at $|q| \ll Q(E)$.
  For instance, if $n_E(q) \sim \exp(Q - q)$, then $\int_0^Q dq\, n_E(q) \sim \int_0^Q dq\, q^2 n_E(q) \sim \exp(Q)$, without a $Q^2$ enhancement in the second integral. 
  This includes the caveat we have made above: if there are large numbers of neutral particles, they correct $\lambda_{\rm grav}(E)$ but not $\lambda_{\rm gauge}(E)$ and hence spoil gauge-gravity unification.

\section{Gauge-gravity unification implies the Weak Gravity Conjecture}\label{sec:unification}

So far we have argued that the sLWGC leads naturally (with some caveats) to a single scale at which gauge forces and gravity become strong, suggesting a unification of forces. We now consider the converse case, where we demand such a unification and explore its consequences. More generally, we will assume that gauge couplings become strong \emph{at or below} the quantum gravity scale, allowing for the possibility of gauge theories which emerge from non-gravitational dynamics, as demonstrated, e.g., by Seiberg duality. Stated alternately, we assume that there cannot be any weakly coupled gauge bosons at the quantum gravity scale.\footnote{We explore some situations where these assumptions fail in~\S\ref{sec:caveats}.}

\subsection{Basic argument} \label{subsec:basicunification}

The requirement that gauge forces become strong ($\lambda_{\rm gauge} \sim 1$) at or below the quantum gravity scale is the requirement that
\be
\frac{1}{e^2} \sim \Lambda_{\rm gauge}^{D-4} \sum_{i | m_i < \Lambda_{\rm gauge}} q_i^2\,, \qquad \mbox{for $\Lambda_{\rm gauge} \lesssim \Lambda_{\rm QG}$.}  \label{eq:strongatLambda}
\ee
To derive the ordinary WGC from this, consider the particle of largest charge-to-mass ratio $(q/m)_{\rm max} \dfr z_{\rm max}$ among all the particles with mass below $\Lambda_{\rm gauge}$. For every $i$ we have $q_i^2 < z_{\rm max}^2 m_i^2$ and so
\begin{align}
\frac{1}{e^2} &\lesssim \Lambda_{\rm gauge}^{D-4} z_{\rm max}^2 \sum_{i | m_i < \Lambda_{\rm gauge}} m_i^2 \nonumber \\
&\lesssim \Lambda_{\rm gauge}^{D-2} z_{\rm max}^2 \, N(\Lambda_{\rm gauge})  \nonumber \\
&\lesssim \Lambda_{\rm QG}^{D-2} z_{\rm max}^2 \, N(\Lambda_{\rm QG})  \nonumber \\
&\lesssim z_{\rm max}^2 M_{\rm Pl}^{D-2}.
\label{eq:WGCproof}
\end{align}
In the second line we use $m_i^2 < \Lambda_{\rm gauge}^2$ to place an upper bound on the sum, in the third line we use $\Lambda_{\rm gauge} \lesssim \Lambda_{\rm QG}$, and in the last line we apply the species bound $N(\Lambda_{\rm QG}) \Lambda_{\rm QG}^{D-2}  \lesssim M_{\rm Pl}^{D-2}$. Rearranging the last inequality, we have $e^2 z_{\rm max}^2 M_{\rm Pl}^{D-2} \gtrsim 1$, which has the form of the original Weak Gravity Conjecture. It is slightly strengthened, since the superextremal particle we have found is below the quantum gravity cutoff. (However, see \S\ref{sec:caveats} below for exceptions in which this argument does not apply.)

From the constraint (\ref{eq:strongatLambda}) we can also obtain statements about the spectrum as a whole. For instance, we can rewrite it as
\be
\frac{1}{e^2} \sim \Lambda_{\rm gauge}^{D-4}\, N(\Lambda_{\rm gauge})\, \langle q^2 \rangle_{\Lambda_{\rm gauge}} \lesssim \frac{1}{\Lambda_{\rm gauge}^2} M_{\rm Pl}^{D-2} \langle q^2 \rangle_{\Lambda_{\rm gauge}} \,,
\ee
again using the species bound. We can rearrange this result in the suggestive form
\begin{align}
\Lambda_{\rm gauge}^2 &\lesssim e^2 \langle q^2\rangle_{\Lambda_{\rm gauge}} M_{\rm Pl}^{D-2}. \label{eqn:averagechargebound}
\end{align}
This is in itself an interesting WGC-like statement that bounds the strong coupling scale in terms of the {\em average} charge of particles with mass below that scale. Since every weakly coupled particle has $m < \Lambda_{\rm gauge}$, all of their masses are bounded in terms of the average charge; for instance (\ref{eqn:averagechargebound}) implies
that the particles lighter than $\Lambda_{\rm gauge}$ are, on average, superextremal.

\subsection{Comparisons at lower energies} \label{sec:lambdasequal}

We have seen that gauge-gravity unification, in the sense defined above, has very interesting consequences. We can obtain even stronger statements if we assume that the strengths of gauge and gravitational interactions unify \emph{below} the quantum gravity scale. However, to do so we need to specify what exactly this means.

We argued in~\S\ref{sec:loops} that $\lambda_{\rm gauge}(E)$ and $\lambda_{\rm grav}(E)$---defined in~(\ref{eq:lambdagauge}) and (\ref{eq:lambdagrav}), respectively---are useful heuristics which estimate the fractional size of gauge theory and gravity loop corrections at a scale $E$ \emph{coming from light particles} (particles with mass below $E$). When $\lambda_{\rm gauge} \gtrsim 1$ or $\lambda_{\rm grav} \gtrsim 1$ the corresponding loop expansion breaks down, though of course it could break down at a lower scale for other reasons. However, at scales where the $\lambda$s are small, they have no clear physical interpretation. They do not even represent the fractional size of \emph{all loop corrections}, but only those coming from particles lighter than $E$; the contributions from heavy particles and/or higher dimensional operators are typically much larger, though still suppressed by powers of $E/\Lambda$.

Nonetheless, in situations where we expect gauge and gravitational forces to unify below the quantum gravity scale, we find $\lambda_{\rm gauge}(E) \sim \lambda_{\rm grav}(E)$ beginning at the expected unification scale. There are two principal examples of this: (1) Kaluza-Klein theory, where the graviphoton shares a common origin with the graviton at the compactification scale, and (2) perturbative string theory, where gauge bosons and gravitons share a common origin as excitations of the string. In the former case, the number of KK modes up to a scale $E$ is $N(E) \sim E R$ where $R$ is the compactification radius. Thus, by a familiar calculation (this is essential the same situation as that in \S\ref{sec:U1example}):
\begin{align}
\lambda_{\rm grav}(E) &\sim \frac{E^{D-2}}{M_{\rm Pl}^{D-2}} \left(E R\right)\,, & \lambda_{\rm gauge}(E) &\sim e^2 E^{D-4} \left(E R\right)^3\,.
\end{align}
However, $1/e^2 = (1/2) R^2 M_{\rm Pl}^{D-2}$, hence $\lambda_{\rm grav}(E) \sim \lambda_{\rm gauge}(E)$. Below the compactification scale $1/R$, $\lambda_{\rm gauge} = 0$ (there are no charged particles), but $\lambda_{\rm grav} \ne 0$, so the matching begins near the compactification scale, exactly where we expect the forces to unify. More complicated KK examples behave in the same way, as can be seen using, e.g., the general arguments of \S\ref{sec:generalargument}.

The case of perturbative string theory is more complicated, since gauge fields can have several different origins, from both open and closed strings, and in the latter case from both the NSNS and RR sectors. Since the graviton lives in the NSNS closed string sector, we expect gauge fields from this sector to unify with it at the string scale. In \S\ref{sec:weakstrings} we will argue that indeed $\lambda_{\rm gauge} \sim \lambda_{\rm grav}$ at the string scale for NSNS sector gauge bosons (except those with no charged particles at or below this scale;
see~\S\ref{subsec:heavyspectra} for related caveats.)

With this motivation, such as it is, we proceed to compare $\lambda_{\rm gauge}(E)$ and $\lambda_{\rm grav}(E)$ at scales parametrically below the quantum gravity scale and derive the consequences of certain simple assumptions. First, suppose that $\lambda_{\rm gauge}(E) \gtrsim \lambda_{\rm grav}(E)$ at some particular scale $E \lesssim \Lambda_{\rm QG}$.
This means that
\begin{align}
G_N N(E) E^2 \lesssim e^2 \sum_{i | m_i \lesssim E} q_i^2 \lesssim e^2 z^2_{\rm max}(E) \sum_{i | m_i \lesssim E} m_i^2 \lesssim e^2 z^2_{\rm max}(E)\, N(E) E^2\,,
\end{align}
by essentially the same reasoning as in~\S\ref{subsec:basicunification}, where $z_{\rm max}(E)$ is the largest charge-to-mass ratio among the particles lighter than $E$.
Dividing by $N(E) E^2$, we conclude that there is a superextremal particle lighter than $E$, and the WGC is satisfied.

A more intriguing statement arises if we assume 
\be
\lambda_{\rm gauge}(E) \sim \lambda_{\rm grav}(E) \quad {\rm for} \quad E \gtrsim E_0 \,, \label{eq:equaloverrange}
\ee
which is the heuristic notion of gauge-gravity unification at weak coupling that we motivated above, with unification scale $E_0 \ll \Lambda_{\rm QG}$.
This means that for $E_0 \lesssim E \lesssim \Lambda_{\rm QG}$ we have
\begin{align}
e^2 \sum_{i | m_i \lesssim E} q_i^2 &\sim G_N N(E)\, E^2.
\end{align}
Here $N(E)$ is a monotonically increasing function of $E$. Let us assume that there are sufficiently many particles above $E_0$ that we can approximate these functions as continuous. If we differentiate both sides with respect to $E$, we can rewrite the derivative of the left-hand side in terms of $d(\sum q^2)/dE = d(\sum q^2)/dN\cdot dN/dE$ and divide through by $dN/dE$ to obtain:
\begin{align}
e^2 \frac{d (\sum q^2)}{dN} &\sim G_N E^2 \left(1 + 2 \frac{d\log E}{d\log N}\right),
\end{align}
where $d\log E / d\log N \geq 0$. Notice that $\frac{d(\sum q^2)}{dE} \Delta E$ has the interpretation as the total squared charge contributed by particles with mass in a range $\Delta E$ near $E$, and as a result $d(\sum q^2)/dN$ can be interpreted as the {\em average} $q^2$ of the particles with mass near $E$. In other words, the condition (\ref{eq:equaloverrange}) implies that for all states with mass approximately $E$, we have
\begin{align} \label{eqn:sLWGClike}
e^2 \langle q^2 \rangle_{m \approx E} \gtrsim G_N m^2,
\end{align}
which is to say that the average particle of mass near $m$ is superextremal. Hence the relation (\ref{eq:equaloverrange}) implies the existence of a {\em tower} of superextremal resonances at energies above $E_0$. This is an sLWGC-like statement. 

The close relationship between (\ref{eq:equaloverrange}) and the sLWGC, at least parametrically, suggests the intriguing possibility that there is some sharp property of, e.g., the high energy behavior of scattering amplitudes in quantum gravities that has the same close relationship to the sLWGC. As we discussed in the introduction, the sLWGC itself is somewhat poorly defined in theories with strong coupling, since it refers to single particles but these may be unstable. An alternate definition in terms of the S-matrix could address this issue, but at present it remains unclear whether any sharpened version of the heuristic $\lambda(E)$ can be extracted from the S-matrix. We leave further exploration of this idea to the future.

Note that if we define a variant $\widetilde{\lambda}(E)$ which includes the fractional size of loops of heavy particles as well as the light particles accounted for in $\lambda(E)$---as discussed further in Appendix \ref{app:heavyloops}---then $\widetilde{\lambda}_{\rm gauge}(E) \gtrsim \widetilde{\lambda}_{\rm grav}(E)$ leads to a different conclusion than above: there must be a superextremal particle with mass between $E$ and $\Lambda_{\rm QG}$. Moreover, $\widetilde{\lambda}_{\rm gauge}(E) \sim \widetilde{\lambda}_{\rm grav}(E)$ for $E\gtrsim E_0$ does not have the same strong implications. Because heavy particles typically give the dominant contribution to $\widetilde{\lambda}(E)$, this parametric matching does not directly constrain lighter particles, and no sLWGC-like statement follows. None of our previous results were sensitive to the distinction between $\lambda(E)$ and $\widetilde{\lambda}(E)$, which illustrates the more speculative nature of the present subsection: gauge-gravity unification at weak coupling is a concept with no obvious definition, and we could have chosen a different one, such as $\widetilde{\lambda}_{\rm gauge}(E) \sim \widetilde{\lambda}_{\rm grav}(E)$.

Nonetheless, $\widetilde{\lambda}(E)$ does not have the same connection to unification in the simple examples that we discussed above. While $\widetilde{\lambda}_{\rm gauge}(E) \sim \widetilde{\lambda}_{\rm grav}(E)$ in KK theory, this continues below the compactification scale even though at low energies the common origin of the graviphoton and graviton is not evident.
In perturbative string theory, it is difficult to even define $\widetilde{\lambda}(E)$, in part because the two-point function is an off-shell quantity. The results we get depend on whether we count states above the string scale (and how we count them). 
Both of these examples illustrate that $\widetilde{\lambda}(E)$ is a UV-sensitive quantity, whereas $\lambda(E)$ depends only on the light spectrum and infrared couplings. 
While the precise physical interpretation of $\lambda(E)$ remains unclear, it is arguably both a better measure of unification than $\widetilde{\lambda}(E)$ as well as a better behaved quantity in effective field theory.

\subsection{Product groups}

The above discussion applies equally well to an arbitrary gauge group as to the case of a single gauge boson or a simple gauge group. In particular, the one-loop correction~(\ref{eq:vacuumpol}) to the gauge boson propagator is unaffected by the presence of other gauge group factors, except that we must allow for kinetic mixing between photons, as we did above in \S\ref{sec:productgroup}.
\be 
\mathcal{L}_{\rm kin} = - \frac{1}{4} \tau_{a b} (F^a)_{\mu \nu} (F^b)^{\mu \nu} \,,
\ee
for some positive definite gauge coupling matrix $\tau_{a b}$ (generalizing $1/e^2$ in the single-photon case). Loop corrections give an energy-dependent correction to the gauge boson propagator scaling as
\be
\tau^{\rm (1-loop)}_{a b}(E) \sim \sum_{i | m_i < E} q_{ia} q_{ib} E^{D-4}. \label{eq:mixingloop}
\ee
The direct analogue of $\lambda_{\rm gauge}(E)$ is a matrix $\lambda^a_{~c}(E) = \tau^{a b}\tau^{\rm (1-loop)}_{b c}(E)$, but a more straightforward approach relies on a choice of direction $n^a$ in charge space.

Specifically, we can adapt our preceding arguments to the kinetically mixed case as follows: the derivation of the WGC from the condition (\ref{eq:strongatLambda}) that a gauge interaction is strong at the scale $\Lambda_{\rm QG}$ carries through with the replacements
\begin{align}
\frac{1}{e^2} \mapsto n^a n^b \tau_{a b},  \qquad
q_i \mapsto n^a q_{i a}. \label{eq:mixingreplacements}
\end{align}
In other words, if any particular linear combination of U($1$)s is strong at the quantum gravity scale, we deduce the existence of a particle charged under that linear combination. If we impose that gauge couplings are strong, in the sense of the condition (\ref{eq:strongatLambda}), for all possible choices of $n^a$, then we obtain the convex hull condition for the product gauge theory. 

Similarly, the arguments of \S\ref{sec:lambdasequal} comparing $\lambda_{\rm grav}(E)$ to $\lambda_{\rm gauge}(E)$ may be rephrased in terms of the condition
\begin{align}
\frac{1}{e^2} &\lesssim \frac{1}{G_N E^2} \langle q^2 \rangle,
\end{align}
a form suitable for making the replacements (\ref{eq:mixingreplacements}). Again, the arguments go through once we select a direction $n^a$ in charge space.

\subsection{Higgsing}\label{sec:higgsing}

It has been pointed out that the WGC and most of its known stronger variants are not automatically preserved under Higgsing (\nospace\cite{prashant2016,Saraswat:2016eaz}, see also~\cite{Heidenreich:2016aqi}). In other words, given an effective field theory which apparently satisfies the variant of the WGC in question and which contains a light charged scalar, the effective field theory obtained by giving a vev to the scalar may not satisfy the same variant of the WGC (or even the WGC itself). Of course, this does not imply the same statement about effective theories with quantum gravity completions: if the WGC variant in question is correct then these must satisfy non-trivial additional constraints which ensure that it remains true after Higgsing. In many concrete examples, this is the case, and the WGC / sLWGC remain true after Higgsing.

Below, we review why the WGC and its lattice variants are not automatically preserved under Higgsing. We then discuss to what extent our arguments above are affected by these subtleties. Other recent discussions of Higgsing and the WGC include~\cite{Ibanez:2017vfl,Furuuchi:2017upe}. Attempts to exploit this kind of loophole for large field axion inflation include~\cite{Hebecker:2015rya,Hebecker:2017lxm}.


We start with the very simple case of two abelian gauge bosons, $A$ and $B$, which are unmixed and do not couple to massless dilatons:
\be
{\cal L } = -\frac{1}{4} \left(\frac{1}{e_A^2}A _{\mu \nu}^2 + \frac{1}{e_B^2} B_{\mu \nu}^2\right).
\ee
Suppose they are Higgsed to the diagonal by a scalar field of charge $(1,-P)$ for integer $P$, so that the linear combination $H_\mu = A_\mu - P B_\mu$
becomes heavy. Then the light field that is not Higgsed is
\be
L_\mu = \frac{e_A^2 B_\mu + P e_B^2 A_\mu}{e_A^2 + P^2 e_B^2}.
\ee
In terms of the light and heavy eigenstates the original fields are
\begin{align}
A_\mu &= P L_\mu + \frac{e_A^2}{e_A^2 + P^2 e_B^2} H_\mu\,, &
B_\mu &= L_\mu - \frac{P e_B^2}{e_A^2 + P^2 e_B^2} H_\mu\, ,
\end{align}
and the kinetic terms become
\be
{\cal L} = -\frac{1}{4} \left(\left(\frac{P^2}{e_A^2} + \frac{1}{e_B^2}\right) L_{\mu \nu}^2 + \frac{1}{e_A^2 + P^2 e_B^2} H_{\mu \nu}^2\right).
\ee
Now, a particle of charge $(q_A, q_B)$ under the original symmetries couples to the linear combination
\be
q_A A_\mu + q_B B_\mu = \left(P q_A + q_B\right) L_\mu +\frac{q_A e_A^2 - P q_B e_B^2}{e_A^2 + P^2 e_B^2} H_\mu.
\ee
As expected, the unbroken gauge field couples to a U($1$) with integer charges $Q=P q_A + q_B$ while the heavy eigenstate in general can couple to irrational charges.

A particle of charge $(q_A, q_B)$ is superextremal with respect to the un-Higgsed theory if
\be
m^2 \leq \gamma \left(e_A^2 q_A^2 + e_B^2 q_B^2\right) M_{\rm Pl}^{D-2}, \label{eq:originalWGC}
\ee
where $\gamma$ is a dimension-dependent factor, see, e.g.,~\cite{Heidenreich:2015nta}.
In the Higgsed theory, it has a diagonal charge $Q = P q_A + q_B$ and couples via the diagonal coupling
\be \label{eqn:eD}
\frac{1}{e_D^2} = \frac{P^2}{e_A^2} + \frac{1}{e_B^2},
\ee
so it is superextremal if $m^2 \leq \gamma e_D^2 Q^2 M_{\rm Pl}^{D-2}$.
Suppose that a particle of charge $Q=P q_A+q_B$ is extremal in the un-Higgsed theory, i.e., saturating~(\ref{eq:originalWGC}). Whether it is superextremal or not in the Higgsed theory depends on $q_A, q_B$.
Putting $q_B = Q-P q_A$ into (\ref{eq:originalWGC}) and completing the square, we find:
\be
m^2 = \gamma e_D^2 Q^2 M_{\rm Pl}^{D-2} + \gamma \frac{e_A^2 e_B^2}{e_D^2} \left(q_A - P \frac{e_D^2}{e_A^2} Q\right)^2 M_{\rm Pl}^{D-2}. \label{eq:originalWGCcompletesquare}
\ee
Thus, the particle is extremal in the Higgsed theory if and only if
\be
q_A = P \frac{e_D^2}{e_A^2} Q\,, \quad \mbox{which is equivalent to} \quad q_B = \frac{e_D^2}{e_B^2} Q \,;
\ee
otherwise, it is subextremal. In other words, extremality is preserved if the charge vector $(q_A, q_B)$ is parallel to $\bigl(P e_D^2/e_A^2,e_D^2/e_B^2\bigr)$. For other charged particles, Higgsing makes them less extremal.

The ordinary WGC (i.e., the convex hull condition~\cite{cheung:2014vva}) is equivalent to the requirement that for every site in the charge lattice $(q_A, q_B) \in \mathbb{Z}^2$ there is a superextremal multiparticle state with charge $(r q_A, r q_B)$ for some rational $r>0$. If $e_A^2 / e_B^2$ is rational then so are $e_D^2/e_A^2$ and $e_D^2/e_B^2$, and
\be
\biggl( \frac{P e_D^2/e_A^2}{\gcd(P e_D^2/e_A^2,e_D^2/e_B^2)},  \frac{e_D^2/e_B^2}{\gcd(P e_D^2/e_A^2,e_D^2/e_B^2)}\biggr) \in \mathbb{Z}^2 \label{eqn:extLattice}
\ee
is a lattice site.\footnote{Note that $\gcd$ is naturally defined for rational arguments via $\gcd\Bigl(\frac{p}{r},\frac{q}{s}\Bigr) = \frac{1}{rs} \gcd(p s, q r)$ so that $x/\gcd(x,y)$ and $y/\gcd(x,y)$ are always coprime integers for any $x,y \in \mathbb{Q}$.}
The ordinary WGC in the un-Higgsed theory then implies that there is a superextremal multiparticle state of charge $(r P e_D^2/e_A^2, r e_D^2/e_B^2)$ for some rational $r>0$. By the above reasoning, this multiparticle state is also superextremal in the Higgsed theory, which implies the existence of a superextremal charged particle, hence the WGC is preserved.

The situation is similar for the sLWGC. When $e_A^2 / e_B^2$ is rational, the lattice vector (\ref{eqn:extLattice}) generates a one-dimensional sublattice of the charge lattice. The intersection of this sublattice with the two-dimensional sublattice of superextremal charged particles required by the un-Higgsed sLWGC is a one-dimensional sublattice, and for each site on this sublattice (corresponding to a one-dimensional sublattice of the Higgsed charged lattice) we obtain a superextremal charged particle in the Higgsed theory, hence the sLWGC is preserved.

\begin{figure}[!h]\begin{center}
\includegraphics[width=0.7\textwidth]{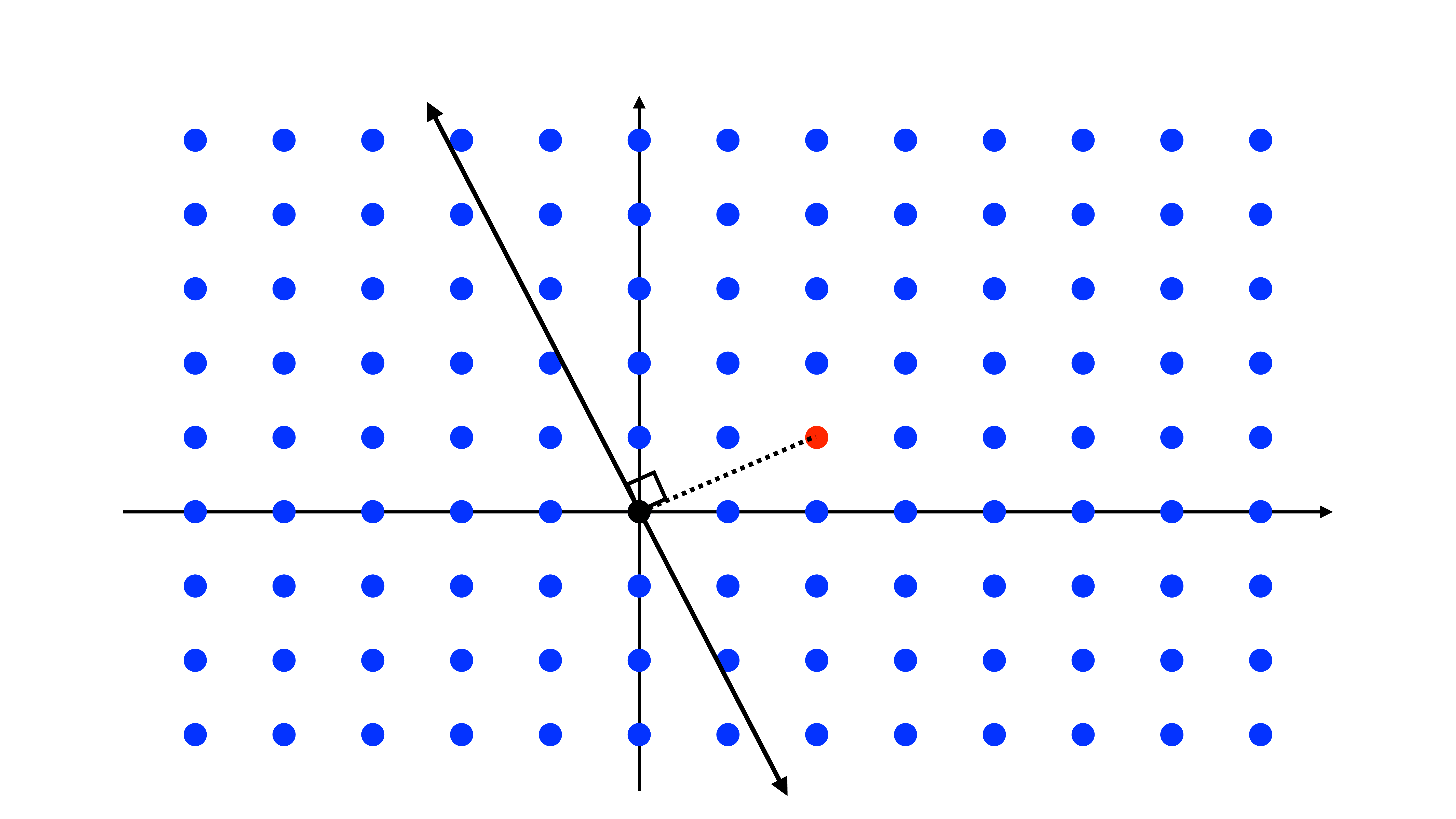}
\end{center}
\caption{Higgsing in a lattice with $e_A^2 / e_B^2$ irrational.  If the direction orthogonal to the Higgsed particle (shown in red) does not intersect any lattice points, then the WGC (and sLWGC) need not be satisfied in the resulting theory.}
\label{fig:Higgsing}
\end{figure}

If on the other hand $e_A^2 / e_B^2$ is irrational, then even if the LWGC is satisfied in the un-Higgsed theory, i.e., if for every $(q_A, q_B) \in \mathbb{Z}^2$ there is a superextremal charged particle, the ordinary WGC in the Higgsed theory may not hold. In particular, if the charged particles are all extremal (or subextremal) in the un-Higgsed theory then there are no superextremal charged particles in the Higgsed theory precisely because there are no charge vectors in the charge lattice parallel to $\bigl(P e_D^2/e_A^2,e_D^2/e_B^2\bigr)$, and the WGC is violated.  This is depicted graphically in Figure \ref{fig:Higgsing}.

\begin{figure}[!h]\begin{center}
\includegraphics[width=0.7\textwidth]{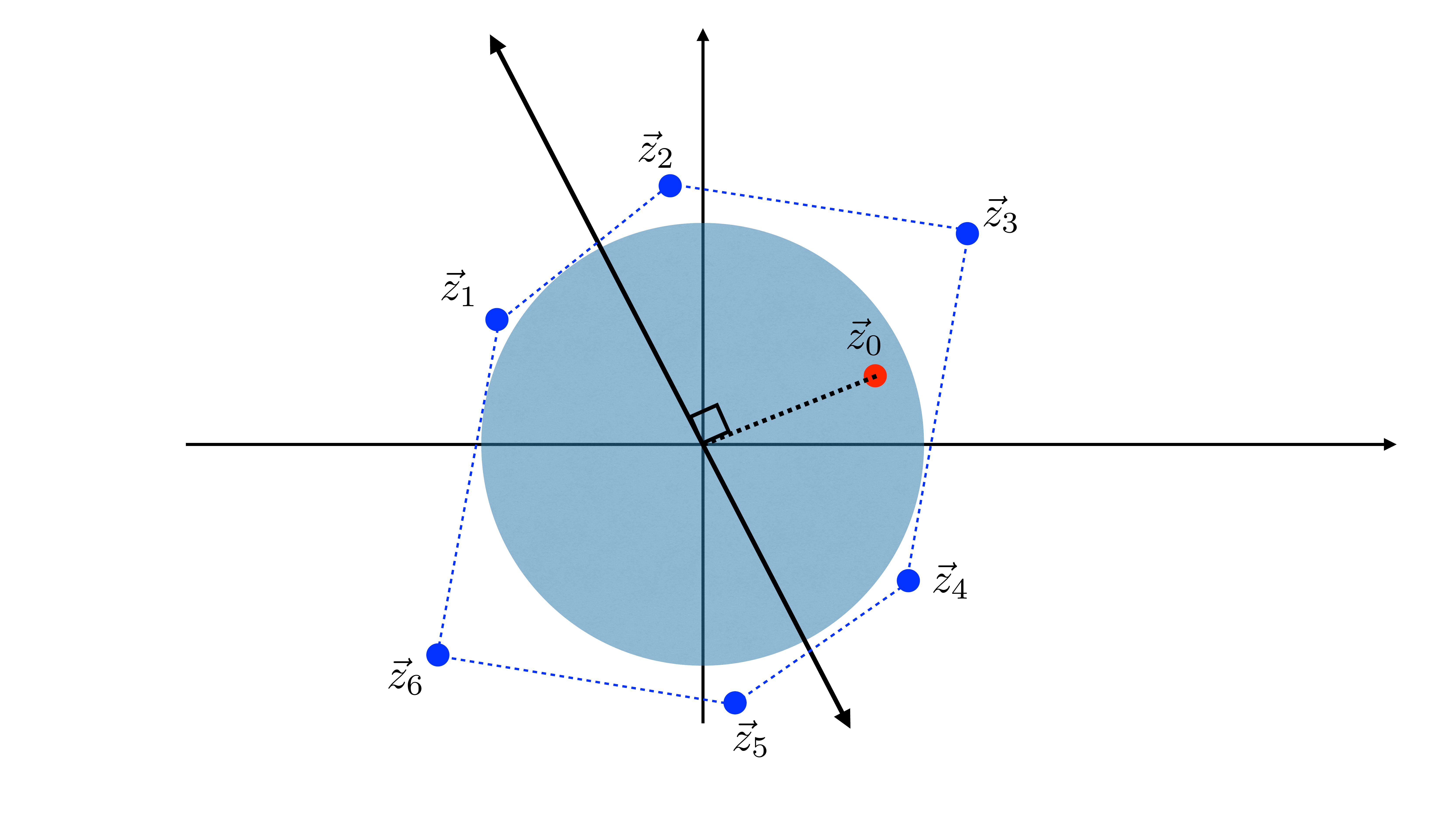}
\end{center}
\caption{Preservation of the WGC under Higgsing for a theory with a finitely generated convex hull.  Since the convex hull condition is satisfied in the direction $\vec{e}_\perp$ orthogonal to the Higgsed particle (shown in red, with charge-to-mass vector $\vec{z}_0$), we necessarily have either $ |\vec{z}_1 \cdot \vec{e}_\perp| \geq 1$ or $ |\vec{z}_2 \cdot \vec{e}_\perp | \geq 1$.  This ensures that one of these two particles will still satisfy the convex hull condition after Higgsing.}
\label{fig:CHCHiggsing}
\end{figure}

However, if the convex hull condition is satisfied by a finite number of particles\footnote{There are concrete examples of (supersymmetric) quantum gravities for which the convex hull condition cannot be satisfied by a finite number of particles, see~\cite{Heidenreich:2015nta,Heidenreich:2016aqi}.} in the un-Higgsed theory then the WGC is automatically satisfied in the Higgsed theory. This is because the above argument produces multiparticle states which are arbitrary close to extremal if we take $q_A, q_B$ large with $q_A / q_B$ a rational approximant to $P e_B^2/e_A^2$. If a finite number of particles generate all of these multiparticle states then at least one of these particles must be superextremal, and the WGC is satisfied.  This is depicted graphically in Figure \ref{fig:CHCHiggsing}.

It is not too hard to generalize this argument to the case of $N > 2$ photons and/or kinetic mixing between the photons. It is convenient to canonically normalize:
\be \label{eqn:nphotonsL}
{\cal L } = -\frac{1}{4} \sum_a (F_{\mu \nu}^a)^2 - \frac{m_A^2}{q^2} \Bigl(\sum_a q_a A_\mu^a\Bigr)^2.
\ee
where $q_a \in \Gamma_Q$ is the charge of the Higgs field and $\Gamma_Q \subset \mathbb{R}^N$ is the charge lattice. We decompose into heavy and light fields gauge fields $H_\mu$ and $L_\mu^a$:
\be
A_{\mu}^a = \hat{q}^a H_{\mu} + e^a_i L_{\mu}^i \,,
\ee
where $\hat{q}_a \df q_a / |q|$ for $|q|^2 \df \delta^{a b} q_a q_b$ and $e^a_i$ is chosen to satisfy $q_a e^a_i = 0$ and $\delta_{a b} e^a_i e^b_j = \delta_{i j}$. The superextremality conditions before and after Higgsing are
\begin{equation}
m^2 \le \gamma \delta^{a b} Q_a Q_b M_{\rm pl}^{D-2} \quad \mbox{and} \quad m^2 \le \gamma \delta^{i j} \tilde{Q}_i \tilde{Q}_j M_{\rm pl}^{D-2} \,,
\end{equation}
respectively, where $\tilde{Q}_i = e^a_i Q_a$ is the charge after Higgsing. Suppose a particle, charge $Q_a$, is extremal before Higgsing, then
\be
m^2 = \gamma \delta^{i j} \tilde{Q}_i \tilde{Q}_j M_{\rm pl}^{D-2} + \gamma (Q_a \hat{q}^a)^2 M_{\rm pl}^{D-2} \,,
\ee
where we use $\delta^{i j} e^a_i e^b_j = \delta^{a b} - \hat{q}^a \hat{q}^b$. Thus, the particle is extremal after Higgsing if and only if $Q_a \hat{q}^a = 0$; otherwise it is subextremal. By the same arguments as above, if the plane orthogonal to $q_a$ contains an $N-1$ dimensional sublattice of the charge lattice $\Gamma_Q$ then the WGC and the sLWGC are each preserved under Higgsing, whereas if not then in general stronger constraints are needed in the un-Higgsed theory to satisfy the (sL)WGC in the Higgsed theory. If the convex hull condition is satisfied by a finite number of particles before Higgsing, it is still satisfied after Higgsing.\footnote{Our results disagree slightly with~\cite{Saraswat:2016eaz}, which concluded that the ordinary WGC and the sLWGC are automatically preserved under Higgsing. As shown above, this is not the case in general, though the ordinary WGC is automatically preserved if the convex hull condition is satisfied by a finite number of particles.}

Note that the condition that the charge lattice $\Gamma_Q$ intersects the plane orthogonal to $q_a$ in an $N-1$ dimensional sublattice generalizes the requirement that $e_A^2/e_B^2$ is rational, since in that simple example the charge lattice is generated by $(e_A, 0)$ and $(0,e_B)$ with $q_a = (e_A, - P e_B)$, so we require non-trivial solutions to $(m e_A, n e_B) \cdot (e_A, -P e_B) = 0$, i.e., $e_A^2/e_B^2 = P n/m$, for rational $m, n$. As in this simple example, in general charge lattices which satisfy this property are dense in the set of all charge lattices.

This means that the WGC is not necessarily preserved under Higgsing, assuming the original theory exactly saturated the WGC bound.  The same argument applies to the sLWGC: a theory which saturates the sLWGC can in principle be Higgsed to a theory that violates it.  This is not a counterexample to the WGC or sLWGC, however: it simply shows that stronger constraints must be imposed on the original theory to ensure that these bounds are not violated after Higgsing.

It is also worth noting that although the WGC can in principle be violated, it will still be approximately true in the Higgsed theory.  In the above example, we may choose $q_A$, $q_B$ such that $q_A / q_B \approx P e_B^2 / e_A^2$ to arbitrarily good precision.  If the WGC is satisfied in the un-Higgsed theory, then there must exist some (possibly multiparticle) state with these charges (or a multiple thereof), and this will reduce upon Higgsing to a (possibly multiparticle) state that approximately satisfies the WGC bound.  The same statement is true for the sLWGC, except we demand that these multiparticle states be single particle states or resonances. 

\medskip

We now consider the effect of Higgsing on our arguments about UV cutoffs. 
If the scale of at which the gauge group is Higgsed is well below the quantum gravity scale, $m_A \ll \Lambda_{\rm QG}$, then from a UV perspective we can treat the gauge group as unbroken, and we still expect a tower of charged states to appear near the WGC scale, $e M_{\rm Pl}^{(D-2)/2}$. Heavier particles in such a tower generally dominate $\lambda_{\rm gauge}$ and $\lambda_{\rm grav}$, hence the conclusions about UV cutoffs are the same as if the gauge group were unbroken---even if $m_A$ lies above the WGC scale---so long as $m_A \ll \Lambda_{\rm QG}$.\footnote{However, we need to be cautious about identifying any massive vector with an enhanced gauge group. One case in which this is obviously incorrect is KK theory with a higher dimensional photon, for which there is a tower of graviphoton-charged massive vectors, but no corresponding nonabelian gauge group.}

In particular, gauge-gravity unification in the sense of~\S\ref{sec:warmup} is unaffected by Higgsing. We could reach the same conclusion by ignoring the massive gauge bosons entirely and focusing on some $\U(1)$ in the Cartan of the unbroken group. If the sLWGC is satisfied in the un-Higgsed theory, the arguments given above imply that it must be at least approximately satisfied in the Higgsed theory.  The general argument of~\S\ref{sec:generalargument} can then be applied, regardless of additional multiplicities which arise from sLWGC constraints coming from the enhanced gauge group in the UV. 

Conversely, if we assume that the gauge forces in the un-Higgsed theory become strong at or below the quantum gravity scale, we can apply the arguments of~\S\ref{subsec:basicunification} to this un-Higgsed theory.  (\ref{eq:WGCproof}) then ensures that the WGC will be (approximately) satisfied for this theory.  Furthermore, the masses of the superextremal particles will be below $\Lambda_{\rm QG}$, and there are only a finite number of such particles.  This means that the convex hull condition will be (approximately) satisfied by a finite number of particles in the un-Higgsed theory, which implies that it will also be (approximately) satisfied in the Higgsed theory.

As before, we reach the same conclusion if we consider a $\U(1)$ in the Cartan of the unbroken gauge group, ignoring the broken generators.  If we assume that this $\U(1)$ becomes strongly coupled below $\Lambda_{\rm QG}$, (\ref{eq:WGCproof}) again ensures that the WGC will be (approximately) satisfied for this U($1$).  If we further assume that $\lambda_{\rm gauge}(E) \sim \lambda_{\rm grav}(E)$ for this U($1$) over some range of energies $E \gtrsim E_0$, as in~\S\ref{sec:lambdasequal}, (\ref{eqn:sLWGClike}) ensures the existence of a tower of superextremal resonances with energy above $E_0$.

Thus, the relationship between unification in the sense of~\S\ref{sec:warmup} and the (sL)WGC is largely unaffected by subtleties related to Higgsing. There is, however, an important caveat to keep in mind when considering bounds on the quantum gravity scale, such as \eqref{eq:LambdaQGU1}, \eqref{eqn:simpleGbound}, or \eqref{eqn:productbound}: the sublattice index can change upon Higgsing. Hence, even if $k\sim 1$ in the UV theory, our arguments do not exclude $k\gg 1$ in the infrared theory, leading to weaker constraints on the UV cutoff.

To illustrate this, consider the two-photon example discussed above. Assume for simplicity that the UV theory contains an LWGC-saturating tower of extremal particles. For fixed $Q = P q_A + q_B$ there is an approximately extremal particle in the IR theory (i.e., with $m^2/m^2_{\rm ext} \leq 1+\varepsilon^2$ for $\varepsilon \lesssim 1$) whenever
\begin{equation}
\frac{P - \varepsilon e_A/e_B}{P^2+e_A^2/e_B^2} Q \leq q_A \leq \frac{P + \varepsilon e_A/e_B}{P^2+e_A^2/e_B^2} Q 
\end{equation}
has an integer solution. If $P \lesssim e_A/e_B$ then $q_A=0$ is a solution for any $Q$, and $k_{\rm eff} \simeq 1$ (the LWGC is approximately satisfied after Higgsing). On the other hand, if $P \gg e_A/e_B$ then $q_A = Q/P$ is a solution for any $Q \in P \mathbb{Z}$ (with $\varepsilon \geq e_A/(P e_B)$) and $k_{\rm eff} \simeq P$ (the sLWGC is approximately satisfied with sublattice index $P$).

Thus, for $P \gg 1$ and $P \gg e_A/e_B$, it is possible for the sublattice index to be parametrically larger in the Higgsed theory.\footnote{These are the same as the conditions that the IR WGC scale $e_D M_{\rm Pl}^{(D-2)/2}$ is parametrically below the UV WGC scales $e_A M_{\rm Pl}^{(D-2)/2}$ and $e_B M_{\rm Pl}^{(D-2)/2}$, as in~\cite{prashant2016,Saraswat:2016eaz}, though the sublattice index is not necessarily the same as the ratio between these two scales.} Accounting for the change in sublattice index, the ``infrared'' constraint
\begin{equation} \label{eqn:IRconstraint}
\Lambda_{\rm QG} \lesssim (k_{\rm eff} e_D)^{\frac{1}{D-1}} M_{\rm Pl}^{\frac{3(D-2)}{2(D-1)}} \,,
\end{equation}
follows automatically in the UV theory. To see this, note that $e_D \simeq e_A/P$ for $P \gg e_A/e_B$ and $e_D \simeq e_B$ for $P \ll e_A/e_B$ from \eqref{eqn:eD}, where $k_{\rm eff} \simeq P$ and $k_{\rm eff} \simeq 1$ in these two limits, respectively. Thus, \eqref{eqn:IRconstraint} follows from either
\begin{equation} \label{eqn:UVconstraint}
\Lambda_{\rm QG} \lesssim e_A^{\frac{1}{D-1}} M_{\rm Pl}^{\frac{3(D-2)}{2(D-1)}}  \qquad \mbox{or} \qquad \Lambda_{\rm QG} \lesssim e_B^{\frac{1}{D-1}} M_{\rm Pl}^{\frac{3(D-2)}{2(D-1)}} 
\end{equation}
in the two limits.\footnote{When the UV WGC scales $e_A M_{\rm Pl}^{(D-2)/2}$ and $e_B M_{\rm Pl}^{(D-2)/2}$ are both below $\Lambda_{\rm QG}$, the UV theory enforces the stronger constraint $\Lambda_{\rm QG} \lesssim ( e_A e_B )^{1/D} M_{\rm Pl}^{2(D-2)/D} $, but~\eqref{eqn:UVconstraint} holds regardless.}

Of course, since it is not possible to determine the sublattice index in the deep infrared, the constraint \eqref{eqn:IRconstraint} is of no practical use unless we can assume that $k_{\rm eff}$ is not too large. Thus, it would be very interesting to determine whether Higgsing can lead to a very large (or even parametrically large) sublattice index in a real quantum gravity. We now give a suggestive argument that this is unlikely to occur.

As before, suppose that the LWGC is saturated in the UV theory. In the above example we needed a Higgs field with parametrically large charge to obtain a parametrically large $k_{\rm eff}$. To quantify how large this charge is, observe that if the Higgs field were extremal, it would have mass
\begin{equation}
m_{\rm Higgs; ext}^2 = \gamma \frac{e_A^2 e_B^2}{e_D^2} M_{\rm Pl}^{D-2} \,.
\end{equation}
If we demand that this lies below the quantum gravity scale, $\Lambda_{\rm QG}$, then we obtain the constraint
\begin{equation}
\frac{e_A e_B}{e_D} M_{\rm Pl}^{\frac{D-2}{2}} \lesssim \Lambda_{\rm QG} \,,
\end{equation}
where we drop the order-one factor $\gamma$. Combining this with the UV constraint
\begin{equation} \label{eqn:UVconstraint2}
\Lambda_{\rm QG} \lesssim ( e_A e_B )^{\frac{1}{D}} M_{\rm Pl}^{\frac{2(D-2)}{D}} 
\end{equation}
from \eqref{eqn:productbound}, we obtain
\begin{equation}
(e_A e_B)^{\frac{2 (D-1)}{D}} M_{\rm Pl}^{\frac{4(D-1)(D-2)}{D}} \lesssim e_D^2 M_{\rm Pl}^{3 (D-2)} \,,
\end{equation}
hence
\begin{equation}
\Lambda_{\rm QG} \lesssim ( e_A e_B )^{\frac{1}{D}} M_{\rm Pl}^{\frac{2(D-2)}{D}}  \lesssim e_D^{\frac{1}{D-1}} M_{\rm Pl}^{\frac{3 (D-2)}{2 (D-1)}} \,,
\end{equation}
and the naive $k\simeq 1$ constraint is enforced in the infrared theory. 

Note that this is not quite the same as enforcing that $k_{\rm eff} \sim 1$. Rather, we merely showed that under these assumptions the constraint \eqref{eqn:IRconstraint} holds with $k_{\rm eff}$ set to $1$. To illustrate the difference, consider, e.g., the case $e_A = e_B = e$ and $D=4$. Then, the constraint that $m_{\rm Higgs; ext} \lesssim \Lambda_{\rm QG}$ is
\begin{equation}
e^2 (P^2+1) \lesssim e \qquad \Rightarrow \qquad P \lesssim e^{-1/2} \,,
\end{equation}
so for $e \ll 1$ we can have $k_{\rm eff} \simeq P \gg 1$, but nonetheless $e_D \simeq e/P \gtrsim e^{3/2}$, so that
\begin{equation}
\Lambda_{\rm QG} \lesssim e^{1/2} M_{\rm Pl} \lesssim e_D^{1/3} M_{\rm Pl} \,,
\end{equation}
and the IR $k\sim1$ constraint is enforced. The reason for this discrepancy is that there are more near-extremal charged states in the infrared theory than the minimal ones required by the sublattice index $k_{\rm eff}$; in this example, for instance, there are $O(n)$ charged particles with charge $q \simeq k_{\rm eff} n$.

It is not difficult to generalize this line of reasoning to the case of $N$ photons and arbitrary kinetic mixing, using the notation of \eqref{eqn:nphotonsL} and following. The UV constraint on $\Lambda_{\rm QG}$ from \eqref{eqn:productbound} can be written as
\begin{equation} \label{eqn:UVconstraintNphoton}
\Lambda_{\rm QG} \lesssim |\Gamma_Q|^{\frac{1}{N+D-2}} M_{\rm Pl}^{\frac{(N+2)(D-2)}{2(N+D-2)}} \,, 
\end{equation}
for $k\sim 1$, where $\Gamma_Q$ is the UV charge lattice and $|\Gamma_Q|$ is the volume of the fundamental domain of $\Gamma_Q$. We can assume that the Higgs charge $q_a$ is primitive---i.e., not a non-unit multiple of another charge in the charge lattice $\Gamma_Q$---since otherwise we can choose a Higgs field with a smaller charge and the same effect. In this case, $|\Gamma_{\tilde{Q}}| = |\Gamma_Q|/|q|$, where $\Gamma_{\tilde{Q}}$ is the charge lattice after Higgsing. The assumption that $m_{\rm Higgs; ext} \lesssim \Lambda_{\rm QG}$ becomes
\begin{equation}
\gamma |q| M_{\rm Pl}^{\frac{D-2}{2}} \simeq (|\Gamma_Q|/|\Gamma_{\tilde{Q}}|) M_{\rm Pl}^{\frac{D-2}{2}} \lesssim \Lambda_{\rm QG} \,,
\end{equation}
so that combining with \eqref{eqn:UVconstraintNphoton} gives
\begin{equation}
\Lambda_{\rm QG} \lesssim |\Gamma_Q|^{\frac{1}{N+D-2}} M_{\rm Pl}^{\frac{(N+2)(D-2)}{2(N+D-2)}} \lesssim |\Gamma_{\tilde{Q}}|^{\frac{1}{N+D-3}} M_{\rm Pl}^{\frac{(N+1)(D-2)}{2(N+D-3)}} \,,
\end{equation}
and the IR $k\sim 1$ constraint follows from the UV $k\sim 1$ constraint. Thus, to parametrically violate these constraints using Higgsing we need $m_{\rm Higgs; ext} \gg \Lambda_{\rm QG}$.

This can be interpreted in two ways. Firstly, we need $m_{\rm Higgs} \ll \Lambda_{\rm QG}$ in order to describe the Higgsing in effective field theory, hence to violate the $k\sim 1$ constraints the Higgs field must be very superextremal. Secondly---following the arguments of~\S\ref{subsec:basicunification}, in particular \eqref{eqn:averagechargebound}---if weakly coupled gauge theory and gravity emerge from the same strong coupling scale $\Lambda_{\rm QG}$ then to violate the $k\sim 1$ constraints the Higgs field must have a charge much larger than the average charge of other particles within the effective field theory. In fact, in typical examples only a few light particles will be very superextremal, with the rest of the spectrum near-extremal or subextremal, implying that the Higgs field has a charge which is much larger than \emph{almost every other particle} in the effective field theory.

Although these suggestive arguments do not rule anything out, they show that violating the $k\simeq 1$ constraints on $\Lambda_{\rm QG}$ through Higgsing in a way consistent with our other assumptions about emergence from a UV cutoff requires the Higgs field to have peculiar properties such as a charge that is much larger than most or all of the other particles in the effective field theory. For this reason---and because we know of no quantum gravities with a very large sublattice index---we expect that the the UV cutoff bounds such as \eqref{eq:LambdaQGU1}, \eqref{eqn:simpleGbound}, or \eqref{eqn:productbound} with $k \sim 1$ are never parametrically violated.

\section{Caveats}\label{sec:caveats}

\subsection{String theory at weak coupling} \label{sec:weakstrings}

We have discussed examples in which the spectrum of particles approximately saturates sLWGC bounds. This is characteristic of Kaluza-Klein theories, for example, and related quantum gravities such as large volume compactifications of M-theory. Of course, other examples of weakly coupled gauge theory can arise in string theory at $g_s \ll 1$.

Does gauge-gravity unification arise in such theories? If we naively compute $\lambda_{\rm gauge}(E)$ and $\lambda_{\rm grav}(E)$
 in a weakly coupled string theory for energies $E$ above the string scale, we find that both grow very rapidly, a simple consequence of the Hagedorn density of states $\rho(E) \sim \exp(E/T_H)$. However, states with higher charge come with lower multiplicities, and as a consequence $\langle q^2 \rangle \propto E$, even though $q_{\rm max}^2 \propto E^2$. Thus, well above the string scale $\lambda_{\rm gauge}(E) \ll \lambda_{\rm grav}(E)$.

We illustrate this in the simple example of ten-dimensional heterotic string theory, with spectrum determined by the conditions
\begin{equation}
\frac{\alpha'}{4} m^2 = N_L + \frac{1}{2} Q^2 - 1 = N_R \,.
\end{equation}
Here $N_{L,R} \in \mathbb{Z}_{\ge 0}$ count left and right-moving oscillators, and each comes with an associated multiplicity $d_L(N_L)$ and $d_R(N_R)$, equal to the multiplicity at the $N$th level of the open bosonic string and the open superstring, respectively. 
Thus, the number states at a given mass-level $\frac{\alpha'}{4} m^2 = N$ is
\begin{equation}
d(N) = \sum_{Q \in \Gamma}^{Q^2 \le 2 (N+1)} d_L(N + 1 - Q^2/2)\, d_R(N) \,. 
\end{equation}
To estimate this, we use the asymptotic formulae \cite[\S 2.3, 5.3]{Green:1987sp}
\begin{equation}
d_L(n) \sim \frac{e^{4 \pi \sqrt{n}}}{n^{27/4}} \,, \qquad d_R(n) \sim \frac{e^{\pi \sqrt{8 n}}}{n^{11/4}} \,,
\end{equation}
up to order-one constants.
Thus,
\begin{equation}
d(N) \sim d_R(N) \int_{Q^2 \le 2 N} \frac{e^{4 \pi \sqrt{N+1-Q^2/2}}}{(N+1-Q^2/2)^{27/4}}\, d^{16} Q \sim d_R(N) \int \frac{e^{4 \pi \sqrt{N} - \frac{\pi}{\sqrt{N}} Q^2}}{N^{27/4}}\, d^{16} Q \sim \frac{e^{2 \pi (2 + \sqrt{2}) \sqrt{N}} }{N^{11/2}}\,,
\end{equation}
where in the second step we use the fact that the integrand is dominated by the region $Q^2 \lesssim O(N^{1/2})$. This agrees with, e.g., \cite[\S 6.4]{Green:1987sp}.

Thus,
\begin{equation}
\lambda_{\rm grav}(E) \sim g_s^2 \frac{e^{2 \pi (2 + \sqrt{2}) \sqrt{N}}}{N^{3/2}} \,, \qquad N =\frac{\alpha'}{4} E^2\,.
\end{equation}
To compute $\lambda_{\rm gauge}$, we follow the same steps but count states weighted with $Q^2_1$ for some particular Cartan generator $Q_1$. By a straightforward calculation we obtain $\langle Q^2_1 \rangle \simeq \frac{1}{2\pi} \sqrt{N}$ as well as
\begin{equation}
\lambda_{\rm gauge}(E) \sim g_s^2 \frac{e^{2 \pi (2 + \sqrt{2}) \sqrt{N}}}{N^2} \,, \qquad N =\frac{\alpha'}{4} E^2\,.
\end{equation}
Therefore, both $\lambda_{\rm grav}$ and $\lambda_{\rm gauge}$ grow rapidly above the string scale, but $\lambda_{\rm gauge}$ grows slightly less rapidly.

The same conclusion should follow for NSNS charges in an arbitrary string theory. A rough argument is as follows: the modular invariance argument of~\cite{Heidenreich:2016aqi,Montero:2016tif} implies that the multiplicities depend only on $t_{L,R} \df \Delta_{L,R} - \frac{1}{2} Q_{L,R}^2$ (and some additional discrete data). Thus, if the multiplicities in the neutral sector, $Q_L = Q_R = 0$, are Hagedorn then $d_L(t_L) \sim e^{a_L \sqrt{t_L}}$ and $d_R(t_R)\sim e^{a_R \sqrt{t_R}}$ are Hagedorn. We find the density of states
\begin{equation}
d(\Delta) = \sum_{Q \in \Gamma}^{Q_{L,R}^2 \le 2 (\Delta+1)} d_L(\Delta- Q_L^2/2)\, d_R(\Delta- Q_R^2/2) \,,
\end{equation}
or
\begin{equation}
d(\Delta) \sim \int e^{a_L \sqrt{\Delta- Q_L^2/2}+a_R \sqrt{\Delta- Q_R^2/2}} d^r Q \sim \int e^{ (a_L+a_R) \sqrt{\Delta} -\frac{a_L}{4 \sqrt{\Delta}} Q_L^2 -\frac{a_R}{4 \sqrt{\Delta}} Q_R^2} d^r Q \sim e^{(a_L+a_R) \sqrt{\Delta}} \,,
\end{equation}
to leading order. Redoing this calculation weighted by $Q_1^2$ for some left-moving charge $Q_1$, we find
\begin{equation}
\langle Q_1^2 \rangle \sim \frac{2}{a_L} \sqrt{\Delta} \,, \qquad \Delta = \frac{\alpha'}{4} E^2 \,,
\end{equation}
and likewise for right-moving charges, hence $\lambda_{\rm gauge}(E) \ll \lambda_{\rm grav}(E)$ far above the string scale, as before.

What should we make of this? It is clear that in string theory there {\em is} a sense in which both gauge theory and gravity are emergent. What has really broken down is our ability to use simple, field-theoretic one-loop arguments to discuss the relative strength of gauge theory and gravity. One way to argue this is that the particles running in loops, for mass well above the string scale, are not {\em particles} at all: they are extended objects, and their couplings should involve form factors. It is a familiar property of closed string worldsheet perturbation theory that naive quantum field theoretic expectations about the behavior of loops are modified due to modular invariance.

It is unclear what energy scale we should call $\Lambda_{\rm QG}$ in weakly coupled string theory. It is tempting to say that it is the string scale, since quantum field theory breaks down there. Such an identification has been argued for in the context of the species bound, with an effective number of species $1/g_s^2$ \cite{Dvali:2009ks, Dvali:2010vm}. In other words, the explosive Hagedorn growth of the density of states may translate into an effectively finite number of degrees of freedom from the point of view of black hole evaporation or of loop corrections to the Planck mass.

On the other hand, above we have taken $\Lambda_{\rm QG}$ to be the energy at which a theory can no longer be viewed as weakly coupled in any sense. When $g_s \ll 1$, string theory {\em is} still weakly coupled at the string scale---it is simply not a field theory. Above the string scale it is no longer straightforward---and perhaps not possible at all---to distinguish between gauge forces, gravitational forces, and other interactions; $\lambda_{\rm gauge}$ and $\lambda_{\rm grav}$ as we have defined them become meaningless. Whether some improved notion can be found is beyond the scope of this paper.

Nonetheless, we can still ask whether unification occurs \emph{at or below} the string scale. We again consider ten-dimensional heterotic string theory as an example. We have
\begin{equation}
G_N \sim  g_s^2/ M_s^{8} \,, \qquad g^2 \sim g_s^2 / M_s^{6} \,, 
\end{equation}
which gives
\begin{equation}
\lambda_{\rm grav} \sim g_s^2 (E/M_s)^8 \,, \qquad \lambda_{\rm gauge} \sim g_s^2 (E/M_s)^6 \,,
\end{equation}
below the string scale, up to numerical constants. Thus, well below the string scale $\lambda_{\rm grav} \ll \lambda_{\rm gauge}$, but at the string scale itself $\lambda_{\rm gauge} \sim \lambda_{\rm grav}$, at least parametrically in $g_s \ll 1$.

Similarly, if we compactify heterotic string theory on a rectangular $p$-torus with radii $R_1, \ldots, R_p \gg \ell_s$ and no Wilson lines then
\begin{equation}
G_N \sim  \frac{g_s^2}{M_s^8 R_1 \ldots R_p}  \,, \qquad g^2 \sim \frac{g_s^2}{M_s^{6} R_1 \ldots R_p} \,, \qquad g_i^2 \sim \frac{g_s^2}{M_s^8 R_i^2 R_1 \ldots R_p} \,,
\end{equation}
where $g_i$ denotes the gauge coupling of the KK photon associated to the $R_i$ circle.\footnote{In addition, there are $p$ gauge bosons associated to the reduction of the Kalb-Ramond $B$-field, but the lightest charged particles are well above the string scale; these are discussed further in the next subsection.} 
Below the compactification scale, we find
\begin{equation}
\lambda_{\rm grav} \sim \frac{g_s^2 E^{8-p}}{M_s^8 R_1 \ldots R_p} \,, \qquad \lambda_{\rm gauge} \sim \frac{g_s^2 E^{6-p}}{M_s^{6} R_1 \ldots R_p} \,,
\end{equation}
with no KK modes contributing, and thus nothing charged under the KK photons. 
Above the compactification scale but below the string scale, we find
\begin{equation}
\lambda_{\rm grav} \sim g_s^2 (E/M_s)^8 \,, \qquad \lambda_{\rm gauge} \sim g_s^2 (E/M_s)^6 \,, \qquad \lambda_{\mathrm{KK}, i} \sim g_s^2 (E/M_s)^8 \,,
\end{equation}
by counting KK modes. Here $\lambda_{\rm KK} \sim \lambda_{\rm grav}$ as expected from the general argument of~\S\ref{sec:generalargument}, but still $\lambda_{\rm gauge} \gg \lambda_{\rm grav}$. At the string scale, however, we find $\lambda_{\rm gauge} \sim \lambda_{\rm grav} \sim \lambda_{\rm KK} \sim g_s^2$, and the forces ``unify'' in the sense of~\S\ref{sec:lambdasequal}.

On the other hand, consider type I string theory in ten dimensions, with gauge fields in the open string sector. In this case,
\begin{equation}
G_N \sim  g_s^2/ M_s^{8} \,, \qquad g^2 \sim g_s / M_s^{6} \,, 
\end{equation}
which gives
\begin{equation}
\lambda_{\rm grav} \sim g_s^2 (E/M_s)^8 \,, \qquad \lambda_{\rm gauge} \sim g_s (E/M_s)^6 \,,
\end{equation}
so that at the string scale $\lambda_{\rm gauge} \sim g_s \gg \lambda_{\rm grav} \sim g_s^2$, with similar results upon toroidal compactification and in T-dual toroidal orientifolds of type II string theories with D-branes.

Thus, in these examples
\begin{equation}
\lambda_{\rm grav}(M_s) \sim g_s^2 \,, \qquad  \lambda_{\rm gauge}^{\rm (closed)}(M_s) \sim g_s^2 \,, \qquad \lambda_{\rm gauge}^{\rm (open)}(M_s) \sim g_s \,. \label{eqn:stringuni}
\end{equation}
We expect this simple result to generalize to a broad class of perturbative string theories with charged particles at or below the string scale. In such examples $\lambda_{\rm gauge} \gtrsim \lambda_{\rm grav}$ at the string scale---regardless of whether the gauge bosons come from closed or open strings---implying the Weak Gravity Conjecture up to order-one factors by the arguments of \S\ref{sec:lambdasequal}.

Gauge fields in the RR sector are a potential exception to~(\ref{eqn:stringuni}), but in this case the charged objects are wrapped D-branes. If the cycle in question is large in string units then the wrapped branes are heavy, and don't appear in the low-energy effective field theory (see~\S\ref{subsec:heavyspectra} for further discussion). However, in K3 and Calabi-Yau compactifcations of type II string theory, shrinking two-, three- and four-cycles can appear at singular points in the moduli space while maintaining a large overall volume. These can lead to light RR charged states, as in, e.g., \cite{Strominger:1995cz}. We won't attempt to address the issue of force unification in such cases in this paper, but we discuss the related issue of ultralight charged particles in four dimensions in~\S\ref{subsec:conifolds}.

\subsection{Heavy spectra} \label{subsec:heavyspectra}

Throughout this paper, we have assumed that there are charged particles with masses below the quantum gravity scale. We can heuristically motivate this assumption in a four dimensional weakly coupled gauge theory by noting that the WGC scale $e M_{\rm Pl}$ is below the Planck scale, and the sLWGC suggests that charged particles must appear at or below this scale. However, this argument is too naive, and fails even in the case of two weakly coupled photons $A,B$ with freely adjustable couplings $e_{A,B}$. If $e_A \lesssim e_B^3 \ll e_B$ then the WGC scale $e_B M_{\rm Pl}$ for the second gauge theory lies {\em above} $\Lambda_{\rm QG} \lesssim e_A^{1/3} M_{\rm Pl}$, so there is no reason for  $B$-charged particles to appear below the quantum gravity scale.

In fact, $e M_{\rm Pl}^{(D-2)/2} \gtrsim \Lambda_{\rm QG}$ occurs frequently in real quantum gravities, and in many such cases there are no charged particles below $\Lambda_{\rm QG}$. A simple example of this is the RR photon $C_1$ in ten-dimensional type IIA string theory, for which
\begin{equation}
g^2 \sim 1/M_s^6 \,, \qquad M_{\rm Pl}^8 \sim M_s^8/g_s^2 \,,
\end{equation}
so that
\begin{equation}
g M_{\rm Pl}^4 \sim M_s / g_s \gg M_s \,.
\end{equation}
The lightest charged object is the D$0$ brane which has mass $g M_{\rm Pl}^4 \sim M_s / g_s$ (up to order-one factors), so there are no charged particles below $\Lambda_{\rm QG} \sim M_s$.

There are many other string theory examples of a similar nature where the charged objects are BPS branes wrapped on cycles. For instance, consider type II string theory compactified on a circle of radius $R$. The Kalb-Ramond $B$-field generates a photon whose charged states are wound strings. The lightest of these has mass of order $R M_s^2$, which for a large torus $R \gg \ell_s$ is well above the string scale. A parametrically similar scaling arises in the WGC bound for gauge fields on D$7$ branes in approximately isotropic, large-volume compactifications of the IIB string.

In these examples, the WGC scale is above the string scale and so the simple perturbative field theory arguments we have given throughout the paper do not apply. The underlying gauge theories may still be thought of as emerging, in some sense, from the quantum gravity scale. In the case of the D$0$ brane this becomes manifest in the $g_s \gg 1$ limit where they are simply KK modes; in the case of winding strings, the U($1$) symmetry arises from the $B$-field which is part of the supergravity multiplet, and which is T-dual to a graviphoton in toroidal examples. It may be worthwhile to search for a modified version of our arguments that can apply to examples like these. For now, we simply highlight them as a shortcoming of our approach.

\subsection{Logarithmic running and ultralight particles} \label{subsec:conifolds}

In our previous discussion we have ignored the possibility of large logarithms. This is justified in many cases. Consider for example the one-loop beta function of a KK photon in four dimensions:
\begin{equation}
\frac{1}{e^2(\mu)} \simeq \frac{1}{2} R^2 M_{\rm Pl}^2 - \frac{b}{8 \pi^2} \sum_{n=1}^{\lfloor \mu R \rfloor} n^2 \log \frac{\mu R}{n} \,,
\end{equation}
for some order-one constant $b$, where for simplicity we assume no massless particles in the original five-dimensional theory. We have
\begin{equation}
\sum_{n=1}^{N} n^2 \log \frac{N}{n} \sim \frac{N^3}{9} - \frac{N}{12} + \ldots \,, \qquad \mbox{whereas} \qquad \sum_{n=1}^{N} n^2 \sim \frac{N^3}{3} + \frac{N^2}{2} + \ldots \,,
\end{equation}
at large $N$, so we obtain the same behavior at large $N$---up to order-one constants---regardless of whether we include the log or omit it. This is because for most terms in the sum the logarithm is not large: the KK modes become increasing dense near the cutoff on a logarithmic scale even as they are spaced evenly on a linear scale.

For this reason, we expect that logarithmic corrections can be consistently neglected to leading order in many of our calculations. However, there are certain circumstances in which this is not the case. For instance, in a four dimensional theory with a light charged particle electric forces are screened at large distances. As explained in the introduction, if there are \emph{massless} charged particles then screening continues at arbitrarily large distances and parametrically large black holes can carry a parametrically large charge-to-mass ratio. This precludes an infinite tower of superextremal resonances, and the sLWGC cannot hold in its original form.

In the remainder of this section, we provide some preliminary discussion of how logarithms can be accounted for in our analysis (focusing on the four-dimensional case of most interest), and what this tells us about theories with ultralight charged particles.

Consider a $\U(1)$ gauge theory coupled to gravity in four dimensions, with the one-loop renormalized gauge coupling
\be
\frac{1}{e^2(E)} = \frac{1}{e^2_{\rm IR}} - \sum_{i:\, m_i < E} \frac{b_i}{8\pi^2} q_i^2 \log \frac{E}{m_i}, \label{eq:ebeta1}
\ee
where the $b_i$ are order-one constants and we neglect threshold corrections for simplicity. Requiring that the Landau pole occurs at or below $\Lambda_{\rm QG}$ gives the condition
\be
\frac{1}{e^2_{\rm IR}} \sim \sum_{i:\, m_i < \Lambda_{\rm gauge}} \frac{b_i}{8\pi^2} q_i^2 \log \frac{\Lambda_{\rm gauge}}{m_i}\,, \qquad \Lambda_{\rm gauge} \lesssim \Lambda_{\rm QG} \,,
\ee
analogous to~(\ref{eq:strongatLambda}). By a similar line of reasoning to before,
\begin{align}
\frac{1}{e^2_{\rm IR}} &\lesssim z_{\rm max}^2 \sum_{i:\, m_i < \Lambda_{\rm gauge}} \frac{b_i}{8\pi^2} m_i^2 \log \frac{\Lambda_{\rm gauge}}{m_i} \nonumber \\
&\lesssim z_{\rm max}^2 N(\Lambda_{\rm gauge})\, \Lambda_{\rm gauge}^2 \nonumber \\
&\lesssim z_{\rm max}^2 M_{\rm Pl}^2 \,,
\end{align}
where $z_{\rm max} \df (q/m)_{\rm max}$, on the second line we use the fact that $x^2 \log(1/x) \le 1/(2\E)$ for $0 \le x \le 1$ and drop order-one factors, and on the third line we apply the species bound. Thus, gauge-gravity unification in the sense of~\S\ref{sec:warmup} still implies the WGC up to order-one factors, even when there are large logarithms.

What about the sLWGC? We have already argued that it cannot hold in its original form in situations with very light particles in four dimensions. In~\S\ref{sec:lambdasequal}, we saw that perturbative gauge-gravity unification, in the form
\be
\lambda_{\rm gauge}(E) \sim \lambda_{\rm grav}(E) \qquad \mbox{for $E\gtrsim E_0$}
\ee
has similar consequences to the sLWGC. It is interesting to ask what this means in situations with ultralight particles in four dimensions.

To do so, we need to define $\lambda_{\rm gauge}(E)$ properly in the presence of large logarithms. Recall that in an abelian four-dimensional gauge theory, we had previously
\be
\lambda_{\rm gauge}(E) \df \frac{e^2}{16 \pi^2} \sum_{i:\, m_i < E} q_i^2 \,.  \label{eq:lambdagauge1}
\ee
where now we will be slightly more careful about tracking loop factors.\footnote{We keep the loop factor to be consistent with the loop factor which appears in~(\ref{eq:ebeta1}). We should also keep the loop factor in $\lambda_{\rm grav}$, roughly $\lambda_{\rm grav}(E) = \frac{\kappa^2 E^2}{16\pi^2} \sum_{i:m_i<E} d(i)$, where $\kappa \df \sqrt{8 \pi G} = 1/M_{\rm Pl}$ is the gravitational coupling.}
Since $\lambda_{\rm gauge}$ is intended as a heuristic measure of the size of loop corrections from light particles at a scale $E$, we interpret $e^2$ in~(\ref{eq:lambdagauge1}) as the renormalized gauge coupling at this scale, given by~(\ref{eq:ebeta1})
at one-loop order. We might try to compute the physics at the scale $E$ with couplings renormalized in the deep infrared instead, but then the loop expansion can break down due to large logarithms, hence $\lambda_{\rm gauge}(E)$ computed with the gauge coupling renormalized at $E$ is a better measure of the validity of the loop expansion.

As a consistency check, we verify that with this definition $\lambda_{\rm gauge}(E) \sim 1$ always signals an imminent Landau pole. Writing this out, we obtain
\be
\frac{1}{e^2_{\rm IR}} - \sum_{i:\, m_i < E} \frac{b_i}{8\pi^2} q_i^2 \log \frac{E}{m_i} \sim \frac{1}{16 \pi^2} \sum_{i:\, m_i < E} q_i^2 \,,
\ee
but then:
\be
\frac{1}{e^2_{\rm IR}} \sim \sum_{i:\, m_i < E} \frac{b_i}{8\pi^2}\, q_i^2 \log \frac{E'}{m_i} \,,
\ee
for $E' \sim E \exp(1/2b)$. Thus, the Landau pole is nearby on a log scale.

Suppose that gauge interactions are strong compared to gravitational interactions at some scale $E \lesssim \Lambda_{\rm QG}$, in the sense that $\lambda_{\rm gauge}(E) \gtrsim \lambda_{\rm grav}(E)$ as in~\S\ref{sec:lambdasequal}. Thus,
\begin{equation}
\frac{1}{e^2(E)} \lambda_{\rm gauge}(E) \gtrsim \frac{1}{e^2_{\rm IR}} \lambda_{\rm grav}(E) - \lambda_{\rm grav}(E) \sum_{i:\, m_i < E} \frac{b_i}{8\pi^2} q_i^2 \log \frac{E}{m_i} \,.
\end{equation}
This can be rearranged to give
\begin{equation}
\frac{1}{N(E)} \sum_{i:\, m_i < E} \Biggl[\frac{1+ 2 b_i \lambda_{\rm grav}(E) \log \frac{E}{m_i}}{E^2/m_i^2}\Biggr] \frac{q_i^2}{m_i^2}   \gtrsim \frac{1}{e^2_{\rm IR} M_{\rm Pl}^2}  \,.
\end{equation}
Since $\lambda_{\rm grav}(E) \lesssim 1$ by assumption the prefactor to the log is at most order-one, and the quantity in brackets is order-one or smaller for any $E>m_i$.
 It follows that $\langle q^2/m^2 \rangle_{m  \le E} \gtrsim 1/(e^2_{\rm IR} M_{\rm Pl}^2)$, i.e., the average particle with $m<E$ is superextremal and the WGC is satisfied.

Next, suppose that $\lambda_{\rm gauge}(E) \simeq \lambda_{\rm grav}(E)$ above some scale $E_0 \ll \Lambda_{\rm QG}$. Following the same steps as in~\S\ref{sec:lambdasequal}, we obtain
\begin{equation}
M_{\rm Pl}^2 \frac{d (\sum q^2) }{ d N} \simeq \frac{E^2}{e^2(E)} +2 E^2 \frac{d \log E}{d \log N} \biggl(\frac{1}{e^2(E)}-\sum_{i:\, m_i < E} \frac{b_i}{16\pi^2} q_i^2\biggr) \,, \label{eqn:lambdamatch2}
\end{equation}
where $d \log E / d \log N \ge 0$ as before. Unless $E$ is very close to a Landau pole, hence by assumption $E \sim \Lambda_{\rm QG}$, it is straightforward to check that the second term in the parenthesis must be small compared to the first, and we conclude that
\begin{equation} \label{eqn:sLWGClike2}
e^2(E)\, \langle q^2 \rangle_{m \simeq E} \gtrsim \frac{E^2}{M_{\rm Pl}^2} \,.
\end{equation}
On the other hand, when $E \sim \Lambda_{\rm QG}$ effective field theory begins to break down, so there is no point analyzing the case where the second term in parenthesis in~(\ref{eqn:lambdamatch2}) becomes significant.

Note that (\ref{eqn:sLWGClike2}) is very similar to (\ref{eqn:sLWGClike}), except that the renormalized gauge coupling $e^2(E)$ appears explicitly. Thus, if gauge and gravitational forces unify (in the sense of $\lambda_{\rm gauge}(E) \simeq \lambda_{\rm grav}(E)$) above some scale $E_0 \ll \Lambda_{\rm QG}$, there must be a tower of charged particles above this scale which are ``superextremal'' in the renormalized sense, $m^2 \lesssim q^2 e^2(m) M_{\rm Pl}^2$.  This lends support to a renormalized version of the sLWGC even for 4d theories with massless charged particles involving the renormalized gauge coupling $e(m)$ rather than the infrared value coupling $e_{\rm IR}$, which runs to zero.

Note that if $d \log E / d \log N \lesssim O(1)$---implying at least a power-law growth in the number of states with energy about $E_0$---then $e^2$ cannot change very much above $E_0$ until very near $\Lambda_{\rm QG}$. This is because, on the one hand,
\begin{equation}
\sum_{i:\, m_i < E} \frac{b_i}{8\pi^2} q_i^2 \ll \frac{1}{e^2(E)} \qquad \mbox{for} \qquad E \ll \Lambda_{\rm QG} \,, 
\end{equation}
as we already argued, but also
\begin{equation}
\frac{1}{e^2(E_0)} - \frac{1}{e^2(E)} \lesssim \sum_{i: m_i < E} \frac{b_i}{8 \pi^2} q_i^2 \,,
\end{equation}
since a power law density of states makes the logs small on average. Therefore, $\frac{1}{e^2(E_0)} - \frac{1}{e^2(E)} \ll \frac{1}{e^2(E)}$, or
\begin{equation}
\frac{e^2(E)}{e^2(E_0)} - 1 \ll 1
\end{equation}
and the tower of charged states must also satisfy $m^2 \lesssim q^2 e^2(E_0) M_{\rm Pl}^2$.

On the other hand, $e^2_{\rm IR}$ can be very different than $e^2(E_0)$ if there are ultralight charged particles, and the tower of states above $E_0$ may be very subextremal with respect to $e^2_{\rm IR}$ as a result. Nonetheless, as we showed above, the Weak Gravity Conjecture follows automatically (in the usual form $m^2 \lesssim e_{\rm IR}^2 q^2 M_{\rm Pl}^2$), because by assumption $\lambda_{\rm gauge}(E_0) \simeq \lambda_{\rm grav}(E_0)$. The mechanism for this is conceptually simple: if there are charged particles light enough to substantially renormalize $e^2_{\rm IR}$ versus $e^2(E_0)$ then these charged particles are necessarily superextremal.

By turning around the above arguments, its easy to see that a tower of charged particles beginning at some scale $E_0$ with masses $m^2 \sim q^2 e^2(E_0) M_{\rm Pl}^2$ will lead to gauge-gravity unification. Thus, it is natural to expect that the sLWGC takes this form in the presence of ultralight charged particles, where $E_0$ is roughly the WGC scale, $E_0 \sim e(E_0) M_{\rm Pl}$. In particular, bounds on the quantum gravity scale for gauge theories with ultralight particles should constrain the \emph{renormalized} gauge coupling at the WGC scale. For instance
\be
\Lambda_{\rm QG} \lesssim \bigl[e(E_0)\bigr]^{1/3} M_{\rm Pl} \,, \label{eqn:renormcutoffbound}
\ee
in the case of a single $\U(1)$. If we instead put the infrared gauge coupling into~(\ref{eqn:renormcutoffbound}) it is easy to derive wrong statements.

The above discussion suggests how our arguments might be extended to four dimensional quantum gravities with ultralight particles, and gives a rough idea of what form a modified sLWGC might take in such theories while still implying the ordinary WGC. We leave further discussion of these interesting questions to a future work.

\section{Phenomenological applications}\label{sec:pheno}

Effective field theory breaks down irrevocably above the scale $\Lambda_{\rm QG}$. Thus the bounds that we have derived potentially constrain physics beyond the Standard Model with very small gauge couplings. In \cite{Heidenreich:2016aqi} we already pointed out one implication: if an unbroken U($1$)$_{B-L}$ gauge theory exists, it is constrained to have such a tiny gauge coupling $e \lesssim 10^{-24}$ \cite{wagner:2012ui,heeck:2014zfa} that even the modest U($1$) sLWGC bound $\Lambda_{\rm QG} \lesssim e^{1/3} M_{\rm Pl}$ would tell us that there is no weakly coupled physics above $10^{10}~{\rm GeV}$.

Here we will focus on new physics involving nonabelian gauge theories, for which the bound derived in this paper is stronger than the one stated in \cite{Heidenreich:2016aqi}. If the coupling is sufficiently small, we might rule out interesting physics at high energy scales. In particular, several large energy scales that are often phenomenologically relevant include:
\begin{itemize}
\item The GUT scale, $M_{\rm GUT} \approx 2 \times 10^{16}~{\rm GeV}$.
\item The energy density during inflation, $V_{\rm inf}^{1/4} \approx r^{1/4} \times 3.1 \times 10^{16}~{\rm GeV}$.
\item The Hubble scale during inflation, $H_{\rm inf} \approx r^{1/2} \times 2.4 \times 10^{14}~{\rm GeV}$.
\item The seesaw mass of right-handed neutrinos, $M_{\rm N} \sim y^2 \times 6 \times 10^{14}~{\rm GeV} \times \frac{0.1~{\rm eV}}{m_\nu}$.
\item The string scale $M_{\rm string} \sim g_s M_{\rm Pl}/\sqrt{\cal V}$ where $\cal V$ is the volume of the internal six dimensions in string units.
\item The QCD axion decay constant $f_a$, which is around $10^{12}~{\rm GeV}$ for conventional axion cold dark matter scenarios but could be larger in other scenarios.
\item The SUSY-breaking scale $\sqrt{F_0} \sim \sqrt{m_{3/2} M_{\rm Pl}}$, which is larger than $6 \times 10^{11}~{\rm GeV}$ if we demand that gravitinos decay before BBN (and can be even larger in sequestered scenarios).
\end{itemize}
We should be careful about drawing too-hasty conclusions about which of these scales must be below $\Lambda_{\rm QG}$ in a consistent theory. Any scale that we treat as the {\em mass} scale of a weakly coupled particle, including $M_{\rm GUT}$, $M_{\rm N}$, and $M_{\rm string}$, are bounded above by $\Lambda_{\rm QG}$. However, {\em expectation values of fields} are not obviously constrained in this way; for one familiar example, consider the transplanckian field ranges in large-field inflation, which may have potential tensions with quantum gravity but certainly do not with effective field theory. The Hubble constant during inflation, $H_{\rm inf}$, should also be below $\Lambda_{\rm QG}$ as it corresponds to the curvature scale of space. However, the {\em energy density} in a spacetime is less obviously bounded. On the other hand, in many concrete scenarios, such as a natural inflation model where the potential $V_{\rm inf}$ is generated by confinement \cite{freese:1990rb}, there will be such a bound. Similarly, in many axion theories, for instance of the KSVZ type, there are physical particle masses of order $f_a$ which must be below $\Lambda_{\rm QG}$.

One general consideration for $\Lambda_{\rm QG}$ is the extremely strong experimental constraint on proton decay. Processes like $p \to e^+ \pi^0$ or $p \to K^+ \overline{\nu}$ arise from dimension-six operators of the type $QQQL$ or $u^c u^c d^c e^c$ in the Standard Model. Super-Kamiokande has constrained the lifetime of these processes to be larger than about $10^{34}$ years \cite{Miura:2016krn}. If we write the operators suppressed simply by $\Lambda_{\rm QG}$, we obtain a bound
\begin{align}
\Lambda_{\rm QG} \gtrsim 2 \times 10^{16}~{\rm GeV} \approx M_{\rm GUT}.
\end{align}
This means that in a quantum gravity theory with no additional structure, proton decay requires a large $\Lambda_{\rm QG}$ and is inconsistent with the existence of very weakly coupled gauge theories. In some theories, this bound becomes much stronger. For instance, in the context of theories with approximate supersymmetry, we have dimension-five proton decay from superpotential operators (after using $R$-parity to forbid the dimension-four terms). In such theories the bound would require not only a large $\Lambda_{\rm QG}$ but also a sufficiently large scale of supersymmetry breaking \cite{Weinberg:1981wj,Ellis:1981tv,Murayama:2001ur,Dine:2013nga}. However, it is worth keeping in mind that certain quantum gravity theories may contain special structure that allows $\Lambda_{\rm QG}$ to be much smaller than this naive estimate. In the extreme limit, the proton could even be absolutely stable due to a discrete gauge symmetry like baryon triality \cite{Ibanez:1991hv}. More generally, approximate flavor symmetries (e.g.~arising from massive U(1) gauge bosons obtaining stringy St\"uckelberg masses) could provide additional spurions suppressing the decay rate. Hence we cannot make absolute statements, except that the discovery of a very small gauge coupling in nature could be consistent with the sLWGC only if additional structure exists to protect the proton.

Nonabelian gauge groups with small couplings have been advocated in several cosmological contexts, which we will now summarize.

\subsection{Nonabelian dark radiation interacting with dark matter}

Nonabelian dark radiation interacting with dark matter has the unusual property that, due to the $t$-channel scattering diagram, the scattering rate $\Gamma \sim T^2$. This is the same scaling as the Hubble rate during a radiation-dominated era, and so the scattering does not decouple. This can lead to striking cosmological consequences even for small couplings \cite{buen-abad:2015ova}. Large couplings would predict sizable deviations from $\Lambda$CDM cosmology that have not been observed, so any phenomenologically viable version of this scenario potentially has an sLWGC constraint.

Studies have found that tensions in CMB data (involving the values of the Hubble constant and $\sigma_8$) may be partially relaxed in such a model with gauge couplings $g \sim 2 \times 10^{-4}$ \cite{lesgourgues:2015wza, Cyr-Racine:2015ihg, Buen-Abad:2017gxg}, though early Lyman-alpha data diminishes the significance \cite{Krall:2017xcw}. (Likelihoods are not yet available to test this model with more recent Lyman-alpha data.) If this scenario is true, then for an SU($2$) gauge theory we would have
\be
\Lambda_{\rm QG} \lesssim g^{1/2} M_{\rm Pl} \approx 3 \times 10^{16}~{\rm GeV}.
\ee 
This would be in modest tension with GUT unification or with the value of $V^{1/4}$ in a just-around-the-corner detection of $r$, since both scenarios involve physics very near the scale $\Lambda_{\rm QG}$. Theories with larger SU(N) groups would have stronger constraints. However, similar cosmological phenomenology can be obtained in models with larger couplings but with only a fraction of dark matter interacting with dark radiation \cite{Chacko:2016kgg, Buen-Abad:2017gxg}. Further observations of the matter power spectrum at smaller scales would be needed to distinguish the signatures of these models.

The sLWGC tower would involve particles with nonabelian gauge charges beginning at a mass scale of $g M_{\rm Pl} \sim 5 \times 10^{14}~{\rm GeV}$, which is heavy enough relative to the Hubble scale during inflation as to not be an obvious problem on its own. On the other hand, if $r$ is relatively large these particles, with masses only an order-one factor above the Hubble scale, might leave detectable imprints in non-Gaussianities.

We emphasize that in this case, the tension would be between the interacting nonabelian dark radiation scenario and {\em other}, unrelated physics, like GUTs. Experimental confirmation of this cosmological scenario would not, in itself, disprove the sLWGC.

\subsection{Chromonatural inflation}

There are few known cosmological mechanisms for generating detectable primordial tensor modes. The most familiar is large-field inflation. In recent years another scenario, chromonatural inflation \cite{Adshead:2012kp}, has claimed to generate detectable tensor modes via a different mechanism \cite{Anber:2012du, Adshead:2013qp, Adshead:2013nka}. Tensor modes arise as the product of a classical gauge field background and perturbations in the gauge field. The classical gauge field background spontaneously breaks the product of spatial rotations and nonabelian gauge symmetries to a diagonal, as first suggested in the related theory of gauge-flation \cite{Maleknejad:2011jw}. 

Aside from kinetic terms, the Lagrangian for chromonatural inflation includes
\be
-\mu^4 \left[1 + \cos(\chi/f)\right] - \frac{\lambda}{8 f} \chi F^a_{\mu \nu} {\widetilde F}^{a \mu \nu}.
\ee
Here $\chi$ is an axion field, and its coupling to the gauge fields (which are approximately static) generates an effective friction term that modifies the evolution compared to standard slow-roll inflation. Demanding that the model gives rise to inflation with sufficiently many e-folds and matches the observed values of the scalar power spectrum amplitude ${\cal P}_s \sim 2 \times 10^{-9}$ and of the spectral index $n_s$ tightly constrains the available parameter space \cite{Dimastrogiovanni:2012st, Adshead:2013qp, Adshead:2013nka}. In fact, the minimal version of the model predicts  a primordial tensor signal $r$ that is too large, but a modified Higgsed version of the theory is compatible with data without qualitatively changing the properties of the model \cite{Adshead:2016omu}.

The upshot of the fit to data is that the standard chromonatural inflation benchmark models have $g \sim 10^{-6}$, a small number that is approximately obtained as ${\cal P}_s^{1/2}$ multiplied by order-one numbers. (For a full explanation, we refer readers to the literature.) The model parametrically favors $\mu \sim g^{1/2} M_{\rm Pl}$, the same scaling as $\Lambda_{\rm QG}$ in the (most optimistic) SU($2$) case. For instance, all benchmark points in Table $1$ of \cite{Adshead:2016omu} have $\mu = \sqrt{g/50}M_{\rm Pl},$ and are thus only marginally compatible with $\mu \lesssim \Lambda_{\rm QG}$. A different set of phenomenologically consistent parameters can be obtained if the axion starts very close to the minimum of its potential. In this regime chromonatural inflation matches onto the earlier gauge-flation model \cite{Adshead:2012qe, SheikhJabbari:2012qf, Maleknejad:2012fw}. This regime accommodates larger values of $g \sim 10^{-3}$, but $\mu$ must be taken significantly larger: $\mu \approx 0.1 M_{\rm Pl}$. Such a large $\mu$ is in clear tension with sLWGC constraints on even mildly small gauge couplings. We emphasize, however, that the original gauge-flation scenario did not invoke a $\chi$ field, and so one could consider other possible UV completions that may evade WGC bounds. A different variant of the model has $\chi$ serving as a spectator field while another field $\phi$ drives inflation \cite{Dimastrogiovanni:2016fuu}. Recently it has been claimed that such a theory can produce detectable tensor modes even for very low-scale inflation, with a Hubble scale just above that constrained by BBN \cite{Fujita:2017jwq}. However, accomplishing this requires exponentially small gauge couplings, and so we expect the sLWGC constraint to be even more severe for such a scenario than for the minimal realization of chromonatural inflation. We will not consider the constraint on this alternative scenario in detail here.

Our results suggest that the sLWGC is in modest, but not decisive, tension with chromonatural inflation. Another field theoretic concern with this theory is that if $\chi$ is a compact axion field of period $2\pi f$ (as suggested by the choice of cosine potential), then the coupling $\lambda$ is actually quantized:
\be
\lambda = n \frac{g^2}{4\pi^2}, \quad n \in \mathbb{Z}.
\ee
Given that the fit to inflationary phenomenology prefers $g \sim 10^{-6}$ and $\lambda \sim 100$, this requires a large integer of order $10^{15}$ to appear in the theory! This suggests that a fairly extreme version of axion monodromy must appear in the UV completion of chromonatural inflation. Since the sLWGC puts the UV cutoff of the theory just overhead, it will be difficult to explain the origin of this large dimensionless number from smaller input parameters. We will not undertake this challenge here.

\subsection{Summary of phenomenological consequences}

\begin{figure}[!h]\begin{center}
\includegraphics[width=0.7\textwidth]{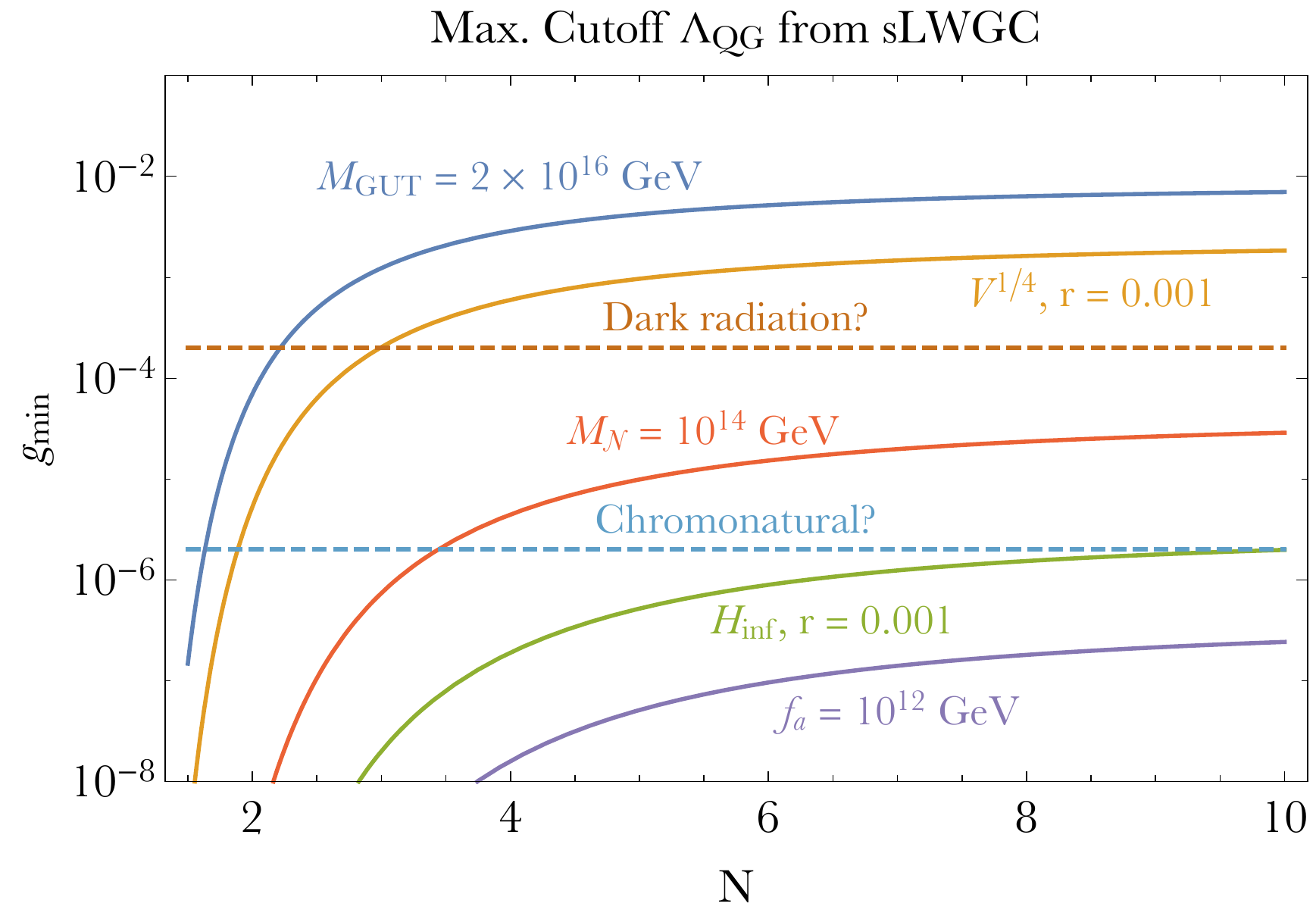}
\end{center}
\caption{(s)LWGC UV cutoff bound as a function of $g$ and $N$ (solid contours, labeled by representative physical scenarios). The dashed lines are benchmark choices of $g$ in two particular cosmological scenarios that favor small nonabelian gauge couplings.}
\label{fig:nonabelianconstraints}
\end{figure}%

In Figure \ref{fig:nonabelianconstraints}, we show contours of the largest $\Lambda_{\rm QG}$ allowed by the sLWGC as a function of $g$ and $N$ for SU($N$) gauge theories. The dashed horizontal lines correspond to the approximate size of couplings of interest for dark matter--dark radiation interactions and chromonatural inflation. We see that the sLWGC has the potential to put a variety of interesting high-scale physics in tension with small gauge couplings that may be of phenomenological interest. On the other hand, some high-scale physics, like a conventional QCD axion with decay constant $f_a \approx 10^{12}~{\rm GeV}$, is relatively safe; we would need a theory to predict a small gauge coupling of order $10^{-7}$ or smaller to have tension between the sLWGC and the PQ-breaking scale.

To be interesting from the viewpoint of sLWGC constraints on $\Lambda_{\rm QG}$, gauge couplings have to be quite small; for instance, merely demanding that the nonabelian gauge theory's confinement scale be smaller than the Hubble scale of our universe today is not sufficient to derive an interesting bound. Cases of interest generally arise in cosmology and come from tight constraints on the size of gauge couplings. For instance, nonabelian gauge preheating \cite{Adshead:2017xll} requires small couplings, of order $10^{-4}$ or smaller, because otherwise the gauge fields' interactions with each other backreact to shut off resonant particle production. On the other hand, this scenario has potential difficulties purely within effective field theory, since it assumes that higher-dimension operators play a crucial role in the scalar coupling to gauge fields but not in the scalar potential.

Another theory that would be interesting to explore from the sLWGC viewpoint is gaugid inflation \cite{Piazza:2017bsd}, which relies not on a nonabelian group but on a U($1$)$^3$ gauge theory. The product of spatial rotations and rotations among the 3 gauge fields is broken to the diagonal. In a theory of free gauge fields, a rotation among the 3 gauge fields is a symmetry (though one without a gauge-invariant Noether current). However, the existence of a charge lattice explicitly breaks the symmetry. Since the scenario relies on higher-dimension operators built out of the gauge fields, one would need some sort of discrete symmetry to ensure that these operators (at least approximately) respect the appropriate symmetry after integrating out the particles whose existence the sLWGC demands. (In other words, in the presence of a charge lattice, the full SO($3$) rotation symmetry must be an accidental symmetry, enforced by some smaller discrete symmetry.) Again, we will defer further consideration of this model for future work.

Finally, we note that the bounds on $\Lambda_{\rm QG}$ can in principle be relaxed if there is a substantial screening effect from light charged particles, since in four dimensions---as we argued in~\S\ref{subsec:conifolds}---the gauge coupling renormalized near the WGC scale $E_0 \sim e(E_0) M_{\rm Pl}$ is what should appear in the bounds, rather than the infrared gauge coupling. However, phenomenological constraints do not depend on the infrared gauge coupling either, but rather the gauge coupling renormalized at some finite scale, albeit possibly a very low scale. Due to the logarithm and the loop factor the screening effect is not very large, even if we choose the largest hierarchy of scales we can imagine:
\be
\frac{b}{8 \pi^2} \log \frac{E}{m} \lesssim \frac{(4/3) n_f}{8 \pi^2} \log \frac{10^{19} \mathrm{\ GeV}}{10^{-33} \mathrm{\ eV}} \sim 2.4\, n_f \,,
\ee
where we consider $n_f$ Dirac fermions for definiteness and put the ratio of the Planck scale to the present-day Hubble scale into the log. This means that unless we have a large number of light charged particles, or light particles with parametrically large charge, the screening effect is negligible for small gauge couplings $e \ll 1$. Even if we choose $n_f$ and/or $q$ to be large, we cannot completely evade constraints. For instance, if we fix $q \sim 1$, then we need $n_f \gtrsim 1/e^2$ to generate significant screening, but then the species bound gives $\Lambda_{\rm QG} \lesssim n_f^{-1/2} M_{\rm Pl} \lesssim e M_{\rm Pl}$, which is a much stronger bound then the one we were trying to evade! If instead we fix $n_f \sim 1$ then we need $q \gtrsim 1/e$ to generate significant screening, but then $q e \gtrsim 1$, and the light particles have $\mathcal{O}(1)$ couplings! Thus, for all practical purposes we can ignore screening effects when placing phenomenological constraints on weak gauge couplings.

\section{Conclusions} \label{sec:conclusions}

We have argued that towers of charged particles generically lead to low cutoffs on both gauge theory and gravity, and that for towers of approximately WGC-saturating particles, these cutoffs are parametrically the same. This suggests that in sufficiently weakly coupled gauge theories, we can concretely understand the emergence of the gauge theory from the quantum gravity scale: the size of the gauge field kinetic term is parametrically determined by loops of the tower of charged particles.

We have also shown some interesting converse statements to this, most notably that if we assume an approximate matching between the gauge theory and gravitational cutoffs then the Weak Gravity Conjecture follows. Yet stronger statements follow if we make additional assumptions concerning ``unification'' between gauge and gravitational forces, though the physical meaning of these assumptions is not complete clear at present.

There are a number of open questions remaining. As emphasized in \S\ref{sec:caveats}, there are simple examples of quantum gravity theories for which our arguments do not apply. These include perturbative heterotic strings at $g_s \ll 1$, D$0$ branes, and winding strings. In these cases we should not necessarily trust the perturbative field theory calculations we have performed, but there may be generalizations of our arguments. Another concern arises from cases with ultralight charged particles in four dimensions, as in the conifold example, which suggests that the sLWGC must be modified in the context of running couplings. These shortcomings should be explored more fully in the future.

Throughout this paper we have treated gauge couplings as constants, but in quantum gravity theories we expect couplings to be determined by the expectation values of moduli fields. The original Swampland conjectures suggest that whenever we find a tower of particles becoming light, we should expect a logarithmic divergence in distance in the moduli space \cite{Vafa:2005ui,Ooguri:2006in}. Recently these moduli space conjectures have appeared in work on the WGC and its connection to scalar fields \cite{Baume:2016psm,Klaewer:2016kiy,Valenzuela:2016yny,Blumenhagen:2017cxt,Palti:2017elp,Hebecker:2017lxm}. One could consider the role of moduli and the Swampland Conjectures from a perturbative viewpoint similar to that taken in this paper: if we treat gauge couplings as background fields, then loop effects of the tower of charged particles can produce kinetic terms for these fields. It would be interesting to explore these effects.

More generally, it continues to be important to seek a proof of the WGC and of potentially stronger forms of the WGC. There is substantial evidence for the Sublattice Weak Gravity Conjecture in perturbative string theory, but at larger coupling the meaning of the conjecture is ambiguous. Our results suggest that reformulating the conjecture in terms of the density of states of a given mass and charge might be promising, as this is the key quantity determining the size of loop effects.

\section*{Acknowledgments}

We thank Ibrahima Bah and Arthur Hebecker for useful discussions, and we thank Jeff Dror for asking about the possible role of proton decay in the context of the quantum gravity scale. The research of BH was supported by Perimeter Institute for Theoretical Physics. Research at Perimeter Institute is supported by the Government of Canada through the Department of Innovation, Science and Economic Development, and by the Province of Ontario through the Ministry of Research, Innovation and Science. MR is supported in part by the DOE Grant {DE-SC}0013607 and the NASA ATP Grant NNX16AI12G. TR is supported by the Carl P. Feinberg Founders' Circle Membership and the NSF Grant PHY-1314311.

\appendix

\section{Further loop contributions}\label{app:moreloops}

In the main text we have focused on the quantities $\lambda_{\rm gauge}(E)$ and $\lambda_{\rm grav}(E)$ which sum the loop corrections to the photon and graviton propagators from particles with mass $m < E$. These stand in for the full loop corrections $\Pi(p^2)$. In this appendix we will carry out two checks of our reasoning. First, we will argue that the contributions from particles with $m > E$ do not change our logic. Second, we will argue that loop corrections to the $n$-point functions do not lead to lower cutoffs than those we have discussed.

\subsection{Loops of heavy particles}\label{app:heavyloops}

Particles with mass $m > E$ run in loops, but have contributions that may be expanded as a series in powers of $E^2$. This suggests that rather than equation (\ref{eq:lambdagauge}) we could have defined
\be
\widetilde{\lambda}_{\rm gauge}(E) \df e^2 E^{D-4} \sum_{i:\, m_i < E} I(i) + e^2 E^{2} \sum_{i:\, E < m_i < \Lambda}  m_i^{D-6} I(i),  \label{eq:lambdagaugemodified}
\ee
now including all particles and not just light ones. Similarly (\ref{eq:lambdagrav}) may be replaced by 
\be
\widetilde{\lambda}_{\rm grav}(E) \df G_N E^{D-2} \sum_{i:\, m_i < E} d(i) + G_N E^2 \sum_{i:\, E < m_i < \Lambda} m_i^{D-4} d(i).\label{eq:lambdagravmodified}
\ee
The question, then, is whether these modified definitions that take into account heavy particles alter our parametric estimates throughout the paper. Notice that we have assumed that we should {\em not} include particles heavier than the effective field theory cutoff $\Lambda$.

Consider the case of a U($1$) tower of particles of charge $q$ and mass $m_q \sim e q M_{\rm Pl}^{(D-2)/2}$. In this case, the second term in $\widetilde{\lambda}_{\rm gauge}$ is
\begin{align}
\delta \widetilde\lambda_{\rm gauge}(E) &\sim e^2 E^2 \sum_{q:\, E < m_q < \Lambda} q^2 (e q M_{\rm Pl}^{(D-2)/2})^{D-6} \nonumber \\
&\sim e^{D-4} E^2 Q(\Lambda)^{D-3} M_{\rm Pl}^{(D-2)(D-6)/2} \sim E^2 \Lambda^{D-3} \frac{1}{e M_{\rm Pl}^{3(D-2)/2}}\,,
\end{align}
where $Q(E) \sim E/(e M_{\rm Pl}^{(D-2)/2})$ is the charge of a particle in the tower with mass near $E$ and we have dropped terms scaling as $Q(E)^{D-3} \ll Q(\Lambda)^{D-3}$ at energies well below the cutoff (assuming $D \geq 4$). Compared to the sum over light particles (\ref{eq:lambdagaugeU1simple}), we see that including this contribution does not change our parametric estimate for the gauge theory cutoff,
\be
\Lambda_{\rm gauge} \sim e^\frac{1}{D-1} M_{\rm Pl}^\frac{3(D-2)}{2(D-1)}.
\ee
Similarly, in the case of $\Lambda_{\rm grav}$ we have
\begin{align}
\delta \widetilde\lambda_{\rm grav} &\sim G_N E^2 \sum_{q \sim Q(E)}^{Q(\Lambda)} (e q M_{\rm Pl}^{(D-2)/2})^{D-4} \sim E^2 \Lambda^{D-3} \frac{1}{e M_{\rm Pl}^{3(D-2)/2}},
\end{align}
which is the same scaling as for $\widetilde\lambda_{\rm gauge}(E)$.

More generally, if we assume a smooth function $N(M)$ characterizing the total number of particles of mass below $M$, with average squared charge $\overline{Q^2(M)}$ for the particles of mass near $M$, we have:
\begin{align}
\delta \widetilde\lambda_{\rm grav}(E) &\sim G_N E^2 \int_E^\Lambda dM\, \frac{dN}{dM} M^{D-4}  \nonumber \\
\delta \widetilde\lambda_{\rm gauge}(E) &\sim e^2 E^2 \int_E^\Lambda dM\, \frac{dN}{dM} \overline{Q^2(M)} M^{D-6}. \label{eq:integralforms}
\end{align}
From these results we immediately see that $\delta \widetilde\lambda_{\rm grav}(E) \sim \delta \widetilde\lambda_{\rm gauge}(E)$ whenever $\overline{Q^2(M)} \sim G_N M^2/e^2$, which is the case for a tower of approximately WGC-saturating particles. In this manner, one can rewrite all the arguments of \S\ref{sec:generalgroup} and \S\ref{sec:unification} in terms of the modified $\lambda_{\rm gauge}(E)$ and $\lambda_{\rm grav}(E)$ including contributions of particles with $m \gg E$. 

Notice that because we only apply these formulas at $E \lesssim \Lambda$, these modifications to the definition of the $\lambda(E)$s actually {\em dominate} over the formulas used elsewhere in the paper. However, they are conceptually somewhat murkier; as discussed in \S\ref{sec:ampgrowth}, for most of the paper we assume that we can study an effective theory at the scale $E$ without necessarily committing ourselves to assumptions about much heavier particles, which are difficult to distinguish from higher-dimension operators.

It is not surprising that our conclusions about the cutoff scale do not change when we include particles with $E \ll m \ll \Lambda$ in loops. The reason is that as $E \to \Lambda$, almost all of the particles we should consider have masses below $E$, for which the formulas used in the bulk of the paper apply. Furthermore, particles of mass {\em near} $E$ can be treated as either heavy or light without any parametric difference in the formulas. We never consider particles parametrically heavier than $\Lambda$, and so the modifications above become irrelevant at energies near the cutoff. Hence, all results in the paper based on ascertaining when $\lambda_{\rm gauge}(E) \sim 1$ or $\lambda_{\rm grav}(E) \sim 1$ should be unaffected by the above modifications.

One place where our arguments require more careful attention to the definition of the $\lambda(E)$s is in \S\ref{sec:lambdasequal}, when we studied the consequences of $\lambda_{\rm gauge}(E) \gtrsim \lambda_{\rm grav}(E)$ at an energy $E$ below the cutoff and of $\lambda_{\rm gauge}(E) \sim \lambda_{\rm grav}(E)$ over a range of energies $E \gtrsim E_0$. These arguments did not assume that the $\lambda(E)$s were ${\cal O}(1)$. We can produce modified versions of these arguments, making use of the form (\ref{eq:integralforms}) where appropriate. However, there is an interesting conceptual difference. The assumption that $\lambda_{\rm gauge}(E) \gtrsim \lambda_{\rm grav}(E)$ is valid at some energy $E$ was argued in \S\ref{sec:lambdasequal} to imply the existence of a particle with $m \lesssim E$ obeying the original WGC (up to order-one factors). With the modified definition, the assumption that $\widetilde\lambda_{\rm gauge}(E) \gtrsim \widetilde\lambda_{\rm grav}(E)$ now implies that the original WGC is obeyed by a particle with mass $E \lesssim m \lesssim \Lambda$. Moreover, the ostensibly stronger statement that $\widetilde\lambda_{\rm gauge}(E) \sim \widetilde\lambda_{\rm grav}(E)$ over a range of energies $E \gtrsim E_0$ is far less constraining that the analogous statement with $\lambda(E)$s, since in typical examples $\widetilde\lambda(E)$ is dominated by particles near the cutoff, and $\widetilde\lambda_{\rm gauge}(E) \sim \widetilde\lambda_{\rm grav}(E)$ for $E \ll \Lambda$ follows if these particles are near extremal, regardless of the light spectrum.

Thus, in this specific context we need to be more careful about which loop corrections we are considering. This is precisely because the notion of gauge-gravity unification at weak coupling has no obvious definition. We argue in~\S\ref{sec:lambdasequal} that the notion $\lambda_{\rm gauge}(E) \sim \lambda_{\rm grav}(E)$ has the right behavior in examples and has nice properties in effective field theory, but do not attempt to motivate it from first principles.

\subsection{Higher-point loop amplitudes} \label{app:higherloops}

Rather than focusing on loop corrections to the off-shell photon and graviton propagators, we can consider the behavior of on-shell $S$-matrix elements. We would like to argue that the cutoff $\Lambda$ at which the loop expansion of the $S$-matrix breaks down is the same as the cutoff we have inferred from two-point functions. We will only discuss the case of a simple U($1$) tower for brevity.

\begin{figure}[!h]\begin{center}
\includegraphics[width=0.7\textwidth]{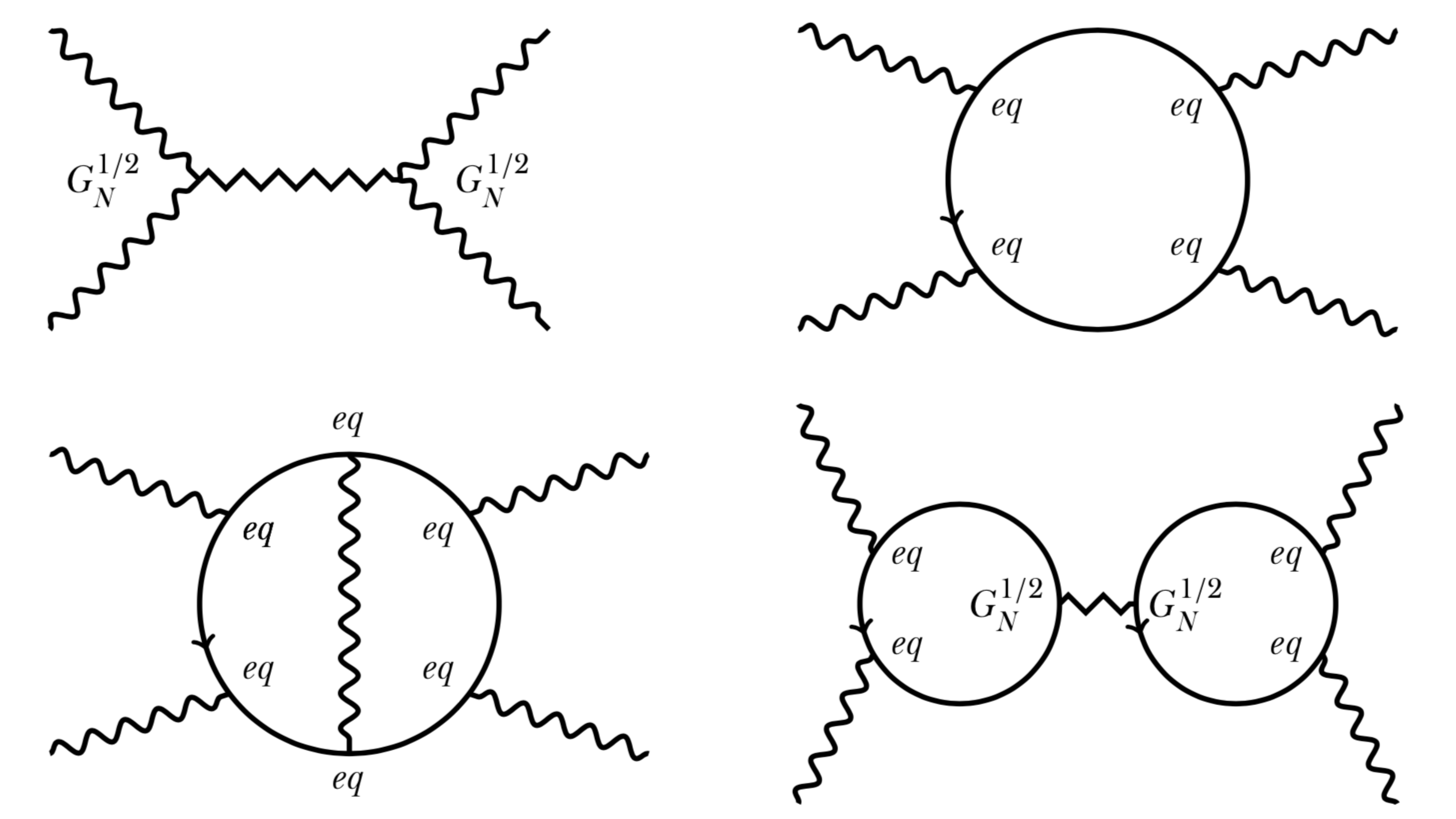}
\end{center}
\caption{Contributions to four-photon scattering: at tree level, through graviton exchange at order $G_N$; at one loop, through electromagnetic interactions with charged particles at order $e^4$; at two loops, from one diagram of order $e^6$ and one of order $e^4 G_N$.}
\label{fig:fourphoton}
\end{figure}

Consider, as an illustration, the $2 \to 2$ photon scattering amplitude. In a $D$-dimensional theory this amplitude has scaling dimension $4 - D$. We show several contributions to the amplitude in Figure \ref{fig:fourphoton}. At tree level, graviton exchange produces an amplitude with parametric scaling $E^2 G_N$ or $E^2 M_{\rm Pl}^{2-D}$. At one loop, there is a contribution from charged particles of mass $m$ and charge $q$. If $E \gg m$, this scales as $e^4 q^4 E^{D - 4}$, while if $E \ll m$, we obtain an Euler-Heisenberg Lagrangian with four field strengths so the amplitude scales as $e^4 q^4 E^4 m^{D - 8}$. If we are interested in behavior near the cutoff $E \sim \Lambda$, we need only consider the $m \lesssim E$ case, and we have
\be
\frac{{\cal M}_{\rm 1-loop}}{{\cal M}_{\rm tree}} \sim \frac{e^4 E^{D-4} \sum q^4}{E^2 G_N} \sim \frac{e^4 E^{D-4} Q(E)^5 }{E^2 G_N} \sim \frac{E^{D-1}}{e M_{\rm Pl}^{3(D-2)/2}}.
\ee
Hence we see that the energy $E$ at which the $1$-loop contribution is of the same order as the tree-level contribution is the familiar scale (\ref{eq:LambdaQGU1})
\be
\Lambda \sim e^{\frac{1}{D-1}} M_{\rm Pl}^\frac{3(D-2)}{2(D-1)}.
\ee

Do higher loops change the result? At two loops we have multiple contributions; from the lower diagrams in Figure \ref{fig:fourphoton} we see that
\be
{\cal M}_{\rm 2-loop} \sim e^6 E^{2D-8} \sum q^6 + e^4 G_N E^{2D - 6} \left(\sum q^2\right)^2.
\ee
Writing $\sum q^6 \sim Q(E)^7$ and $\left(\sum q^2\right)^2 \sim Q(E)^6$, we see that the second term grows more quickly with energy, so that
\be
\frac{{\cal M}_{\rm 2-loop}}{{\cal M}_{\rm tree}} \sim \frac{e^4 G_N E^{2D-6} Q(E)^6}{E^2 G_N} \sim e^4 E^{2D-8} \frac{E^6}{e^6 M_{\rm Pl}^{3(D-2)}}.
\ee
Again, we read off that the scale where this becomes order-one is the familiar scale (\ref{eq:LambdaQGU1}).

Similar results can be extracted for higher-point diagrams. For instance, the tree-level $n$-photon scattering amplitude through graviton exchange scales as $E^2 G_N^{n/2 - 1}$. The one-loop $n$-photon scattering through charged particles scales as $e^n E^{D - n} \sum q^n$. Again writing $\sum q^n \sim Q(E)^{n+1} \sim \left(E/(e M_{\rm Pl}^{(D-2)/2})\right)^{n+1}$, we derive the same cutoff $\Lambda$ for the scale at which the one-loop contribution becomes competitive with the tree-level result. Once again one can readily check that there is a class of 2-loop diagrams (containing two closed loops of charged particles) that become of the same order at the same scale $\Lambda$. Thus, we believe our results to be quite robust against the specific choices of the loop diagrams we have considered to obtain them.

\section{Weyl-invariant sublattices} \label{app:WeylInv}

In this appendix, we classify Weyl invariant sublattices of the weight lattice $\Gamma_G$ of a compact Lie group $G$. We focus on the case where $G$ is simple, but similar techniques can be extended to arbitrary $G$.

We say that a sublattice $\Gamma \subseteq \Gamma_G$ is \emph{primitive} if it is not a multiple of another sublattice. Any sublattice can be written as a multiple of a primitive sublattice, hence it suffices to classify primitive Weyl-invariant sublattices.

We initially assume that $G$ is simply connected. 
A lattice $\Gamma$ is a Weyl-invariant sublattice of $\Gamma_G$ if and only if
\begin{equation} \label{eqn:Weylinvconds}
  \forall  \vec{Q}_{\alpha} \in \Phi \,, \qquad \vec{Q}^\vee_\alpha \cdot \Gamma \subseteq \mathbbm{Z} \quad \mbox{and} \quad
  ( \vec{Q}^\vee_\alpha \cdot \Gamma ) \vec{Q}_\alpha
  \subseteq \Gamma \,,
\end{equation}
where $\vec{Q}^\vee_\alpha \df 2 \vec{Q}_{\alpha}/Q_{\alpha}^2$ is the coroot associated to $\vec{Q}_\alpha$. The first condition requires that $\Gamma$ is a sublattice of the weight lattice, whereas the second is equivalent to Weyl invariance, since the Weyl group reflection associated to $\vec{Q}_\alpha$ takes
\begin{equation} \label{eqn:Weylgroupaction}
\vec{Q} \rightarrow \vec{Q} - ( \vec{Q}^\vee_\alpha \cdot \vec{Q} )  \vec{Q}_{\alpha} \,.
\end{equation}
Note that the coroots are the roots of the Langlands dual
group $G^{\vee}$, whose weight lattice is the dual of the $G$ weight lattice and which is simple if and only if $G$ is simple.

Thus, for each root $\vec{Q}_\alpha$ there is a positive integer $k_\alpha$ such that $\vec{Q}^\vee_\alpha \cdot \Gamma = k_\alpha \mathbbm{Z}$.
Since $G$ is simple, the Weyl group relates any two roots of the same length,
and we can have at most two distinct $k$s, $k_{\rm long}$ and $k_{\rm short}$.
Moreover, the short roots generate the root
lattice, and likewise the short
coroots---corresponding to the long roots---generate the coroot lattice, implying that
$k_{\rm long}$ divides $k_{\rm short}$. Therefore, $\tilde{\Gamma} \df
\Gamma / k_{\rm long}$ with $\tilde{k}_{\rm long} = 1$ and
$\tilde{k}_{\rm short} = k_{\rm short} / k_{\rm long} \in
\mathbbm{Z}$ is a Weyl invariant sublattice, implying that $k_{\rm long} = 1$
if and only if $\Gamma$ is primitive.\footnote{If $G$ is simply laced, we consider the roots to be ``long'' by convention.}

Using the second condition in (\ref{eqn:Weylinvconds}), we find that $\Gamma_{\rm long} \subseteq \Gamma$ for primitive $\Gamma$, where $\Gamma_{\rm long} \subseteq \Gamma_G$ is the sublattice generated by the long roots. Thus, primitive Weyl-invariant sublattices correspond to Weyl invariant subgroups of the finite group $\hat{Z}_G\df\Gamma_G / \Gamma_{\rm long}$. Not all such subgroups correspond to primitive lattices, since $\Gamma_{\rm long} \subseteq \Gamma$ does not imply $k_{\rm long}=1$,\footnote{$\Gamma_{\rm long} \subseteq \Gamma$ does imply either $k_{\rm long}=1$ or $k_{\rm long}=2$ since $\vec{Q}^\vee_\alpha \cdot \vec{Q}_\alpha = 2$; $k_{\rm long}=2$ implies that $\Gamma_{\rm long}/2 \subseteq \Gamma_G$, which holds for $\Sp(2k)$, including $\Spin(2) \cong \SU(2) \cong \Sp(2)$ and $\Spin(5) \cong \Sp(4)$, but not in other simple groups.} but every primitive Weyl-invariant sublattice corresponds to a unique Weyl-invariant subgroup of $\hat{Z}_G$.

To compute $\hat{Z}_G$ and the Weyl group action on it, we note that there is a simply laced Lie group $\hat{G}$ with root lattice $\Gamma_{\rm long}$ and weight lattice $\Gamma_G$, where $G = \hat{G}$ if and only if $G$ is simply laced. Since $\Gamma_{\rm long} = \Gamma_{\rm root}(\hat{G})$, $\hat{Z}_G$ is the center $Z(\hat{G})$ of $\hat{G}$. Precisely when $G$ is not simply laced, the Weyl group $W_G$ acts non-trivially on $\hat{Z}_G$, with the non-trivial action generated by the reflections associated to the short roots. This action is a homomorphism $\phi: W_G \to \Aut(\hat{Z}_G)$, hence it is encoded by $\widehat{W}_G \df W_G / \ker \phi$. The elements of $\widehat{W}_G$ are outer automorphisms of $\hat{G}$.

It is now straightforward to compute $\hat{G}$, $\hat{Z}_G$, and $\widehat{W}_G$ for all simply connected, simple Lie groups $G$:
\begin{equation}
\begin{array}{cc|cccc}
G & \hat{Z}_G & G & \hat{G} & \hat{Z}_G & \widehat{W}_G  \\ \hline
     \SU(n) & \mathbbm{Z}_n & \Sp(2 k) & \Sp(2)^k & \mathbbm{Z}_2^k & S_k \\
     \Spin(4 k) &  \mathbbm{Z}_2 \oplus \mathbbm{Z}_2 & \Spin(4 k + 1) & \Spin(4 k)  & \mathbbm{Z}_2 \oplus \mathbbm{Z}_2 & \mathbb{Z}_2 \\
     \Spin(4 k + 2) & \mathbbm{Z}_4 & \Spin(4 k + 3) & \Spin(4 k+2)  & \mathbbm{Z}_4 & \mathbb{Z}_2 \\
     E_6 & \mathbbm{Z}_3 & G_2 & \SU(3) & \mathbbm{Z}_3 & \mathbb{Z}_2 \\
     E_7 & \mathbbm{Z}_2 & F_4 & \Spin(8) & \mathbbm{Z}_2 \oplus \mathbbm{Z}_2 & S_3 \\
     E_8 & \mathdash 
\end{array}
\end{equation}
On the left are the simply laced groups, for which $\hat{G} = G$, $\hat{Z}_G = Z(G)$, and $\widehat{W}_G$ is trivial. On the right are non-simply-laced groups, for which the $\widehat{W}_G$ action is as follows: for $\Spin(4k+3)$ and $G_2$, $\widehat{W}_G \cong \mathbb{Z}_2$ maps elements of $\hat{Z}_G$ to their inverses, preserving all subgroups. For $\Sp(2k)$ and $\Spin(4k+1)$, $\widehat{W}_G \cong S_n$ permutes the $\mathbb{Z}_2$ factors of $\hat{Z}_G \cong \mathbb{Z}_2^n$. Finally, for $F_4$, $\widehat{W}_G \cong S_3$ permutes the three non-trivial elements of the Klein four-group $\mathbb{Z}_2 \oplus \mathbb{Z}_2$.

Therefore, the Weyl-invariant subgroups of $\hat{Z}_G$ are as follows: for simply laced $G$, $\Spin(4k+3)$, or $G_2$, any subgroup of $\hat{Z}_G$ is Weyl invariant. For $F_4$, only the trivial subgroup and $\hat{Z}_G$ itself are Weyl-invariant. Finally, for $\Sp(2k)$ and $\Spin(4k+1)$, the Weyl invariant subgroups are the permutation invariant subgroups of $\hat{Z}_G \cong \mathbb{Z}_2^n$. There are four of these: $\mathbb{Z}_2^n$ itself, the trivial subgroup, the diagonal subgroup $\mathbb{Z}_2 \subset \mathbb{Z}_2^n$, and the index-two subgroup
\begin{equation}
  H_+ \df \left\{ (a_1, \ldots, a_n) \middle| a_i \in \{ 0, 1 \}, \sum_i
  a_i \equiv 0 \mod 2 \right\} \,,
\end{equation}
where the latter two are equivalent for $n = 2$.

We can summarize the above classification as follows. For each $H \subseteq Z (G)$, there is a primitive Weyl-invariant sublattice corresponding to the weight lattice of $G / H$. In addition, for $G$ not simply laced and $G \ne \Sp(2k)$, $\Gamma_{\rm long}$ is a primitive Weyl-invariant sublattice. Finally, for $G = \Sp(2k)$---for which $\Gamma_{\rm long}$ is twice the weight lattice---there is a primitive Weyl-invariant sublattice of the form $(2\mathbb{Z})^k \cup [(2\mathbb{Z})^k+ (1,\ldots,1)]$ in a basis where the weight lattice is $\mathbb{Z}^k$ and the root lattice is the index-two sublattice $\{ (n_1, \ldots, n_k) | \sum_i n_i \cong 0 \mod 2 \}$. Thus, besides the $G/H$ weight lattices, there is exactly one additional primitive Weyl-invariant sublattice for each non-simply-laced $G$.

When $G$ is not simply connected, the weight lattice $\Gamma_G$ is a sublattice of the weight lattice $\Gamma_{\tilde G}$ of the universal cover $\tilde{G}$, and the smallest multiple of each primitive Weyl-invariant sublattice of $\Gamma_{\tilde G}$ which lies inside $\Gamma_G$ is a primitive Weyl-invariant sublattice of $\Gamma_G$. All primitive Weyl-invariant sublattices take this form, since any primitive sublattice of $\Gamma_G$ is a multiple of a primitive sublattice of $\Gamma_{\tilde G}$.

\bibliographystyle{utphys}
\bibliography{ref}

\providecommand{\href}[2]{#2}\begingroup\raggedright\begin{thebibliography}{10}

\bibitem{Arkanihamed:2006dz}
N.~Arkani-Hamed, L.~Motl, A.~Nicolis, and C.~Vafa, ``{The String landscape,
  black holes and gravity as the weakest force},''
  \href{http://dx.doi.org/10.1088/1126-6708/2007/06/060}{{\em JHEP} {\bfseries
  0706} (2007) 060},
\href{http://arxiv.org/abs/hep-th/0601001}{{\ttfamily arXiv:hep-th/0601001
  [hep-th]}}.

\bibitem{Vafa:2005ui}
C.~Vafa, ``{The String landscape and the swampland},''
\href{http://arxiv.org/abs/hep-th/0509212}{{\ttfamily arXiv:hep-th/0509212
  [hep-th]}}.

\bibitem{Ooguri:2006in}
H.~Ooguri and C.~Vafa, ``{On the Geometry of the String Landscape and the
  Swampland},'' \href{http://dx.doi.org/10.1016/j.nuclphysb.2006.10.033}{{\em
  Nucl.Phys.} {\bfseries B766} (2007) 21--33},
\href{http://arxiv.org/abs/hep-th/0605264}{{\ttfamily arXiv:hep-th/0605264
  [hep-th]}}.

\bibitem{banks:2010zn}
T.~Banks and N.~Seiberg, ``{Symmetries and Strings in Field Theory and
  Gravity},'' \href{http://dx.doi.org/10.1103/PhysRevD.83.084019}{{\em Phys.
  Rev.} {\bfseries D83} (2011) 084019},
\href{http://arxiv.org/abs/1011.5120}{{\ttfamily arXiv:1011.5120 [hep-th]}}.

\bibitem{Hod:2017uqc}
S.~Hod, ``{A proof of the weak gravity conjecture},'' {\em Int. J. Mod. Phys.}
  {\bfseries D26} (2017) 1742004,
\href{http://arxiv.org/abs/1705.06287}{{\ttfamily arXiv:1705.06287 [gr-qc]}}.

\bibitem{Fisher:2017dbc}
Z.~Fisher and C.~J. Mogni, ``{A Semiclassical, Entropic Proof of a Weak Gravity
  Conjecture},''
\href{http://arxiv.org/abs/1706.08257}{{\ttfamily arXiv:1706.08257 [hep-th]}}.

\bibitem{Cottrell:2016bty}
G.~Shiu, P.~Soler, and W.~Cottrell, ``{Weak Gravity Conjecture and Extremal
  Black Hole},''
\href{http://arxiv.org/abs/1611.06270}{{\ttfamily arXiv:1611.06270 [hep-th]}}.

\bibitem{Shiu:2017toy}
G.~Shiu, W.~Cottrell, and P.~Soler, ``{Weak Gravity Conjecture and Black Holes
  in $N = 2$ Supergravity},''
{\em PoS} {\bfseries CORFU2016} (2017) 130.

\bibitem{banks:2006mm}
T.~Banks, M.~Johnson, and A.~Shomer, ``{A Note on Gauge Theories Coupled to
  Gravity},'' \href{http://dx.doi.org/10.1088/1126-6708/2006/09/049}{{\em JHEP}
  {\bfseries 0609} (2006) 049},
\href{http://arxiv.org/abs/hep-th/0606277}{{\ttfamily arXiv:hep-th/0606277
  [hep-th]}}.

\bibitem{Horowitz:2016ezu}
G.~T. Horowitz, J.~E. Santos, and B.~Way, ``{Evidence for an Electrifying
  Violation of Cosmic Censorship},''
  \href{http://dx.doi.org/10.1088/0264-9381/33/19/195007}{{\em Class. Quant.
  Grav.} {\bfseries 33} no.~19, (2016) 195007},
\href{http://arxiv.org/abs/1604.06465}{{\ttfamily arXiv:1604.06465 [hep-th]}}.

\bibitem{Crisford:2017zpi}
T.~Crisford and J.~E. Santos, ``{Violating the Weak Cosmic Censorship
  Conjecture in Four-Dimensional Anti--de Sitter Space},''
  \href{http://dx.doi.org/10.1103/PhysRevLett.118.181101}{{\em Phys. Rev.
  Lett.} {\bfseries 118} no.~18, (2017) 181101},
\href{http://arxiv.org/abs/1702.05490}{{\ttfamily arXiv:1702.05490 [hep-th]}}.

\bibitem{Crisford:2017gsb}
T.~Crisford, G.~T. Horowitz, and J.~E. Santos, ``{Testing the Weak Gravity -
  Cosmic Censorship Connection},''
  \href{http://dx.doi.org/10.1103/PhysRevD.97.066005}{{\em Phys. Rev.}
  {\bfseries D97} no.~6, (2018) 066005},
\href{http://arxiv.org/abs/1709.07880}{{\ttfamily arXiv:1709.07880 [hep-th]}}.

\bibitem{rudelius:2015xta}
T.~Rudelius, ``{Constraints on Axion Inflation from the Weak Gravity
  Conjecture},'' \href{http://dx.doi.org/10.1088/1475-7516/2015/9/020}{{\em
  JCAP} {\bfseries 09} (2015) 020},
\href{http://arxiv.org/abs/1503.00795}{{\ttfamily arXiv:1503.00795 [hep-th]}}.

\bibitem{Montero:2015ofa}
M.~Montero, A.~M. Uranga, and I.~Valenzuela, ``{Transplanckian axions!?},''
  \href{http://dx.doi.org/10.1007/JHEP08(2015)032}{{\em JHEP} {\bfseries 08}
  (2015) 032},
\href{http://arxiv.org/abs/1503.03886}{{\ttfamily arXiv:1503.03886 [hep-th]}}.

\bibitem{Brown:2015iha}
J.~Brown, W.~Cottrell, G.~Shiu, and P.~Soler, ``{Fencing in the Swampland:
  Quantum Gravity Constraints on Large Field Inflation},''
  \href{http://dx.doi.org/10.1007/JHEP10(2015)023}{{\em JHEP} {\bfseries 10}
  (2015) 023},
\href{http://arxiv.org/abs/1503.04783}{{\ttfamily arXiv:1503.04783 [hep-th]}}.

\bibitem{Bachlechner:2015qja}
T.~C. Bachlechner, C.~Long, and L.~McAllister, ``{Planckian Axions and the Weak
  Gravity Conjecture},'' \href{http://dx.doi.org/10.1007/JHEP01(2016)091}{{\em
  JHEP} {\bfseries 01} (2016) 091},
\href{http://arxiv.org/abs/1503.07853}{{\ttfamily arXiv:1503.07853 [hep-th]}}.

\bibitem{Hebecker:2015rya}
A.~Hebecker, P.~Mangat, F.~Rompineve, and L.~T. Witkowski, ``{Winding out of
  the Swamp: Evading the Weak Gravity Conjecture with F-term Winding
  Inflation?},'' \href{http://dx.doi.org/10.1016/j.physletb.2015.07.026}{{\em
  Phys. Lett.} {\bfseries B748} (2015) 455--462},
\href{http://arxiv.org/abs/1503.07912}{{\ttfamily arXiv:1503.07912 [hep-th]}}.

\bibitem{Brown:2015lia}
J.~Brown, W.~Cottrell, G.~Shiu, and P.~Soler, ``{On Axionic Field Ranges,
  Loopholes and the Weak Gravity Conjecture},''
  \href{http://dx.doi.org/10.1007/JHEP04(2016)017}{{\em JHEP} {\bfseries 04}
  (2016) 017},
\href{http://arxiv.org/abs/1504.00659}{{\ttfamily arXiv:1504.00659 [hep-th]}}.

\bibitem{junghans:2015hba}
D.~Junghans, ``{Large-Field Inflation with Multiple Axions and the Weak Gravity
  Conjecture},'' \href{http://dx.doi.org/10.1007/JHEP02(2016)128}{{\em JHEP}
  {\bfseries 02} (2016) 128},
\href{http://arxiv.org/abs/1504.03566}{{\ttfamily arXiv:1504.03566 [hep-th]}}.

\bibitem{Heidenreich:2015wga}
B.~Heidenreich, M.~Reece, and T.~Rudelius, ``{Weak Gravity Strongly Constrains
  Large-Field Axion Inflation},''
  \href{http://dx.doi.org/10.1007/JHEP12(2015)108}{{\em JHEP} {\bfseries 12}
  (2015) 108},
\href{http://arxiv.org/abs/1506.03447}{{\ttfamily arXiv:1506.03447 [hep-th]}}.

\bibitem{Ibanez:2015fcv}
L.~E. Ibanez, M.~Montero, A.~Uranga, and I.~Valenzuela, ``{Relaxion Monodromy
  and the Weak Gravity Conjecture},''
  \href{http://dx.doi.org/10.1007/JHEP04(2016)020}{{\em JHEP} {\bfseries 04}
  (2016) 020},
\href{http://arxiv.org/abs/1512.00025}{{\ttfamily arXiv:1512.00025 [hep-th]}}.

\bibitem{Hebecker:2015zss}
A.~Hebecker, F.~Rompineve, and A.~Westphal, ``{Axion Monodromy and the Weak
  Gravity Conjecture},'' \href{http://dx.doi.org/10.1007/JHEP04(2016)157}{{\em
  JHEP} {\bfseries 04} (2016) 157},
\href{http://arxiv.org/abs/1512.03768}{{\ttfamily arXiv:1512.03768 [hep-th]}}.

\bibitem{Baume:2016psm}
F.~Baume and E.~Palti, ``{Backreacted Axion Field Ranges in String Theory},''
  \href{http://dx.doi.org/10.1007/JHEP08(2016)043}{{\em JHEP} {\bfseries 08}
  (2016) 043},
\href{http://arxiv.org/abs/1602.06517}{{\ttfamily arXiv:1602.06517 [hep-th]}}.

\bibitem{Klaewer:2016kiy}
D.~Klaewer and E.~Palti, ``{Super-Planckian Spatial Field Variations and
  Quantum Gravity},'' \href{http://dx.doi.org/10.1007/JHEP01(2017)088}{{\em
  JHEP} {\bfseries 01} (2017) 088},
\href{http://arxiv.org/abs/1610.00010}{{\ttfamily arXiv:1610.00010 [hep-th]}}.

\bibitem{Ooguri:2016pdq}
H.~Ooguri and C.~Vafa, ``{Non-supersymmetric AdS and the Swampland},''
\href{http://arxiv.org/abs/1610.01533}{{\ttfamily arXiv:1610.01533 [hep-th]}}.

\bibitem{Freivogel:2016qwc}
B.~Freivogel and M.~Kleban, ``{Vacua Morghulis},''
\href{http://arxiv.org/abs/1610.04564}{{\ttfamily arXiv:1610.04564 [hep-th]}}.

\bibitem{Dolan:2017vmn}
M.~J. Dolan, P.~Draper, J.~Kozaczuk, and H.~Patel, ``{Transplanckian Censorship
  and Global Cosmic Strings},''
  \href{http://dx.doi.org/10.1007/JHEP04(2017)133}{{\em JHEP} {\bfseries 04}
  (2017) 133},
\href{http://arxiv.org/abs/1701.05572}{{\ttfamily arXiv:1701.05572 [hep-th]}}.

\bibitem{Hebecker:2017wsu}
A.~Hebecker, P.~Henkenjohann, and L.~T. Witkowski, ``{What is the Magnetic Weak
  Gravity Conjecture for Axions?},''
  \href{http://dx.doi.org/10.1002/prop.201700011}{{\em Fortsch. Phys.}
  {\bfseries 65} no.~3-4, (2017) 1700011},
\href{http://arxiv.org/abs/1701.06553}{{\ttfamily arXiv:1701.06553 [hep-th]}}.

\bibitem{Hebecker:2017uix}
A.~Hebecker and P.~Soler, ``{The Weak Gravity Conjecture and the Axionic Black
  Hole Paradox},'' \href{http://dx.doi.org/10.1007/JHEP09(2017)036}{{\em JHEP}
  {\bfseries 09} (2017) 036},
\href{http://arxiv.org/abs/1702.06130}{{\ttfamily arXiv:1702.06130 [hep-th]}}.

\bibitem{Montero:2017yja}
M.~Montero, A.~M. Uranga, and I.~Valenzuela, ``{A Chern-Simons Pandemic},''
  \href{http://dx.doi.org/10.1007/JHEP07(2017)123}{{\em JHEP} {\bfseries 07}
  (2017) 123},
\href{http://arxiv.org/abs/1702.06147}{{\ttfamily arXiv:1702.06147 [hep-th]}}.

\bibitem{Palti:2017elp}
E.~Palti, ``{The Weak Gravity Conjecture and Scalar Fields},''
  \href{http://dx.doi.org/10.1007/JHEP08(2017)034}{{\em JHEP} {\bfseries 08}
  (2017) 034},
\href{http://arxiv.org/abs/1705.04328}{{\ttfamily arXiv:1705.04328 [hep-th]}}.

\bibitem{Hebecker:2017lxm}
A.~Hebecker, P.~Henkenjohann, and L.~T. Witkowski, ``{Flat Monodromies and a
  Moduli Space Size Conjecture},''
  \href{http://dx.doi.org/10.1007/JHEP12(2017)033}{{\em JHEP} {\bfseries 12}
  (2017) 033},
\href{http://arxiv.org/abs/1708.06761}{{\ttfamily arXiv:1708.06761 [hep-th]}}.

\bibitem{Brennan:2017rbf}
T.~D. Brennan, F.~Carta, and C.~Vafa, ``{The String Landscape, the Swampland,
  and the Missing Corner},''
\href{http://arxiv.org/abs/1711.00864}{{\ttfamily arXiv:1711.00864 [hep-th]}}.

\bibitem{Furuuchi:2017upe}
K.~Furuuchi, ``{Weak Gravity Conjecture From Low Energy Observers'
  Perspective},''
\href{http://arxiv.org/abs/1712.01302}{{\ttfamily arXiv:1712.01302 [hep-th]}}.

\bibitem{Heidenreich:2015nta}
B.~Heidenreich, M.~Reece, and T.~Rudelius, ``{Sharpening the Weak Gravity
  Conjecture with Dimensional Reduction},''
  \href{http://dx.doi.org/10.1007/JHEP02(2016)140}{{\em JHEP} {\bfseries 02}
  (2016) 140},
\href{http://arxiv.org/abs/1509.06374}{{\ttfamily arXiv:1509.06374 [hep-th]}}.

\bibitem{cheung:2014vva}
C.~Cheung and G.~N. Remmen, ``{Naturalness and the Weak Gravity Conjecture},''
  \href{http://dx.doi.org/10.1103/PhysRevLett.113.051601}{{\em Phys.Rev.Lett.}
  {\bfseries 113} (2014) 051601},
\href{http://arxiv.org/abs/1402.2287}{{\ttfamily arXiv:1402.2287 [hep-ph]}}.

\bibitem{gibbons:1982ih}
G.~Gibbons, ``{Antigravitating Black Hole Solitons with Scalar Hair in N=4
  Supergravity},''
\href{http://dx.doi.org/10.1016/0550-3213(82)90170-5}{{\em Nucl.Phys.}
  {\bfseries B207} (1982) 337--349}.

\bibitem{myers:1986un}
R.~C. Myers and M.~Perry, ``{Black Holes in Higher Dimensional Space-Times},''
\href{http://dx.doi.org/10.1016/0003-4916(86)90186-7}{{\em Annals Phys.}
  {\bfseries 172} (1986) 304}.

\bibitem{gibbons:1987ps}
G.~Gibbons and K.-i. Maeda, ``{Black Holes and Membranes in Higher Dimensional
  Theories with Dilaton Fields},''
\href{http://dx.doi.org/10.1016/0550-3213(88)90006-5}{{\em Nucl.Phys.}
  {\bfseries B298} (1988) 741}.

\bibitem{garfinkle:1990qj}
D.~Garfinkle, G.~T. Horowitz, and A.~Strominger, ``{Charged black holes in
  string theory},''
\href{http://dx.doi.org/10.1103/PhysRevD.43.3140,
  10.1103/PhysRevD.45.3888}{{\em Phys.Rev.} {\bfseries D43} (1991) 3140}.

\bibitem{Heidenreich:2016aqi}
B.~Heidenreich, M.~Reece, and T.~Rudelius, ``{Evidence for a Lattice Weak
  Gravity Conjecture},'' \href{http://dx.doi.org/10.1007/JHEP08(2017)025}{{\em
  JHEP} {\bfseries 08} (2017) 025},
\href{http://arxiv.org/abs/1606.08437}{{\ttfamily arXiv:1606.08437 [hep-th]}}.

\bibitem{Montero:2016tif}
M.~Montero, G.~Shiu, and P.~Soler, ``{The Weak Gravity Conjecture in three
  dimensions},'' \href{http://dx.doi.org/10.1007/JHEP10(2016)159}{{\em JHEP}
  {\bfseries 10} (2016) 159},
\href{http://arxiv.org/abs/1606.08438}{{\ttfamily arXiv:1606.08438 [hep-th]}}.

\bibitem{Strominger:1995cz}
A.~Strominger, ``{Massless black holes and conifolds in string theory},''
  \href{http://dx.doi.org/10.1016/0550-3213(95)00287-3}{{\em Nucl. Phys.}
  {\bfseries B451} (1995) 96--108},
\href{http://arxiv.org/abs/hep-th/9504090}{{\ttfamily arXiv:hep-th/9504090
  [hep-th]}}.

\bibitem{Harlow:2015lma}
D.~Harlow, ``{Wormholes, Emergent Gauge Fields, and the Weak Gravity
  Conjecture},'' \href{http://dx.doi.org/10.1007/JHEP01(2016)122}{{\em JHEP}
  {\bfseries 01} (2016) 122},
\href{http://arxiv.org/abs/1510.07911}{{\ttfamily arXiv:1510.07911 [hep-th]}}.

\bibitem{ArkaniHamed:2005yv}
N.~Arkani-Hamed, S.~Dimopoulos, and S.~Kachru, ``{Predictive landscapes and new
  physics at a TeV},''
\href{http://arxiv.org/abs/hep-th/0501082}{{\ttfamily arXiv:hep-th/0501082
  [hep-th]}}.

\bibitem{Distler:2005hi}
J.~Distler and U.~Varadarajan, ``{Random polynomials and the friendly
  landscape},''
\href{http://arxiv.org/abs/hep-th/0507090}{{\ttfamily arXiv:hep-th/0507090
  [hep-th]}}.

\bibitem{Dimopoulos:2005ac}
S.~Dimopoulos, S.~Kachru, J.~McGreevy, and J.~G. Wacker, ``{N-flation},''
  \href{http://dx.doi.org/10.1088/1475-7516/2008/08/003}{{\em JCAP} {\bfseries
  0808} (2008) 003},
\href{http://arxiv.org/abs/hep-th/0507205}{{\ttfamily arXiv:hep-th/0507205
  [hep-th]}}.

\bibitem{Dvali:2007hz}
G.~Dvali, ``{Black Holes and Large N Species Solution to the Hierarchy
  Problem},'' \href{http://dx.doi.org/10.1002/prop.201000009}{{\em Fortsch.
  Phys.} {\bfseries 58} (2010) 528--536},
\href{http://arxiv.org/abs/0706.2050}{{\ttfamily arXiv:0706.2050 [hep-th]}}.

\bibitem{Dvali:2007wp}
G.~Dvali and M.~Redi, ``{Black Hole Bound on the Number of Species and Quantum
  Gravity at LHC},'' \href{http://dx.doi.org/10.1103/PhysRevD.77.045027}{{\em
  Phys. Rev.} {\bfseries D77} (2008) 045027},
\href{http://arxiv.org/abs/0710.4344}{{\ttfamily arXiv:0710.4344 [hep-th]}}.

\bibitem{Anber:2011ut}
M.~M. Anber and J.~F. Donoghue, ``{On the running of the gravitational
  constant},'' \href{http://dx.doi.org/10.1103/PhysRevD.85.104016}{{\em Phys.
  Rev.} {\bfseries D85} (2012) 104016},
\href{http://arxiv.org/abs/1111.2875}{{\ttfamily arXiv:1111.2875 [hep-th]}}.

\bibitem{Heidenreich:2016jrl}
B.~Heidenreich, M.~Reece, and T.~Rudelius, ``{Axion Experiments to Algebraic
  Geometry: Testing Quantum Gravity via the Weak Gravity Conjecture},''
  \href{http://dx.doi.org/10.1142/S0218271816430057}{{\em Int. J. Mod. Phys.}
  {\bfseries D25} no.~12, (2016) 1643005},
\href{http://arxiv.org/abs/1605.05311}{{\ttfamily arXiv:1605.05311 [hep-th]}}.

\bibitem{prashant2016}
P.~Saraswat, ``{Can the Weak Gravity Conjecture Rule Out Effective Field
  Theories? [a talk in Madrid regarding work in progress of de la Fuente,
  Saraswat, and Sundrum]},'' 2016.

\bibitem{Saraswat:2016eaz}
P.~Saraswat, ``{Weak gravity conjecture and effective field theory},''
  \href{http://dx.doi.org/10.1103/PhysRevD.95.025013}{{\em Phys. Rev.}
  {\bfseries D95} no.~2, (2017) 025013},
\href{http://arxiv.org/abs/1608.06951}{{\ttfamily arXiv:1608.06951 [hep-th]}}.

\bibitem{Ibanez:2017vfl}
L.~E. Ibanez and M.~Montero, ``{A Note on the WGC, Effective Field Theory and
  Clockwork within String Theory},''
  \href{http://dx.doi.org/10.1007/JHEP02(2018)057}{{\em JHEP} {\bfseries 02}
  (2018) 057},
\href{http://arxiv.org/abs/1709.02392}{{\ttfamily arXiv:1709.02392 [hep-th]}}.

\bibitem{Green:1987sp}
M.~B. Green, J.~H. Schwarz, and E.~Witten, {\em {Superstring Theory, Volume 1:
  Introduction}}.
\newblock Cambridge Monographs on Mathematical Physics. Cambridge University
  Press,
1987.
\newblock

\bibitem{Dvali:2009ks}
G.~Dvali and D.~Lust, ``{Evaporation of Microscopic Black Holes in String
  Theory and the Bound on Species},''
  \href{http://dx.doi.org/10.1002/prop.201000008}{{\em Fortsch. Phys.}
  {\bfseries 58} (2010) 505--527},
\href{http://arxiv.org/abs/0912.3167}{{\ttfamily arXiv:0912.3167 [hep-th]}}.

\bibitem{Dvali:2010vm}
G.~Dvali and C.~Gomez, ``{Species and Strings},''
\href{http://arxiv.org/abs/1004.3744}{{\ttfamily arXiv:1004.3744 [hep-th]}}.

\bibitem{wagner:2012ui}
T.~A. Wagner, S.~Schlamminger, J.~H. Gundlach, and E.~G. Adelberger,
  ``{Torsion-balance tests of the weak equivalence principle},''
  \href{http://dx.doi.org/10.1088/0264-9381/29/18/184002}{{\em Class. Quant.
  Grav.} {\bfseries 29} (2012) 184002},
\href{http://arxiv.org/abs/1207.2442}{{\ttfamily arXiv:1207.2442 [gr-qc]}}.

\bibitem{heeck:2014zfa}
J.~Heeck, ``{Unbroken $B - L$ symmetry},''
  \href{http://dx.doi.org/10.1016/j.physletb.2014.10.067}{{\em Phys. Lett.}
  {\bfseries B739} (2014) 256--262},
\href{http://arxiv.org/abs/1408.6845}{{\ttfamily arXiv:1408.6845 [hep-ph]}}.

\bibitem{freese:1990rb}
K.~Freese, J.~A. Frieman, and A.~V. Olinto, ``{Natural inflation with pseudo -
  Nambu-Goldstone bosons},''
\href{http://dx.doi.org/10.1103/PhysRevLett.65.3233}{{\em Phys.Rev.Lett.}
  {\bfseries 65} (1990) 3233--3236}.

\bibitem{Miura:2016krn}
{\bfseries Super-Kamiokande} Collaboration, K.~Abe {\em et~al.}, ``{Search for
  proton decay via $p \to e^+\pi^0$ and $p \to \mu^+\pi^0$ in 0.31
  megaton-years exposure of the Super-Kamiokande water Cherenkov detector},''
  \href{http://dx.doi.org/10.1103/PhysRevD.95.012004}{{\em Phys. Rev.}
  {\bfseries D95} no.~1, (2017) 012004},
\href{http://arxiv.org/abs/1610.03597}{{\ttfamily arXiv:1610.03597 [hep-ex]}}.

\bibitem{Weinberg:1981wj}
S.~Weinberg, ``{Supersymmetry at Ordinary Energies. 1. Masses and Conservation
  Laws},''
\href{http://dx.doi.org/10.1103/PhysRevD.26.287}{{\em Phys. Rev.} {\bfseries
  D26} (1982) 287}.

\bibitem{Ellis:1981tv}
J.~R. Ellis, D.~V. Nanopoulos, and S.~Rudaz, ``{GUTs 3: SUSY GUTs 2},''
\href{http://dx.doi.org/10.1016/0550-3213(82)90220-6}{{\em Nucl. Phys.}
  {\bfseries B202} (1982) 43--62}.

\bibitem{Murayama:2001ur}
H.~Murayama and A.~Pierce, ``{Not even decoupling can save minimal
  supersymmetric SU(5)},''
  \href{http://dx.doi.org/10.1103/PhysRevD.65.055009}{{\em Phys. Rev.}
  {\bfseries D65} (2002) 055009},
\href{http://arxiv.org/abs/hep-ph/0108104}{{\ttfamily arXiv:hep-ph/0108104
  [hep-ph]}}.

\bibitem{Dine:2013nga}
M.~Dine, P.~Draper, and W.~Shepherd, ``{Proton decay at $M_{pl}$ and the scale
  of SUSY-breaking},'' \href{http://dx.doi.org/10.1007/JHEP02(2014)027}{{\em
  JHEP} {\bfseries 02} (2014) 027},
\href{http://arxiv.org/abs/1308.0274}{{\ttfamily arXiv:1308.0274 [hep-ph]}}.

\bibitem{Ibanez:1991hv}
L.~E. Ibanez and G.~G. Ross, ``{Discrete gauge symmetry anomalies},''
\href{http://dx.doi.org/10.1016/0370-2693(91)91614-2}{{\em Phys. Lett.}
  {\bfseries B260} (1991) 291--295}.

\bibitem{buen-abad:2015ova}
M.~A. Buen-Abad, G.~Marques-Tavares, and M.~Schmaltz, ``{Non-Abelian dark
  matter and dark radiation},''
  \href{http://dx.doi.org/10.1103/PhysRevD.92.023531}{{\em Phys. Rev.}
  {\bfseries D92} no.~2, (2015) 023531},
\href{http://arxiv.org/abs/1505.03542}{{\ttfamily arXiv:1505.03542 [hep-ph]}}.

\bibitem{lesgourgues:2015wza}
J.~Lesgourgues, G.~Marques-Tavares, and M.~Schmaltz, ``{Evidence for dark
  matter interactions in cosmological precision data?},''
  \href{http://dx.doi.org/10.1088/1475-7516/2016/02/037}{{\em JCAP} {\bfseries
  1602} no.~02, (2016) 037},
\href{http://arxiv.org/abs/1507.04351}{{\ttfamily arXiv:1507.04351
  [astro-ph.CO]}}.

\bibitem{Cyr-Racine:2015ihg}
F.-Y. Cyr-Racine, K.~Sigurdson, J.~Zavala, T.~Bringmann, M.~Vogelsberger, and
  C.~Pfrommer, ``{ETHOS---an effective theory of structure formation: From dark
  particle physics to the matter distribution of the Universe},''
  \href{http://dx.doi.org/10.1103/PhysRevD.93.123527}{{\em Phys. Rev.}
  {\bfseries D93} no.~12, (2016) 123527},
\href{http://arxiv.org/abs/1512.05344}{{\ttfamily arXiv:1512.05344
  [astro-ph.CO]}}.

\bibitem{Buen-Abad:2017gxg}
M.~A. Buen-Abad, M.~Schmaltz, J.~Lesgourgues, and T.~Brinckmann, ``{Interacting
  Dark Sector and Precision Cosmology},''
  \href{http://dx.doi.org/10.1088/1475-7516/2018/01/008}{{\em JCAP} {\bfseries
  1801} no.~01, (2018) 008},
\href{http://arxiv.org/abs/1708.09406}{{\ttfamily arXiv:1708.09406
  [astro-ph.CO]}}.

\bibitem{Krall:2017xcw}
R.~Krall, F.-Y. Cyr-Racine, and C.~Dvorkin, ``{Wandering in the Lyman-alpha
  Forest: A Study of Dark Matter-Dark Radiation Interactions},''
  \href{http://dx.doi.org/10.1088/1475-7516/2017/09/003}{{\em JCAP} {\bfseries
  1709} no.~09, (2017) 003},
\href{http://arxiv.org/abs/1705.08894}{{\ttfamily arXiv:1705.08894
  [astro-ph.CO]}}.

\bibitem{Chacko:2016kgg}
Z.~Chacko, Y.~Cui, S.~Hong, T.~Okui, and Y.~Tsai, ``{Partially Acoustic Dark
  Matter, Interacting Dark Radiation, and Large Scale Structure},''
  \href{http://dx.doi.org/10.1007/JHEP12(2016)108}{{\em JHEP} {\bfseries 12}
  (2016) 108},
\href{http://arxiv.org/abs/1609.03569}{{\ttfamily arXiv:1609.03569
  [astro-ph.CO]}}.

\bibitem{Adshead:2012kp}
P.~Adshead and M.~Wyman, ``{Chromo-Natural Inflation: Natural inflation on a
  steep potential with classical non-Abelian gauge fields},''
  \href{http://dx.doi.org/10.1103/PhysRevLett.108.261302}{{\em Phys. Rev.
  Lett.} {\bfseries 108} (2012) 261302},
\href{http://arxiv.org/abs/1202.2366}{{\ttfamily arXiv:1202.2366 [hep-th]}}.

\bibitem{Anber:2012du}
M.~M. Anber and L.~Sorbo, ``{Non-Gaussianities and chiral gravitational waves
  in natural steep inflation},''
  \href{http://dx.doi.org/10.1103/PhysRevD.85.123537}{{\em Phys. Rev.}
  {\bfseries D85} (2012) 123537},
\href{http://arxiv.org/abs/1203.5849}{{\ttfamily arXiv:1203.5849
  [astro-ph.CO]}}.

\bibitem{Adshead:2013qp}
P.~Adshead, E.~Martinec, and M.~Wyman, ``{Gauge fields and inflation: Chiral
  gravitational waves, fluctuations, and the Lyth bound},''
  \href{http://dx.doi.org/10.1103/PhysRevD.88.021302}{{\em Phys. Rev.}
  {\bfseries D88} no.~2, (2013) 021302},
\href{http://arxiv.org/abs/1301.2598}{{\ttfamily arXiv:1301.2598 [hep-th]}}.

\bibitem{Adshead:2013nka}
P.~Adshead, E.~Martinec, and M.~Wyman, ``{Perturbations in Chromo-Natural
  Inflation},'' \href{http://dx.doi.org/10.1007/JHEP09(2013)087}{{\em JHEP}
  {\bfseries 09} (2013) 087},
\href{http://arxiv.org/abs/1305.2930}{{\ttfamily arXiv:1305.2930 [hep-th]}}.

\bibitem{Maleknejad:2011jw}
A.~Maleknejad and M.~M. Sheikh-Jabbari, ``{Gauge-flation: Inflation From
  Non-Abelian Gauge Fields},''
  \href{http://dx.doi.org/10.1016/j.physletb.2013.05.001}{{\em Phys. Lett.}
  {\bfseries B723} (2013) 224--228},
\href{http://arxiv.org/abs/1102.1513}{{\ttfamily arXiv:1102.1513 [hep-ph]}}.

\bibitem{Dimastrogiovanni:2012st}
E.~Dimastrogiovanni, M.~Fasiello, and A.~J. Tolley, ``{Low-Energy Effective
  Field Theory for Chromo-Natural Inflation},''
  \href{http://dx.doi.org/10.1088/1475-7516/2013/02/046}{{\em JCAP} {\bfseries
  1302} (2013) 046},
\href{http://arxiv.org/abs/1211.1396}{{\ttfamily arXiv:1211.1396 [hep-th]}}.

\bibitem{Adshead:2016omu}
P.~Adshead, E.~Martinec, E.~I. Sfakianakis, and M.~Wyman, ``{Higgsed
  Chromo-Natural Inflation},''
  \href{http://dx.doi.org/10.1007/JHEP12(2016)137}{{\em JHEP} {\bfseries 12}
  (2016) 137},
\href{http://arxiv.org/abs/1609.04025}{{\ttfamily arXiv:1609.04025 [hep-th]}}.

\bibitem{Adshead:2012qe}
P.~Adshead and M.~Wyman, ``{Gauge-flation trajectories in Chromo-Natural
  Inflation},'' \href{http://dx.doi.org/10.1103/PhysRevD.86.043530}{{\em Phys.
  Rev.} {\bfseries D86} (2012) 043530},
\href{http://arxiv.org/abs/1203.2264}{{\ttfamily arXiv:1203.2264 [hep-th]}}.

\bibitem{SheikhJabbari:2012qf}
M.~M. Sheikh-Jabbari, ``{Gauge-flation Vs Chromo-Natural Inflation},''
  \href{http://dx.doi.org/10.1016/j.physletb.2012.09.014}{{\em Phys. Lett.}
  {\bfseries B717} (2012) 6--9},
\href{http://arxiv.org/abs/1203.2265}{{\ttfamily arXiv:1203.2265 [hep-th]}}.

\bibitem{Maleknejad:2012fw}
A.~Maleknejad, M.~M. Sheikh-Jabbari, and J.~Soda, ``{Gauge Fields and
  Inflation},'' \href{http://dx.doi.org/10.1016/j.physrep.2013.03.003}{{\em
  Phys. Rept.} {\bfseries 528} (2013) 161--261},
\href{http://arxiv.org/abs/1212.2921}{{\ttfamily arXiv:1212.2921 [hep-th]}}.

\bibitem{Dimastrogiovanni:2016fuu}
E.~Dimastrogiovanni, M.~Fasiello, and T.~Fujita, ``{Primordial Gravitational
  Waves from Axion-Gauge Fields Dynamics},''
  \href{http://dx.doi.org/10.1088/1475-7516/2017/01/019}{{\em JCAP} {\bfseries
  1701} no.~01, (2017) 019},
\href{http://arxiv.org/abs/1608.04216}{{\ttfamily arXiv:1608.04216
  [astro-ph.CO]}}.

\bibitem{Fujita:2017jwq}
T.~Fujita, R.~Namba, and Y.~Tada, ``{Does the detection of primordial
  gravitational waves exclude low energy inflation?},''
  \href{http://dx.doi.org/10.1016/j.physletb.2017.12.014}{{\em Phys. Lett.}
  {\bfseries B778} (2018) 17--21},
\href{http://arxiv.org/abs/1705.01533}{{\ttfamily arXiv:1705.01533
  [astro-ph.CO]}}.

\bibitem{Adshead:2017xll}
P.~Adshead, J.~T. Giblin, and Z.~J. Weiner, ``{Non-Abelian gauge preheating},''
  \href{http://dx.doi.org/10.1103/PhysRevD.96.123512}{{\em Phys. Rev.}
  {\bfseries D96} no.~12, (2017) 123512},
\href{http://arxiv.org/abs/1708.02944}{{\ttfamily arXiv:1708.02944 [hep-ph]}}.

\bibitem{Piazza:2017bsd}
F.~Piazza, D.~Pirtskhalava, R.~Rattazzi, and O.~Simon, ``{Gaugid inflation},''
  \href{http://dx.doi.org/10.1088/1475-7516/2017/11/041}{{\em JCAP} {\bfseries
  1711} no.~11, (2017) 041},
\href{http://arxiv.org/abs/1706.03402}{{\ttfamily arXiv:1706.03402 [hep-th]}}.

\bibitem{Valenzuela:2016yny}
I.~Valenzuela, ``{Backreaction Issues in Axion Monodromy and Minkowski
  4-forms},'' \href{http://dx.doi.org/10.1007/JHEP06(2017)098}{{\em JHEP}
  {\bfseries 06} (2017) 098},
\href{http://arxiv.org/abs/1611.00394}{{\ttfamily arXiv:1611.00394 [hep-th]}}.

\bibitem{Blumenhagen:2017cxt}
R.~Blumenhagen, I.~Valenzuela, and F.~Wolf, ``{The Swampland Conjecture and
  F-term Axion Monodromy Inflation},''
  \href{http://dx.doi.org/10.1007/JHEP07(2017)145}{{\em JHEP} {\bfseries 07}
  (2017) 145},
\href{http://arxiv.org/abs/1703.05776}{{\ttfamily arXiv:1703.05776 [hep-th]}}.

\end{thebibliography}\endgroup

\end{document}